\documentclass[useAMS,usenatbib]{mnras}

\usepackage[english,english]{babel}
\usepackage{amsmath}
\usepackage{amssymb,amsfonts,textcomp}
\usepackage{array}
\usepackage{comment}
\usepackage{supertabular}
\usepackage{multirow}
\usepackage{placeins}
\usepackage{hhline}
\usepackage[usenames]{color}
\usepackage{soul}
\usepackage{textcomp}
\usepackage{graphicx}
\usepackage{float}
\usepackage{dcolumn}
\usepackage{bm}
\usepackage{comment}
\usepackage{color}
 \usepackage{float}
\usepackage{times}

\newcommand{\mP}{{\cal P}}

\newcommand{\dd}{\hbox{d}}

\definecolor{Blue}{rgb}{0,0.08,0.65}
\definecolor{Green}{rgb}{0.2,0.55,0.35}
\definecolor{grey}{rgb}{0.75,0.75,0.75}
\definecolor{Orange}{rgb}{1.0,0.5,0.15}
\definecolor{brown}{rgb}{0.7,0.25,0.0}
\definecolor{Pink}{rgb}{1.0,0.5,0.5}
\definecolor{darkerred}{rgb}{0.8,0,0}
\definecolor{darkerblue}{rgb}{0,0,0.8}
\definecolor{darkergreen}{rgb}{0,0.5,0}

\def\Cora#1{\noindent{\color{darkergreen}#1}}

\def\CoraToDo#1{\noindent{\color{darkergreen}[{\bf To Do: #1}]}}

\def\Cora#1{#1}
\def\CoraToDo#1{}

\def\citefuture#1{\noindent{\color{blue}#1}}

\begin{document}

\title[Counts-in-cells Cosmology]{Fisher for complements: Extracting cosmology and neutrino mass  from the counts-in-cells PDF}

\author[C. Uhlemann et al.]{
\parbox[t]{\textwidth}
{Cora Uhlemann$^{1,2}$, Oliver Friedrich$^{3,4}$, Francisco Villaescusa-Navarro$^{5,6}$,\\ Arka Banerjee$^{7,8,9}$, Sandrine Codis$^{10}$}
\vspace*{10pt}\\
$^{1}$ {Centre for Theoretical Cosmology, DAMTP, University of Cambridge, CB3 0WA Cambridge, United Kingdom}\\
$^{2}$ {Fitzwilliam College, University of Cambridge, CB3 0DG Cambridge, United Kingdom}\\
$^3$ {Kavli Institute for Cosmology, University of Cambridge, CB3 0HA Cambridge, United Kingdom}\\
$^{4}$ {Churchill College, University of Cambridge, CB3 0DS Cambridge, United Kingdom}\\
$^5$ Department of Astrophysical Sciences, Princeton University, Peyton Hall, Princeton NJ 08544-0010, USA\\
$^6$ Center for Computational Astrophysics, Flatiron Institute, 162 5th Avenue, 10010, New York, NY, USA\\
$^7$ Kavli Institute for Particle Astrophysics and Cosmology, Stanford University, 452 Lomita Mall, Stanford, CA 94305, USA\\
$^8$ Department of Physics, Stanford University, 382 Via Pueblo Mall, Stanford, CA 94305, USA\\
$^9$ SLAC National Accelerator Laboratory, 2575 Sand Hill Road, Menlo Park, CA 94025, USA\\
$^{10}$ CNRS \& Sorbonne Universit\'e, UMR 7095, Institut d'Astrophysique de Paris, 75014, Paris, France\\
}

\maketitle
\begin{abstract}
We \Cora{comprehensively analyse} the cosmology dependence of counts-in-cell statistics. We focus on the shape of the one-point probability distribution function (PDF) of the matter density field at mildly nonlinear scales. Based on large-deviation statistics, we parametrise the cosmology dependence of the matter PDF in terms of the linear power spectrum, the growth factor, the spherical collapse dynamics, and the nonlinear variance. We extend our formalism to include massive neutrinos, finding that the \Cora{total} matter PDF is highly sensitive to the total neutrino mass $M_\nu$ and can disentangle it from the clustering amplitude $\sigma_8$.

Using more than a million PDFs extracted from the Quijote simulations, we determine the response of the matter PDF to changing parameters in the $\nu\Lambda$CDM model and successfully cross-validate the theoretical model and the simulation measurements. We present the first $\nu\Lambda$CDM Fisher forecast for the matter PDF at multiple scales and redshifts, and its combination with the matter power spectrum. We establish that the matter PDF and the matter power spectrum are highly complementary at mildly nonlinear scales. The matter PDF is particularly powerful for constraining the matter density $\Omega_m$, clustering amplitude $\sigma_8$ and the total neutrino mass $M_\nu$. Adding the \Cora{mildly nonlinear} matter PDF to the \Cora{mildly nonlinear} matter power spectrum improves constraints on $\Omega_m$ by a factor of 5 and $\sigma_8$ by a factor of 2 \Cora{when considering the three lowest redshifts}. 
In our joint analysis of the matter PDF and matter power spectrum at three redshifts, the total neutrino mass is constrained to better than 0.01 eV with a total volume of 6 (Gpc/h)$^3$. 
We discuss how density-split statistics can be used to translate those encouraging results for the matter PDF into realistic observables in galaxy surveys.

 \end{abstract}

\section{Introduction}
\label{sec:intro}
The $\Lambda$CDM model of a universe filled with a cosmological constant and cold dark matter has proved to be an extraordinarily successful paradigm. This concordance model is capable of explaining a large variety of cosmological observations, from the anisotropies of the cosmic microwave background \citep{Planck18} to the spatial distribution of galaxies at low redshift. This ``Standard Model" of cosmology can be extended by a few additional parameters representing fundamental physics quantities, e.g. the total neutrino mass, the equation-of-state of dark energy, as well as the amplitudes and shapes of primordial non-Gaussianity. Within current observational limits, the latter two extensions are consistent with a cosmological constant and Gaussian initial conditions, respectively. By contrast, we already know that at least two neutrino families must have a nonzero mass in order to explain the observations of neutrino oscillations \citep{Becker-Szendy92,Fukuda98,Ahmed04}. However, both the total mass of the three neutrino mass eigenstates, as well as the hierarchy of these mass states are still unknown, and can provide crucial hints to physics beyond the Standard Model of particle physics. The minimal total neutrino mass in the presence of known mass splittings is $M_\nu=\sum m_\nu\geq 0.056$ eV for a normal hierarchy and $M_\nu\geq 0.095$ eV for an inverted hierarchy \citep{Lesgourgues06}. Cosmological observations from the cosmic microwave background (CMB) already provide upper bounds on the sum of neutrino masses. Upcoming Stage-IV CMB polarization experiments and large-scale structure surveys like Euclid \citep{Euclid}, LSST \citep{LSST}, DESI \citep{DESI}, \Cora{PFS \citep{PFS} and WFIRST \citep{WFIRST}} will seek to detect the signature of total neutrino mass in galaxy clustering and weak lensing statistics conclusively \citep{Abazajian15}.

Constraining the value of all the fundamental physical parameters mentioned above is one of the most important goals of modern precision cosmology. For that reason, many different cosmological missions are going to survey the sky to collect data that allow to constrain the value of the cosmological parameters as accurately and precisely as possible. 
Unfortunately, a large fraction of the raw data available from these surveys cannot be converted into tighter constraints on cosmological parameters, because the information is embedded on \Cora{small scales and in non-Gaussian observables for which accurate theory predictions are challenging \citep{Scoccimarro99,Cooray02,Rimes05,Neyrinck06,Nishimichi16}.} These scales are typically in the mildly to fully nonlinear regime due to gravitational collapse, and for this reason, analytical predictions based on perturbation theory are invalid. At high redshifts, the matter density fluctuations  are close to a Gaussian random field, which is fully characterised by its power spectrum or two-point correlation function. However, as non-linear gravitational clustering proceeds, the density field becomes non-Gaussian. The information that initially was contained in the power spectrum, leaks into higher-order moments of the density field \citep{Peebles,Bernardeau02}. Thus, if the analysis of large-scale structure survey data is limited to the power spectrum, a significant amount of information is unused.
In the non-linear regime, it is unknown what fraction of the information is contained in each statistic. In this paper, we focus our attention on one of the simplest statistics of a three-dimensional field: the probability distribution function (hereafter PDF) of the  matter density field smoothed on a given scale. 

Empirically, it has been found that one-point matter density PDFs are close to lognormal  \citep{ColesJones91,Kayo01}, with further improvements by skewed lognormal models \citep{Colombi94,Repp18} or generalised normal distributions \citep{Shin17}. While the lognormal model is only a crude approximation, it highlights that by limiting the cosmological analysis to two-point statistics one inevitably misses a large amount of information encoded in a non-Gaussian field. This idea has been formalised by considering the power spectrum of log-densities \citep{Neyrinck09,Seo11,Wolk15} and more generally sufficient statistics \citep{Carron14}. 

In order to unlock additional information in upcoming large-scale surveys like Euclid, we need accurate predictions for non-Gaussian statistics {\it and} their dependence on cosmology. Having multiple complementary large-scale structure probes is particularly important for breaking degeneracies when jointly constraining fundamental physics such as neutrino masses, modified gravity and dynamical dark energy \citep{Font-Ribera14,Sahlen19}.
Counts-in-cells statistics like density PDFs are ideal candidates for this purpose, as they can be easily measured in surveys and their cosmology dependence can be accurately predicted. Recently, this idea has been applied to surveys like DES \citep{DES} and KiDS \citep{KiDS} using galaxy troughs and ridges \citep{Gruen16troughs,Brouwer18troughs}, moments of galaxy density and lensing convergence \citep{Bel14,Petri15,Clerkin16,Salvador18,Gatti19} and density-split statistics \citep{Friedrich18, Gruen18}. \Cora{In particular, \cite{Gruen18} have shown} that density-split statistics from joint counts- and lensing-in-cells yields cosmological constraints competitive with two-point function measurements. \Cora{At the same time, density-split statistics recover additional information about higher-order moments of the density field and the relation between galaxy and matter density. Their combined use of galaxy counts and lensing allowed them to connect models of the matter density PDF to photometric data of the galaxy density field, demonstrating that the methodology presented here can be carried over to real data analyses.} 

In this work, we combine insights from an analytical model for the matter PDF based on large-deviation statistics and spherical collapse \citep{Bernardeau14,Uhlemann16} with measurements from the large suite of the Quijote simulations \citep{Quijote}. We quantify, for the first time, the amount of cosmological information encoded in the matter density PDF at multiple scales and redshifts, and compare it to the one from the matter power spectrum. In our analysis, we take into account the full covariance between density PDFs measured at different scales and its cross-covariance with the power spectrum. To perform the full analysis we extracted more than a million matter density PDFs from the Quijote suite for different cosmologies and made them publicly available.\footnote{The PDFs can be accessed as part of the public data release of the Quijote simulations, see \href{https://github.com/ franciscovillaescusa/Quijote-simulations}{github.com/ franciscovillaescusa/Quijote-simulations}.} We consider the $\nu\Lambda$CDM model, which extends $\Lambda$CDM by including the sum of neutrino masses, $M_\nu$, as parameter. In particular, we use \Cora{what we call} derivative simulations, which vary exactly one parameter in the $\nu\Lambda$CDM model compared to the fiducial model with magnitudes given in Table~\ref{tab:models}.

Our paper is structured as follows: in Section~\ref{sec:matterPDF}, we describe a theoretical model for the PDF of matter densities in spheres and discuss the physical ingredients that determine the resulting shape of the PDF. In Section~\ref{sec:matterPDFnu}, we generalise the PDF model to account for the presence of massive neutrinos. We then cross-validate our theoretical predictions with measurements from the Quijote simulation suite in Section~\ref{sec:validation}. In Section~\ref{sec:Fisher} we present a Fisher analysis that demonstrates the constraining power of the matter PDF for $\Lambda$CDM parameters and total neutrino mass. Section~\ref{sec:conclusion}  provides a conclusion and an outlook to further work.

\begin{table}
\centering
\begin{tabular}{l|c|c|c|c|c|c}
& $\sigma_8$ & $\Omega_m$ & $\Omega_b$& $n_s$ & $h$ & $M_\nu [eV]$ \\\hline
fid & 0.834 & 0.3175 & 0.049 & 0.9624 & 0.6811 & 0 \\
$\Delta$ & 0.015 & 0.01 & 0.002 & 0.02 & 0.02 &  0.1, 0.2, 0.4 \\
\end{tabular}
\caption{Cosmological model parameters for the set of $\nu\Lambda$CDM Quijote simulations under consideration in this paper.}
\label{tab:models}
\end{table}     

\section{Physical ingredients for the PDF of matter densities in spheres}
\label{sec:matterPDF}

Large deviation statistics provides a means to compute the probability distribution function (PDF) of nonlinear matter densities in spheres (that is to say density smoothed with a top-hat kernel). In the present paper, we limit ourselves to Gaussian initial conditions, but primordial non-Gaussianities can also be implemented in the formalism \cite[][\citefuture{Friedrich et al. 2019}]{Uhlemann18pNG}. For Gaussian initial conditions, the PDF $\mP(\delta_L)$ of the linear matter density contrast $\delta_L$ in a sphere of radius $r$ is fully specified by the linear variance at that scale
\begin{align}
\label{eq:PDFlin}
\mathcal P_r^{\rm ini}(\delta_L) &= \sqrt{\frac{1}{2\pi\sigma_L^2(r)}} \exp\left[-\frac{ \delta_L^2}{2\sigma^2_{\rm L}(r)}\right]\,.
\end{align}
The linear variance at scale $r$ is obtained from an integral over the linear power spectrum with spherical top-hat filter in coordinate space
\begin{equation}
\sigma^2_{\rm L}(r) 
= \int \, \frac{\dd k}{2\pi^2} P_{\rm L}(k)  k^2 W^2_{\rm 3D}(k r)\,,
\label{eq:defSigma2lin}
\end{equation}
where  $W_{\rm 3D}(k)$ is the the Fourier transform of the 3D spherical top-hat kernel
 \begin{equation}
 \label{eq:filter}
W_{\rm 3D}(k)=3\sqrt{\frac{\pi}{2}}\frac{J_{3/2}(k)}{k^{3/2}}\,,
\end{equation}
and $J_{3/2}(k)$ is the Bessel function of the first kind of order $3/2$. 

To describe the impact of nonlinear gravitational dynamics on the shape of the initially Gaussian matter PDF, it is \Cora{informative} to look at \Cora{the exponential decay of the PDF with increasing density contrast.} 
The decay-rate function is the negative argument of the exponent in equation~\eqref{eq:PDFlin} and reads
\begin{align}
\label{eq:PsiL}
\Psi_r^{\rm ini}(\delta_L)&= \frac{ \delta_L^2}{2\sigma^2_{\rm L}(r)}\,.
\end{align}

According to the contraction principle of large deviation statistics \citep{LDPinLSS}, the exponential decay of the PDF of final densities (at scale $R$ and redshift $z$) can be obtained from the initial one by inserting the most likely mapping between linear and nonlinear densities in spheres and their radii into
\begin{align}
\label{eq:PsiNL}
 \Psi_R(\rho)&
= \frac{\sigma^2_{\rm L}(R)}{\sigma_{\rm NL}^2(z,R)} \frac{\delta_L(\rho)^2}{2\sigma^2_{\rm L}(R\rho^{1/3})}\,,
\end{align}
where $\sigma_{\rm NL}$ is the nonlinear variance of the density at scale $R$ and redshift $z$. Thanks to the symmetry of the statistics and statistical isotropy, the most probable evolution of densities in spheres $\delta_L(\rho)$ can be accurately approximated by spherical collapse and the initial and final radii are related by mass conservation $r=R\rho^{1/3}$. This argument can be made more precise by writing the final PDF as a path integral over all possible histories relating final densities to linear densities. As shown in \cite{Bernardeau94,Valageas02}, the dominant contribution to this integral comes from spherical collapse, which is a saddle point of the corresponding functional integrals. Recently, non-perturbative effects not captured by the saddle point have been analysed analytically in one dimension \citep{vanderWoude017} and estimated using a path-integral approach based on perturbation theory with a renormalisation of small-scale physics \citep{Ivanov19}. In practice, non-perturbative effects mostly renormalise the nonlinear variance entering equation~\eqref{eq:PsiNL} and higher order reduced cumulants are still reliably predicted by spherical collapse.

From the decay-rate function in equation~\eqref{eq:PsiNL} one can compute the cumulant generating function via a Legendre transform. Then, one obtains the final PDF from the cumulant generating function via an inverse Laplace transform that can be computed numerically \citep{Bernardeau14,Bernardeau15,Friedrich18}. For a standard $\Lambda$CDM universe, there are only three ingredients that enter this theoretical model for the matter PDF,
\begin{enumerate}
    \item the scale-dependent linear variance and the linear growth,
    \item the mapping between initial and final densities in spheres,
    \item and the nonlinear variance\,.
\end{enumerate}
We will discuss each of these ingredients in the following three subsections. Then we generalise the formalism to include the impact of massive neutrinos in Section~\ref{sec:matterPDFnu}. 

While in general, the transformation from the decay-rate of the PDF~\eqref{eq:PsiNL} to the PDF itself has to be evaluated numerically, one can find an excellent analytical approximation using a saddle-point technique. As shown in \cite{Uhlemann16}, the inverse Laplace transform can be evaluated 
with a saddle-point approximation for the log-density $\mu=\ln\rho$, the expression reads
\begin{subequations}
\label{eq:PDFpred}
\begin{align}
\label{eq:PDFfromPsilog}
\mP_{R}(\rho) = \sqrt{\frac{\Psi''_R(\rho)+\Psi'_R(\rho)/\rho}{2\pi}} \exp\left(-\Psi_R(\rho)\right)\,.
\end{align}
Because of the use of the logarithmic variable, one has to ensure the correct mean density $\langle\rho\rangle=1$ by specifying the mean of the log-density $\langle\ln\rho\rangle$. This can be implemented by properly rescaling the `raw' PDF~\eqref{eq:PDFfromPsilog}
\begin{align}
\label{eq:PDFnorm}
\hat\mP_{R}(\rho) &= \mP_{R}\left(\rho \cdot \frac{\langle\tilde\rho\rangle}{\langle 1\rangle}\right)  \cdot \frac{\langle\tilde\rho\rangle}{\langle 1\rangle^2}\,,
\end{align}
\end{subequations}
where $\langle f(\tilde\rho) \rangle=\int d\tilde\rho\,, f(\tilde\rho) \mP(\tilde\rho)$. Note that the normalization is only necessary because the inverse Laplace transform is not computed explicitly, which would automatically preserve the normalisation and ensure a correct mean. Since the saddle-point approximation makes use of the log-density, the nonlinear variance that enters the decay-rate function~\eqref{eq:PsiNL} is the one of the logarithmic density $\mu=\ln \rho$.\footnote{Note that the relevant variable is the logarithm of the smoothed density $\rho$ in a sphere of radius $R$, not the smoothed logarithm of the density.} If one computes the PDF numerically using an inverse Laplace transform (as was for instance performed in \cite{Bernardeau15}, the nonlinear variance of the density enters directly. Since the final matter density PDFs obtained from both approaches agree very well \citep{Uhlemann16}, the shape of the density PDF allows to translate the variances to each other.

\subsection{The scale-dependence of the linear variance}
\label{sec:scaledepvar}
The amplitude of the linear density fluctuation at different scales are calculated from the CAMB \citep{CAMB} linear power spectrum at $z=0$. To obtain good agreement with the finite resolution simulations, the integral for the linear variance should be cut at the Nyquist frequency $k_{\rm Ny}=\pi\cdot N_{\rm mesh}/L_{\rm box}$, which is around $1.6 h/$Mpc for the box size $L=1000$ Mpc/h and $N_{\rm mesh}=512$.
In Figure~\ref{fig:sigmalincomparison} we show how different cosmologies with fixed $\sigma_8$ impact the scale-dependence of the linear variance. While changes in $\sigma_8$ simply modify the overall amplitude, changes in $\Omega_m$, $\Omega_b$, $n_s$ and $h$ modify the variance in a scale-dependent way. 
As expected, a change in the primordial spectral index $n_s$ results in a constant shift of the logarithmic derivative $d\log \sigma^2_L(R)/d\log R$. In contrast, changes in the matter and baryon densities, $\Omega_m$ and $\Omega_b$, as well as the Hubble parameter $h$ induce an additional scale-dependent running of the spectral index.

While it is possible to numerically determine the precise $\Lambda$CDM parameter dependence of the linear power spectrum from CAMB \citep{CAMB} or CLASS \citep{CLASS}, a closed-form expression for forecasts and data analysis is desirable. In Figure~\ref{fig:sigmalinEisensteinHu} in Appendix~\ref{app:linPSvar} we show that the linear variance computed from the Eisenstein-Hu formula for the linear power spectrum \citep{EisensteinHu98} is accurate at about 0.5\% for the fiducial model and the derivative simulations.

\subsubsection*{Impact on the matter PDF and its cumulants}

\begin{figure}
\includegraphics[width=1.\columnwidth]{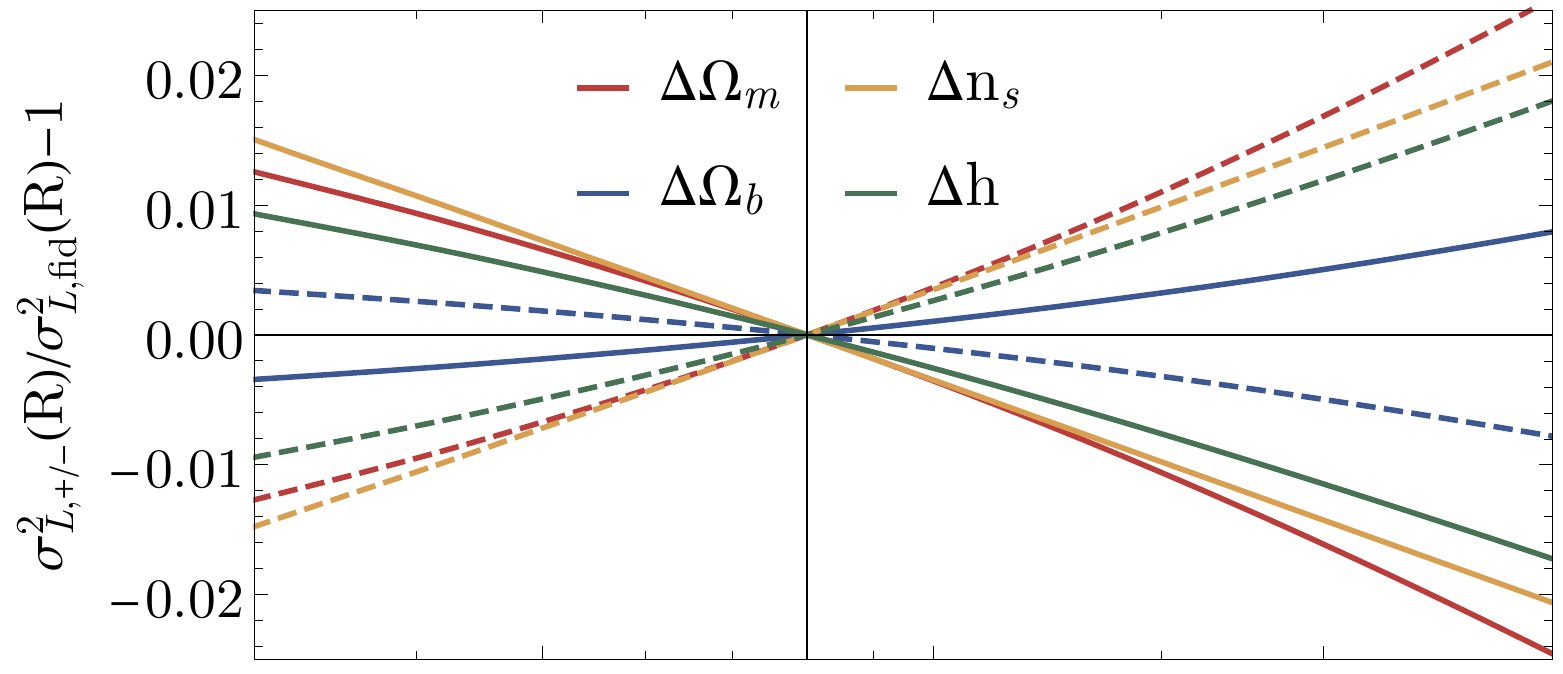}
\includegraphics[width=1.\columnwidth]{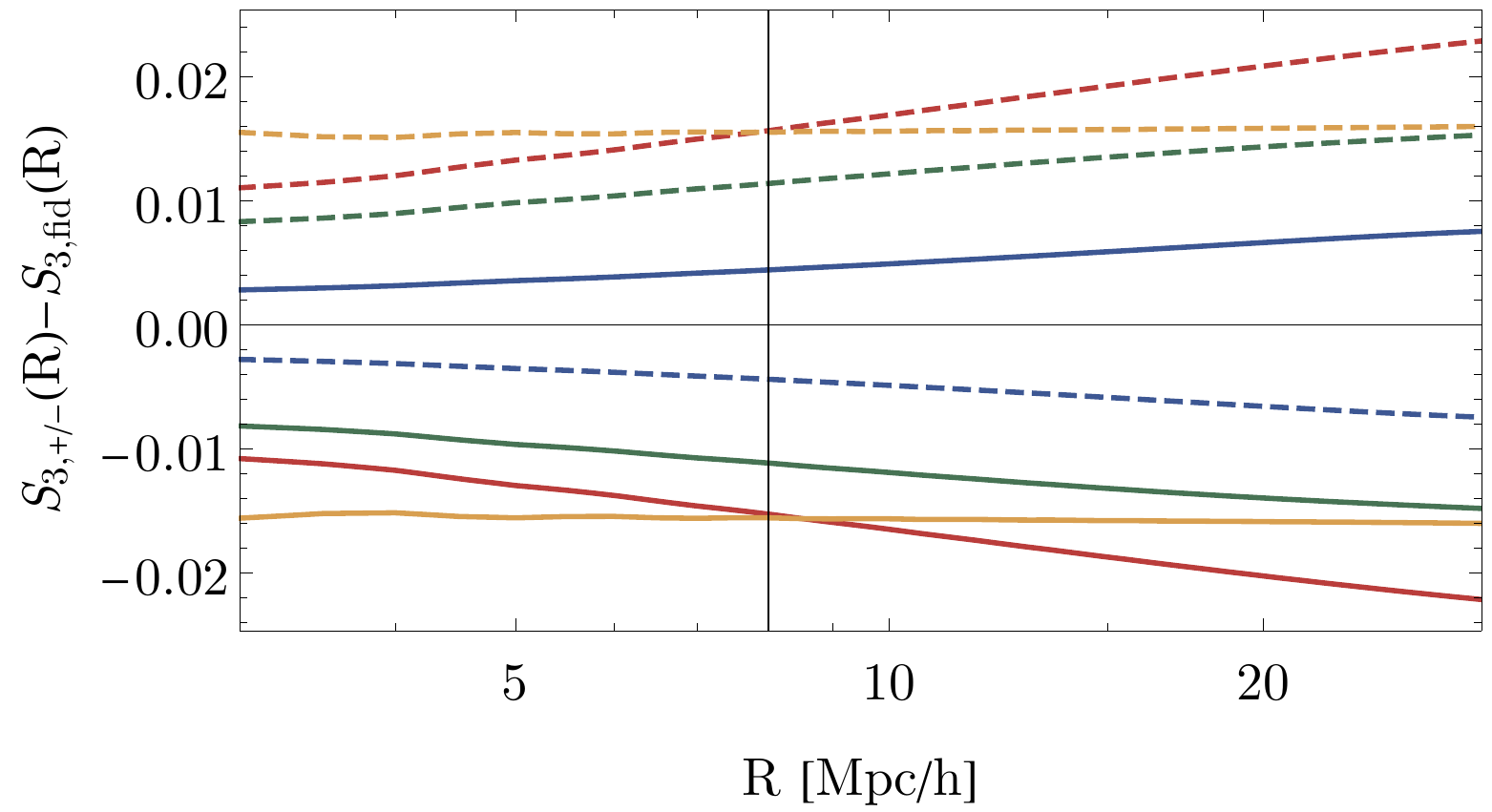}

   \caption{Comparison between 
   the linear variance $\sigma_L^2(R)$ \Cora{as computed from CAMB} at $z=0$ for $\Lambda$CDM cosmologies with varying $\Omega_m$ (red), $\Omega_b$ (blue), $n_s$ (yellow) and $h$ (green) with positive sign (solid lines) and negative sign (dashed lines) as indicated in Table~\ref{tab:models} with fixed $\sigma_8$. The lower panel shows the differences in the predicted reduced skewness $S_3$ for the different cosmologies according to equation~\eqref{eq:S3pred}.} 
   \label{fig:sigmalincomparison}
\end{figure}

The scale-dependence of the linear variance determines the exponential decay of the PDF according to equation~\eqref{eq:PsiNL}. With different densities $\rho$, one scans the linear variance at scales $R\rho^{1/3}$ in a range of values around the radius $R$. One can also understand this behaviour from the tree order perturbation theory prediction for the reduced skewness, $S_3$, of the density at scale $R$. In an EdS universe, this quantity is determined by the first logarithmic derivative of the linear variance \citep{Bernardeau94skewkurt}
\begin{align}
\label{eq:S3pred}
    S_3(R) = \frac{\langle\delta^3(R)\rangle}{\langle\delta^2(R)\rangle^2} =\frac{34}{7} + 
    \frac{d\log \sigma^2_L(R)}{d\log R}\,.
\end{align}
\Cora{Note that differences between the reduced skewness amplitude before smoothing in EdS, which predicts $34/7$ and the $\Lambda$CDM spherical collapse prediction are below 0.0017 for the  changes in $\Omega_m$ considered here. Hence, the main change in the reduced skewness is indeed caused by a difference in the scale-dependent variance.}
\Cora{Note that the clustering amplitude $\sigma_8$ cancels in the logarithmic derivative and does not change the reduced skewness.}
We show how \Cora{the other $\Lambda$CDM parameters impact the ratio of linear variances and hence the differences in the reduced skewness in Figure~\ref{fig:sigmalincomparison}. The lower panel demonstrates how changing cosmological parameters offsets the reduced skewness. When focusing on a single radius $R$, one can only detect the overall offset, but not distinguish between the cosmological parameters. In particular, the skewness at one scale} cannot distinguish a constant tilt caused by $n_s$ from a running of the tilt induced by $\Omega_m$, $\Omega_b$ and $h$. This degeneracy is partially broken by considering the full PDF, whose shape is also sensitive to a combination of the reduced kurtosis $S_4$ and higher order cumulants, which depend on higher order logarithmic derivatives. In the presence of irreducible noise such as cosmic variance, the amount of this additional information is however limited. \Cora{This is why it is important to jointly consider matter density PDFs at multiple radii. In this study, we limit ourselves to modelling the PDFs at different radii individually, but include their cross-covariance, which captures some part of the joint one-point PDF.}

\subsubsection*{Growth of density fluctuations}
In the linear regime and for a $\Lambda$CDM model, the amplitude of density fluctuations grows with the growth rate such that
\begin{equation}
\label{eq:siggrowth}
    \sigma^2(z,R)=D^2(z)\sigma_L^2(R)\,.
\end{equation} 
 For a flat $\Lambda$CDM universe, the linear growth of structure depends only on the matter density $\Omega_m$ and a closed form is known \citep{Matsubara95}
 \begin{align}
 \label{eq:Dsimp}
 D(z) = \frac{\sqrt{1 - \Omega_m + \Omega_m (1 + z)^3}\, {}_{2}F_1\left[\frac{5}{6}, \frac{3}{2}, \frac{11}{6}, \frac{-1 + \Omega_m}{\Omega_m (1 + z)^3}\right])}{(1 + z)^{5/2}\, {}_{2}F_1\left[\frac{5}{6}, \frac{3}{2}, \frac{11}{6},\frac{-1 + \Omega_m}{\Omega_m}\right]}\,,
 \end{align}
 where ${}_{2}F_1$ is the hypergeometric function. At redshift $z=1$, a change in the matter density $\Omega_m$ of $\pm$ 3\% around the fiducial value leads to $\mp$ 1\% difference in the square of  the growth function. From the overall amplitude of density fluctuations in equation~\eqref{eq:siggrowth}, we see that for the matter PDF at a single nonzero redshift $z>0$, the linear growth function controlled by $\Omega_m$ becomes degenerate with the overall amplitude $\sigma_8$\Cora{, which is not the case at $z=0$, because $D(z=0)=1$}. This degeneracy can be broken by performing an analysis in multiple redshift slices, which also helps to disentangle the matter density $\Omega_m$ from the spectral index $n_s$, as we demonstrate in a Fisher forecast shown in Figure~\ref{fig:degeneracybreakingdiffz}.
 
 In Appendix~\ref{app:linvarapprox}, we discuss the impact of a  dark energy equation of state beyond a cosmological constant. In Figure~\ref{fig:Dofzcomparison} in Section~\ref{sec:conclusion}, we compare the dependence of the growth-rate on the matter density $\Omega_m$ for a flat $\Lambda$CDM universe to a change in the dark energy equation of state parameter $w_0$ and to the modifications induced by the presence of massive neutrinos. 

\subsection{Spherical collapse dynamics}
In an Einstein-de Sitter (EdS) universe, there is a parametric solution for spherical collapse dynamics \citep{Peebles}, which relates the linear density contrast $\delta_L$ to the nonlinear density $\rho_{\rm NL}$
\begin{subequations}
\label{eq:SphericalCollapseEdS}
\begin{equation}
\delta_L \geq 0 \; : \;\; \left\{ \begin{array}{rl}
\rho_{\rm NL} & = {\displaystyle \frac{9}{2} \; \frac{(\theta-\sin
\theta)^2}{(1-\cos \theta)^3} } \\ \\ 
\delta_L & = {\displaystyle
\frac{3}{20} \; \left[ 6 (\theta-\sin \theta) \right]^{2/3} } \end{array}
\right.
\label{Fcol1}
\end{equation}
and 
\begin{equation}
\delta_L < 0 \; : \;\; \left\{ \begin{array}{rl}
\rho_{\rm NL} & = {\displaystyle \frac{9}{2} \; \frac{(\sinh
\eta-\eta)^2}{(\cosh \eta-1)^3} } \\ \\ 
\delta_L & = {\displaystyle -
\frac{3}{20} \; \left[ 6 (\sinh \eta-\eta) \right]^{2/3} } \end{array}
\right. \,,
\label{Fcol2}
\end{equation}
\end{subequations} 
where $\theta\in [0,2\pi]$ is the development angle and $\eta$ its counterpart for an open universe and both parameters can be eliminated from the relation. \Cora{Note that these formulae can be extended to any background with zero cosmological constant \citep[see Appendix A of][]{Bernardeau02}.}
Let us note that the initial and final radii of the sphere are related by mass conservation $r^3 = R^3 \rho$. 

For simplicity, one can rely on an approximate explicit parameterisation for spherical collapse in an EdS universe, given by
\begin{equation}
\label{eq:SCnu}
\rho_{{\rm NL},\nu}(\delta_L)=\left(1-\frac{\delta_L}{\nu}\right)^{-\nu}\,,
\end{equation}
where $\delta_L$ is the linear density at redshift zero. The parameter $\nu$ controls the amplitude of the skewness before smoothing $3(1+1/\nu)$ and can be matched to the prediction in equation~\eqref{eq:S3pred}, yielding $\nu=21/13$. Originally, this parametric form has been suggested in \cite{Bernardeau94}, with approximately $\nu=1.5$\Cora{, which becomes exact for $\Lambda$=0 in the limit of $\Omega_m\rightarrow 0$ and drives the shape of the PDF in low density regions}. In excursion-set inspired models, usually the critical linear density for collapse is used, setting $\nu=\delta_c=1.686$ \citep{LamSheth08}. 

\begin{figure}
\centering
\includegraphics[width=0.95\columnwidth]{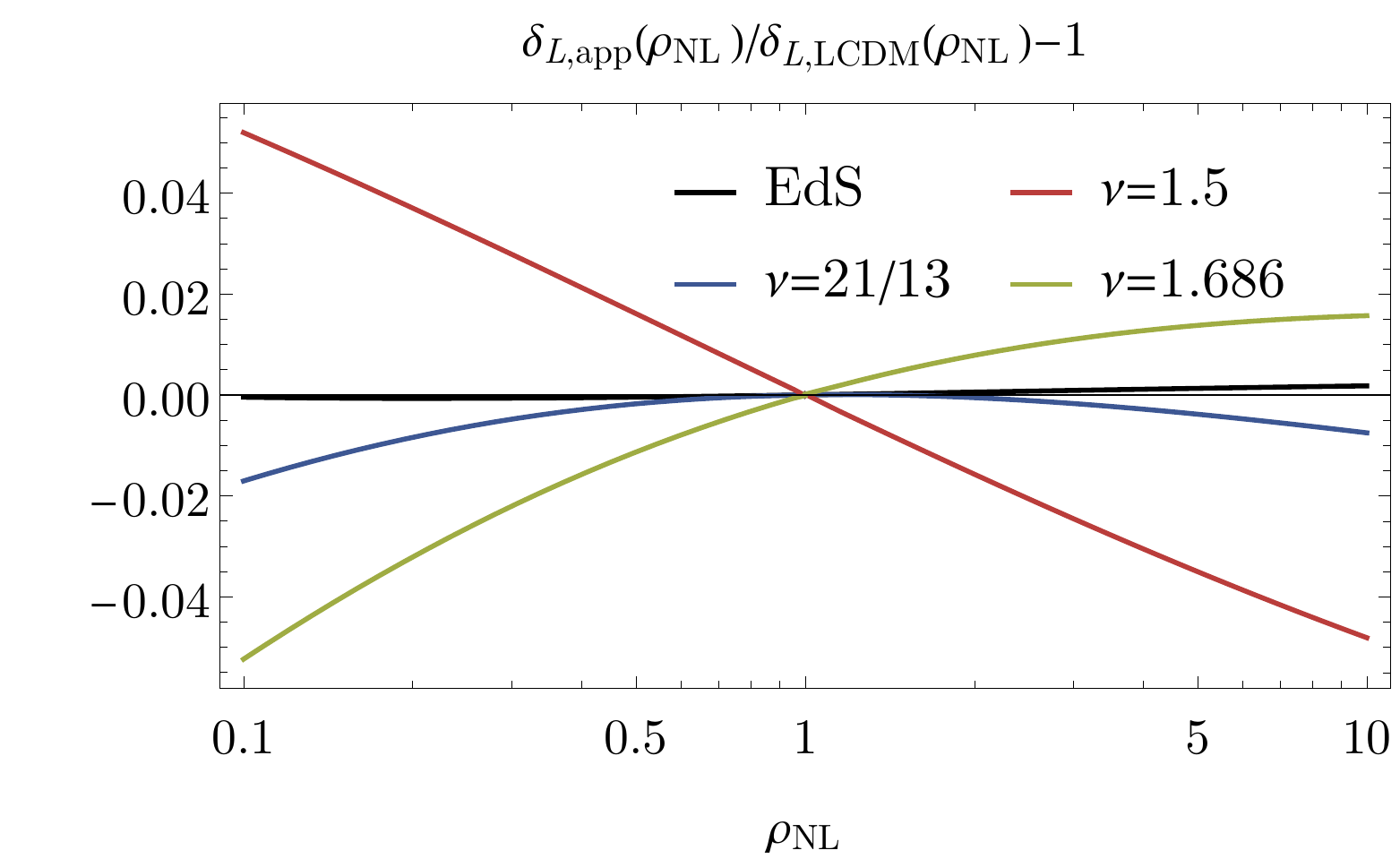}
   \caption{Comparison between the parametric spherical collapse dynamics for EdS given by equation~\eqref{eq:SphericalCollapseEdS} (black) and the $\nu$-parametrization given by equation~\eqref{eq:SCnu} 
   for the parameters $\nu=1.5$ (red), $\nu=21/13$ (blue) and $\nu=1.686$ (green) compared to the numerical solution for spherical collapse in a $\Lambda$CDM universe.}
   \label{fig:SCcomparison}
\end{figure}

In Figure~\ref{fig:SCcomparison}, we compare the parametric EdS form~\eqref{eq:SphericalCollapseEdS} (black line) and the $\nu$-parameterisation~\eqref{eq:SCnu} with different parameters (blue, green and red lines) to the numerical solution of the $\Lambda$CDM spherical collapse dynamics, \Cora{described in \citet[see Appendix A]{Friedrich18}}. We find that the parametric solution for EdS approximates the numerical $\Lambda$CDM solution extremely well, with sub-percent residuals in the range of relevant densities $\rho\in [0.1,10]$. For the fiducial cosmology ($\Omega_m=0.3175$), the deviations are less than 0.2\%. As expected, the agreement improves with increasing $\Omega_m$. But even for a matter density of $\Omega_m\simeq 0.21$, differences stay below 0.3\%.  Therefore, for the purpose of constraining cosmology with the bulk of the matter PDF, the cosmology dependence of spherical collapse can be neglected. In what follows, we rely on the parametric EdS spherical collapse solution~\eqref{eq:SphericalCollapseEdS} for the theoretically predicted PDF. 

Note that spherical collapse is potentially affected more seriously by dynamical dark energy \citep{Mota04,Abramo07,Pace10,Mead17} or modified gravity  \citep{Schaefer08,Barreira13,Kopp13,Cataneo16}, which are beyond the scope of this work. \Cora{The impact of massive neutrinos on spherical collapse is discussed in Section~\ref{sec:matterPDFnu}.}

\subsection{The nonlinear variance}
In full analogy to its linear counterpart, the nonlinear variance is defined in terms of the nonlinear power spectrum
\begin{equation}
\sigma^2_{\rm NL}(z,R) = \int\frac{\dd^{3} k}{(2\pi)^{3}}P_{\rm NL}(z,k) W_{\rm 3D}(k R)^2\,.
\label{eq:defSigma2nonlin}
\end{equation}
While the reduced cumulants are well predicted by spherical collapse dynamics, the nonlinear variance cannot be inferred accurately enough. In the following we discuss how one can efficiently parameterise the cosmology dependence of this quantity, and how accurately this quantity can be predicted using the nonlinear matter power spectrum from fitting functions (\textsc{halofit} from \citet{halofit} and \textsc{respresso} from \citet{Nishimichi17}) or perturbative techniques (2-loop SPT and RegPT from \cite{Taruya2012,Osato19}).

\subsubsection{Cosmology dependence}
For lognormal fields with unit mean, the variance of the logarithm of the smoothed density is related to the density variance as \citep{ColesJones91}
\begin{equation}
\label{eq:siglog}
\sigma_{\ln \rho}^2(z,R) = \ln\left[1 + \sigma^2_{\rho}(R,z)\right]\,.
\end{equation}
To estimate the impact of changing cosmology, we will use the linear density variance in this relation.\footnote{As discussed in \cite{ReppSzapudi16}, one could introduce a free parameter in this simple approximation in order to improve the matching on small scales. Note that \cite{ReppSzapudi16} use this approach for obtaining expression the power spectrum of the log-density. Consequently, their relationships are formulated for the smoothed log-density rather than the logarithm of the smoothed density that we consider here.}
In the absence of massive neutrinos, we use a factorisation of the linear variance into the growth function and the scale-dependent linear variance extrapolated to today, $\sigma_{\rm L}(R,z)=D(z)\sigma_{\rm L}(R)$.

This simple relation proves useful for parameterising the cosmology dependence of the nonlinear log-variance, which is induced by changes in the linear variance (computed from the linear power spectrum) and the growth rate $D(z)$. We use this relation to predict the scaling of the nonlinear variance for cosmologies with changed cosmological parameters from the measured variance at the fiducial cosmology
\begin{equation}
\label{eq:siglogcos}
    \sigma_{\ln \rho,\rm cos}^2(z,R) = \frac{\ln\left[1 + \sigma^2_{\rm L, cos}(R,z)\right]}{\ln\left[1 + \sigma^2_{\rm L, fid}(R,z)\right]} \sigma_{\ln \rho, \rm fid}^2(z,R) \,.
\end{equation}
We have checked that this yields residuals smaller than $0.1$\% for radii $R=10, 15, 20$ Mpc$/h$ at all redshifts and will use the approximation~\eqref{eq:siglogcos} for predicting the change of nonlinear log-variances with cosmology.

\subsubsection{Calibration for fiducial cosmology}

The nonlinear variance can be predicted from a given nonlinear power spectrum according to equation~\eqref{eq:defSigma2nonlin}. We compare the result from fitting functions (halofit and respresso) and perturbative techniques (SPT and RegPT) with the variance obtained from the measured nonlinear power spectrum. With this comparison, we circumvent potential convergence issues that might affect the nonlinear variance measured from the PDF at small scales, as discussed in Section~\ref{sec:convergence}. We compare the nonlinear power spectra using the different approaches in Figure~\ref{fig:PNLcomparison} in Appendix~\ref{app:linvarapprox}. At low redshifts $z=0,0.5,1$, the halofit power spectrum is accurate at a few percent level in the whole range up to $k\simeq 0.4$, which is relevant for obtaining the variance down to $R=10$ Mpc$/h$. When using the halofit fitting function for the nonlinear power spectrum to predict the nonlinear variance at $R=15$ Mpc$/h$, we find about $1\%$ and $2\%$ disagreement at redshifts $z=0.5$ and $z=0$. At the smaller scale $R=10$ Mpc/h, we obtain residuals of $1.5\%$ and $2.5\%$ for $z=0.5$ and $z=0$, respectively. 
The cosmology dependence of the nonlinear variance\Cora{, measured by the residuals of the ratio of variances in different cosmologies,} is predicted at sub-percent level for all derivative simulations with changed $\Lambda$CDM parameters according to Table~\ref{tab:models}, where the largest deviations of are found for $\Omega_m$ and $n_s$.

The nonlinear power spectrum generated from the response function approach \citep{Nishimichi16,Nishimichi17}\footnote{computed using the publicly available \href{http://www-utap.phys.s.u-tokyo.ac.jp/~nishimichi/public_codes/respresso/index.html}{Respresso Python package}} is extremely close to the measured power spectrum, having sub-percent residuals throughout. The response function approach is aided by a few sets of simulations, one of which is for a Planck 2015 cosmology with a very similar set of parameters as chosen in Quijote. While this could explain the spectacular agreement for fiducial cosmology, also the predictions for the dependence on changed $\Lambda$CDM parameter according to Table~\ref{tab:models} are sub-percent throughout.

Obtaining accurate nonlinear variances from perturbative results for the nonlinear power spectrum is difficult, since a broad range of scales up to $k\simeq 0.4 h/$Mpc enters the integration from equation~\eqref{eq:defSigma2nonlin} when decreasing radius towards $R=10$ Mpc$/h$, as we demonstrate in Figure~\ref{fig:sigmaNLintegrand} in Appendix~\ref{app:linvarapprox}. RegPT and SPT at 2-loop order \footnote{computed using the publicly available \href{http://www2.yukawa.kyoto-u.ac.jp/~atsushi.taruya/regpt_code.html}{Eclairs code}} have residuals typically larger than $4\%$ at redshift $z=0$, with improvements at higher redshifts and larger radii. While RegPT gives a more accurate result than SPT out to wavenumbers of about $k\simeq 0.2 h/$Mpc, its exponential cutoff causes predictions for the nonlinear variance to seem worse than SPT. \Cora{Note that within the framework of the effective field theory of large-scale structure \citep{Baumann12,Carrasco12}, the nonlinear variance gets renormalised in order to account for the short-scale effects \citep{Ivanov19}.} 

\begin{table}
\centering
\begin{tabular}{l|c|c|c|c|c|c}
fid, HR & $R$ [Mpc/h] & z=0 & z=0.5 & z=1 &z=2 &z=3 \\\hline 
& 10 & 0.619 & 0.498 & 0.404 & 0.285 & 0.218\\
 $\sigma_\mu$ & 15 & 0.486 & 0.384 & 0.308 & 0.215 & 0.164\\
& 20 & 0.388 & 0.304 & 0.242 & 0.169 & 0.129\\\hline
 &10 & 0.743 & 0.560 & 0.436 & 0.296 & 0.223\\
$\sigma_\rho$  & 15 & 0.532 & 0.406 & 0.319 & 0.219 & 0.166\\
  & 20 & 0.408 & 0.313 & 0.247 & 0.171 & 0.13\\
\end{tabular}
\caption{Variances of the density $\rho$ and the log-density $\mu=\ln\rho$ for different radii $R$ [Mpc/$h$], redshifts $z$ and cosmological models (see Table~\ref{tab:models}) as measured from the mean of the 100 realisations of the high resolution fiducial simulation.}
\label{tab:varianceShin}
\end{table}     

\section{The matter PDF with massive neutrinos}
\label{sec:matterPDFnu}

\subsection{Basic effects of massive neutrinos}
Before we start to discuss the effect of massive neutrinos on the matter PDF, let us review the relevant basics following \cite{Lesgourgues06}. The neutrino abundance is related to the total matter density \Cora{and can be approximated} as 
\begin{align}
\label{eq:nuOmfromM}
    \Omega_\nu = \frac{M_\nu}{93.14 h^2 \text{eV}}\,,\, f_\nu=\frac{\Omega_\nu}{\Omega_{m}}\,.
\end{align}
For the simulations considered here with $M_\nu=0.1,0.2,0.4$ eV, we have $f_\nu=0.0075, 0.015, 0.030$. Due to their large thermal velocities, neutrinos do not cluster on scales below their (physical) free-streaming length, which is defined in analogy to the Jeans length and depends on their mass according to 
\begin{align}
    \label{eq:freestreaming}
   \lambda_{\rm fs}(z) = \frac{7.7 (1 + z)}{\sqrt{\Omega_\Lambda + \Omega_m (1 + z)^3}}\frac{1 \text{eV}}{m_\nu} \text{Mpc}/h\,,
\end{align}
where $m_\nu$ is the mass of the considered neutrino species. The corresponding free-streaming wavenumber is $k_{\rm fs}=2\pi a/\lambda_{\rm fs}$. In this study, we consider degenerate neutrino masses, and therefore, we have $m_\nu=M_\nu/3$. For total neutrino masses of $M_\nu=0.1,0.2,0.4$ eV, the free streaming lengths are $\lambda_{\rm fs}(z=0)=231, 115, 57$ Mpc$/h$. While the free streaming scale sets the scale below which neutrinos are not clumping under the influence of gravity at some redshift $z$, there is an additional scale of interest in the problem. The maximum free streaming scale in comoving units achieved at any redshift is much larger, as it is related to the time when neutrinos become non-relativistic
\begin{equation}
    k_{\rm nr} \simeq 0.018\, \Omega_m^{1/2} \left(\frac{m_\nu}{1\text{eV}}\right)^{1/2} h/ \text{Mpc}\,.
\end{equation}
The associated comoving length scale $\lambda_{\rm nr}=2\pi/k_{\rm nr}$ determines above which scales massive neutrinos behave like an ordinary clustering dark matter component. Below this scale, the neutrino power spectrum is damped compared to the cold dark matter one. For the total neutrino masses considered here, those comoving length scales are beyond the size of the simulation box.

In the presence of massive neutrinos, we need to extend the list of physical ingredients for the matter PDF from Section~\ref{sec:matterPDF} by an additional element. In order to predict the total matter PDF, we need to specify the nonlinear matter density of the clustering component. After discussing this key change below, we will describe the imprint of massive neutrinos on the other standard ingredients, namely spherical collapse and the variances. 

\subsection{Matter density of clustering component}
\begin{figure}
\includegraphics[width=\columnwidth]{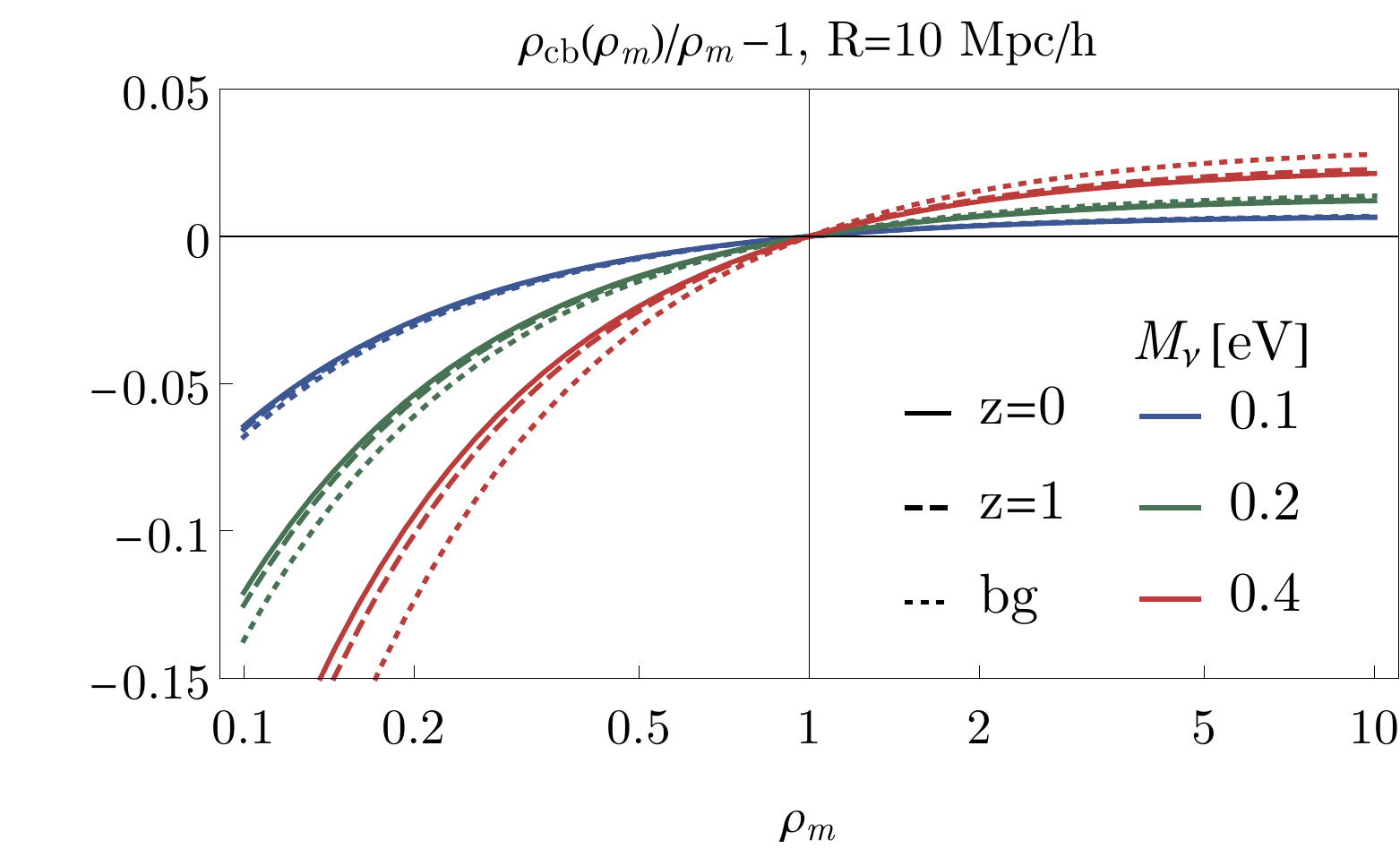}
\caption{Effect of massive neutrinos on the ratio of the normalised densities of CDM plus baryons (cb) and total matter (m) \Cora{in a given massive neutrino cosmology. We show predictions from} equation~\eqref{eq:rhomofrhonu} considering only the neutrino background (dotted lines) or also the scale-dependent neutrino clustering according to equations~\eqref{eq:rhomappingnu} for radius $R=10$ Mpc$/h$ at redshift $z=1$ (dashed lines) and $z=0$ (solid lines) for total neutrino mass $M_\nu=0.1$eV (blue), $0.2$eV (green) and $0.4$eV (red).}
   \label{fig:changerho}
\end{figure}

Let us denote normalised densities by $\rho=1+\delta$ and physical densities by $\tilde \rho$, such that $\rho=\tilde \rho/ \bar\rho$.
The normalized total density $\rho_m$ can be expressed in terms of the normalised matter density in cold dark matter and baryons, $\rho_{cb}$, and the neutrino density $\rho_\nu$, and the relative abundances of the two species
\begin{subequations}
\begin{align}
\label{eq:rhomofrhonu}
\rho_{m}&=\frac{\tilde \rho_{cb+\nu}}{\bar\rho_{cb+\nu}} \simeq\frac{\tilde \rho_{cb} +
\tilde\rho_\nu}{\bar\rho_{cb} + \bar \rho_\nu} 
= \rho_{cb}\frac{\Omega_{cb}}{\Omega_{m}} + \rho_\nu\frac{\Omega_\nu}{\Omega_{m}} \\
\label{eq:deltaofdeltanu}
\delta_{m} &=\delta_{cb} \frac{\Omega_{cb}}{\Omega_{m}} + \delta_\nu\frac{\Omega_\nu}{\Omega_{m}} \,.
\end{align}

A first estimate for the effect of neutrinos masses can be obtained by considering a uniform background density constituted by the massive neutrinos, thus setting $\rho_\nu=1$. We illustrate this most simplistic relationship between the total normalized matter density and the normalized matter density in CDM and baryons as dotted line in Figure~\ref{fig:changerho}. We see that the effect is biggest for underdensities, where the presence of massive neutrinos causes one to  probe CDM+baryon densities that are effectively even lower than the total matter density, hence rarer and less probable. To take into account that neutrinos do cluster on scales larger than their free-streaming length, we assume that a portion of massive neutrinos cluster like the cold dark matter and baryon component. We approximate the clustering portion using the ratio between the linear variances computed from the CAMB power spectra for massive neutrinos and cold dark matter and baryons
\begin{equation}
\label{eq:deltanuofdeltacb}
    \delta_\nu \simeq \frac{\sigma_{{\rm L},\nu}(R,z)}{\sigma_{\rm L,cb}(R,z)} \delta_{\rm cb}\,.
\end{equation}
\label{eq:rhomappingnu}
\end{subequations}
This corresponds to considering a simplified scale-dependent bias between neutrinos and cold dark matter plus baryons, and a tight correlation between the linear fields. Hence, the total matter density is a biased version of the cold dark matter and baryon density and the impact of this bias on the PDF is given by a change of variables
\begin{equation}
\label{eq:PDFcbtom}
    \mathcal P_{\rm m}(\rho_{\rm m}) = \mathcal P_{\rm cb}(\rho_{\rm cb}(\rho_{\rm m})) \frac{d\rho_{\rm cb}}{d\rho_{\rm m}}\,.
\end{equation}

In Figure~\ref{fig:changerho}, we show the impact of this neutrino clustering on the ratio between the cold dark matter and baryon component to the total matter for two redshifts, $z=1$ (dashed lines) and $z=0$ (solid lines). We see that the presence of massive neutrinos lowers the clustering density in underdense regions. This shifts the underdense tail of the total matter PDF to slightly higher densities, as seen in the upper panel of Figure~\ref{fig:diffPDFsimvstheo_Mnu}. Additionally, there is a few percent effect on the shape around the peak, that is not visible in a log-log plot but shown in the lower panels of Figure~\ref{fig:diffPDFsimvstheo_Mnu}. For the prediction, we inferred the impact of massive neutrinos on the nonlinear log-variance of cold dark matter plus baryons using equation~\eqref{eq:siglogcos}.

Since the weak lensing convergence is a projected version of the total matter density contrast~\eqref{eq:deltaofdeltanu}, we expect that the scale-dependent neutrino clustering according to equation~\eqref{eq:deltanuofdeltacb} is directly related to the imprint of massive neutrinos in the weak lensing PDF recently measured in simulations \citep{Liu19WLPDF}. In fact, the residuals between the PDFs with and without massive neutrinos shown in Figure~\ref{fig:resPDFsimvstheo_Mnu} look qualitatively very similar. However, we note that part of the observed signature could be due to a change in the nonlinear variance that is driven by $\sigma_8$, which is not fixed in the MassiveNu simulations used in \cite{Liu19WLPDF}. 

\subsection{Spherical collapse}
For realistic total neutrino masses that are in agreement with current bounds, the effect of massive neutrinos on spherical collapse is typically sub-percent. \cite{LoVerde14} demonstrated that the main net effect of massive neutrinos with total mass $M_\nu <0.5$eV is to increase the collapse threshold by at most $1\%$. In our theoretical model for spherical collapse, the impact of this change can be estimated by changing the parameter $\nu\propto \delta_c$ entering the spherical collapse approximation from equation~\ref{eq:SCnu}. We have checked that this effect remains below $1\%$ in the entire 2-sigma region around the mean in logarithmic scale.

When focusing on nonlinear objects like halos and voids that intrinsically live in the tails of the PDF, the impact of massive neutrinos on their formation can be more significant. On the one hand, massive neutrinos lower the abundance of massive dark matter halos that host galaxy clusters \citep{IchikiTakada12,LoVerde14, Paco_11,Paco_13} by delaying the collapse time, which is only slightly counteracted by the nonlinear clustering of neutrinos with the halo. On the other hand, neutrinos do not evacuate voids as efficiently as CDM due to their thermal velocities \citep{Massara15}. This is why the scale-dependent bias for voids defined in the total matter field in massive neutrino cosmologies \citep{Banerjee16} is stronger than the scale-dependent halo bias at fixed neutrino mass and rarity of the fluctuation. Unfortunately, this signal is difficult \Cora{to use} currently since it is challenging to robustly detect voids in total matter through lensing measurements. Additionally, if one attempts to look for these effects by defining voids using tracer populations like galaxies or massive clusters, it has been shown that the massive neutrino effects sensitively depends on the choice of tracers \citep{Kreisch18}.

\subsection{Scale-dependent linear variance and nonlinear variance}
\begin{figure}
\includegraphics[width=1.\columnwidth]{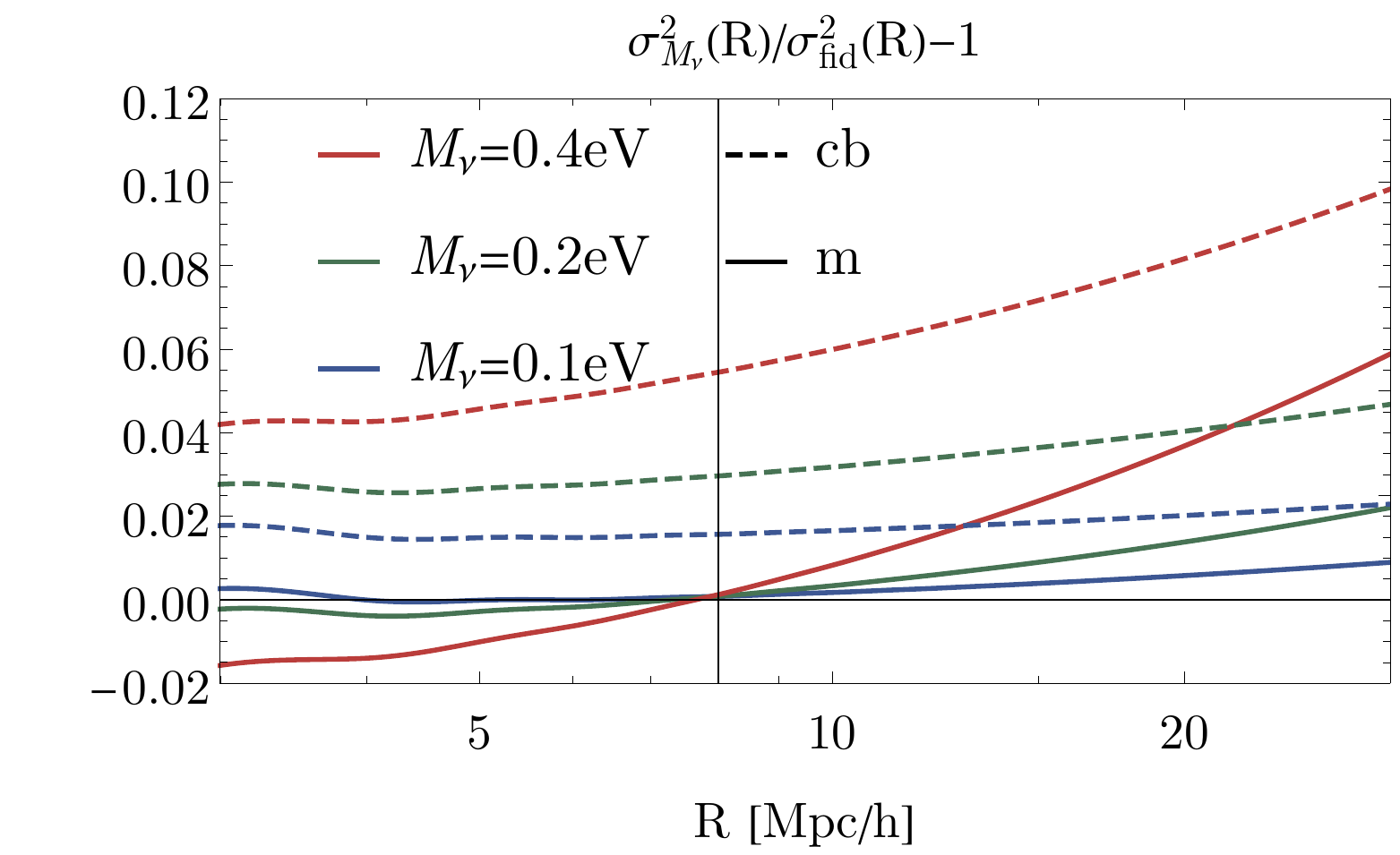}
   \caption{Comparison between 
   the linear variance $\sigma_L^2(R)$ computed from the linear power spectrum of total matter (solid lines) and cold dark matter plus baryons (dashed lines) for cosmologies with total neutrino mass $M_\nu=0.1$eV (blue), $0.2$eV (green) and $0.4$eV (red) with fixed $\sigma_8$. Note that the offset of the curves for cold dark matter and baryons is because their clustering amplitude needs to be enhanced to compensate for the lack of neutrino clustering on large scales to yield the desired $\sigma_8$ for total matter.} 
   \label{fig:sigmalincomparisonMnu}
\end{figure}

In Figure~\ref{fig:sigmalincomparisonMnu} we show the scale-dependence in the linear variance induced by the presence of massive neutrinos, for the total matter component (solid lines) and the clustering matter component (dashed lines) at redshift $z=0$. For the theoretical predictions that follow, we use the linear variance computed for the clustering matter component and approximate the impact of massive neutrinos on the nonlinear log-variance according to equation~\eqref{eq:siglogcos}. Note that the nonlinear variance could also be predicted from halofit \citep{halofit}, its extensions \citep{Bird12} or perturbative models including massive neutrinos \citep{Saito09}.

In Figure~\ref{fig:Dofzcomparison} in Section~\ref{sec:conclusion}, we demonstrate the impact of massive neutrino on the growth of structure at the scales of interest here $R\gtrsim 10$ Mpc$/h$.
For the case of massive neutrinos, the density of the clustering matter component (CDM+baryons) can be estimated from the density of the total matter according to equation~\eqref{eq:rhomofrhonu}.

\begin{table}
\begin{center}
\begin{tabular}{c|cc|cc}
$M_{\nu}$ [eV] &  $\sigma_{\mu,cb}$  & $\sigma_{\mu,m}$ & $\sigma_{\rho,cb}$ & $\sigma_{\rho,m}$\\ \hline
0 & \multicolumn{2}{c|}{0.6144} & \multicolumn{2}{c}{0.7399}\\
0.1 & 0.6183 & 0.6124 & 0.7468 & 0.7414\\
0.2 & 0.6222 & 0.6113 & 0.7541 & 0.7438\\
0.4 & 0.6297 & 0.6107 & 0.7686 & 0.7501\\
\end{tabular}
\end{center}
\caption{Measured variances for the log-density $\mu=\log \rho$ and the density $\rho$ using all matter (m) or the cold dark matter and baryon component (cb) at $z=0$ for a sphere of radius $R=10$ Mpc$/h$.}
\label{tab:variances}
\end{table}

\subsection{Summarised recipe and result}

Combining the recipes outlined above, one can compute the total matter density PDF in the presence of massive neutrinos from the linear power spectrum for cold dark matter plus baryon, their nonlinear log-variance, standard spherical collapse and a mapping between the total clustering density and the cold dark matter plus baryon density according to equations~\eqref{eq:rhomappingnu}. 

 In Figure~\ref{fig:diffPDFsimvstheo_Mnu} we show the predicted and measured effect of massive neutrinos (with fixed matter clustering amplitude $\sigma_8$). The impact of massive neutrinos at scales around 10 Mpc$/h$ is largest in the underdense regions, where it is at the 10\%  level ($\rho\simeq 0.2$) and at the few percent level for overdense ($\rho\simeq 5$) regions. This effect is significantly stronger than cosmic variance, as we highlight in a corresponding residual plot shown in Figure~\ref{fig:resPDFsimvstheo_Mnu} in Appendix~\ref{app:PDFplots}.

\section{Validating theory and simulations}
\label{sec:validation}

\subsection{Measurements from the Quijote simulations}
\label{sec:Quijote}
The Quijote simulations are a large suite of full N-body simulations designed for two main purposes: 1) to quantify the information content of cosmological observables and 2) to provide enough data to train machine learning algorithms. They contain 43100 simulations spanning more than 7000 cosmological models in the $\{\Omega_{\rm m}, \Omega_{\rm b}, h, n_s, \sigma_8, M_\nu, w\}$ hyperplane. At a single redshift, the total number of particles in the simulations excess 8.5 trillions, over a combined volume of 43100 $({\rm Gpc}/h)^3$. The simulations follow the gravitational evolution of $N_p^3$ particles ($2\times N_p^3$ for simulations with massive neutrinos) over a comoving volume of 1 $({\rm Gpc}/h)^3$ starting from redshift $z=127$. Initial conditions for the pure $\Lambda$CDM simulations (without massive neutrinos) are generated from 2LPT, while the simulations to assess the impact of massive neutrinos are run using initial conditions from the Zeldovich approximation taking into account the scale-dependent growth factor and growth rate present in these models. Simulations were run using the \textsc{SPH-TreePM  Gadget-III code}. We refer the reader to \cite{Quijote} for further details on the Quijote simulations. Different values of particle number $N_p$ are provided: $N_p=256$ (low resolution), $N_p=512$ (fiducial resolution) and $N_p=1024$ (high resolution).

In this section, we focus on the PDF of the matter field, smoothed with a spherical top-hat at different scales, computed from the Quijote simulations. The PDFs have been computed from the Quijote simulations as follows. First, particle positions and masses are assigned to a regular grid with \Cora{$N_m$ (typically $512^3$)} cells using the Cloud-in-Cell (CIC) mass assignment scheme. Next, the value of the normalised density field $\rho=1+\delta$ in each grid cell is computed by dividing the mass of each cell by the average cell mass. We then smooth the density field with a top-hat filter of radius $R$. This procedure is done in Fourier-space, by first computing the Fourier transform of the density field and then multiplying it by the Fourier transform of the filter on the regular grid itself, to avoid numerical artifacts on small scales. Finally, the smoothed field is estimated by computing the inverse Fourier transform of the previous quantity. The PDF is measured in 99 logarithmically spaced bins of normalised density between $10^{-2}$ and $10^2$ by calculating the fraction of cells that lie in a given bin and dividing by the bin width. 

\subsection{Matter PDF resolution effects}
\label{sec:convergence}
We tested the convergence of the PDF measurements with respect to two resolution parameters --- the number of particles $N_p$, and the number of mesh cells $N_m$. The fiducial resolution is $N_p=N_m=512^3$, giving typical initial inter-particle and grid spacing of about $2$ Mpc$/h$. The high-resolution PDF is obtained from $N_p=N_m=1024^3$, where the initial inter-particle and grid spacing are about $1$ Mpc$/h$. To disentangle the impact of the particle number and mesh resolution effects, we compared all combinations of $N_p,N_m \in \{512^3,1024^3\}$.

\begin{figure}
\includegraphics[width=\columnwidth]{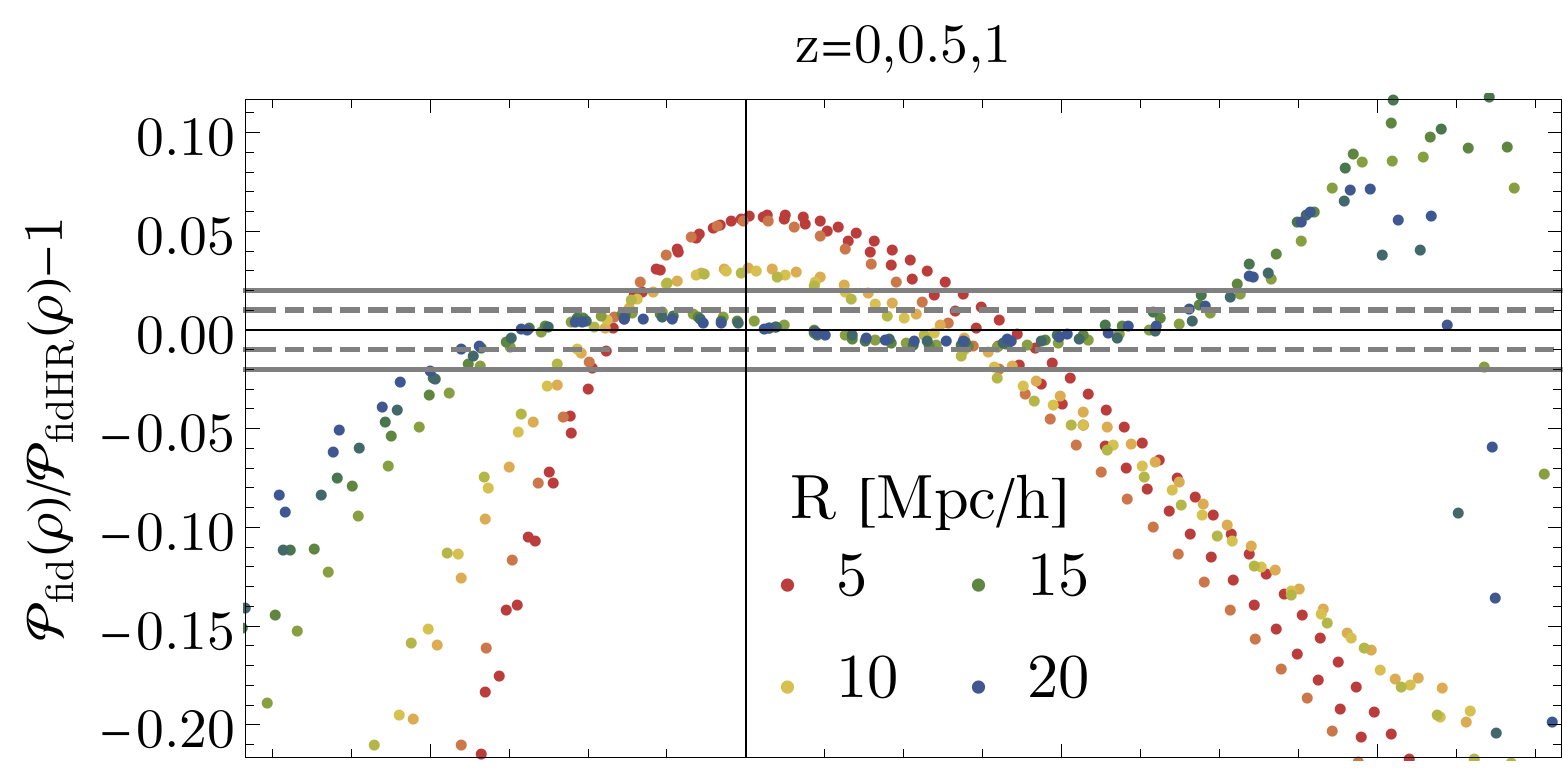}
\includegraphics[width=\columnwidth]{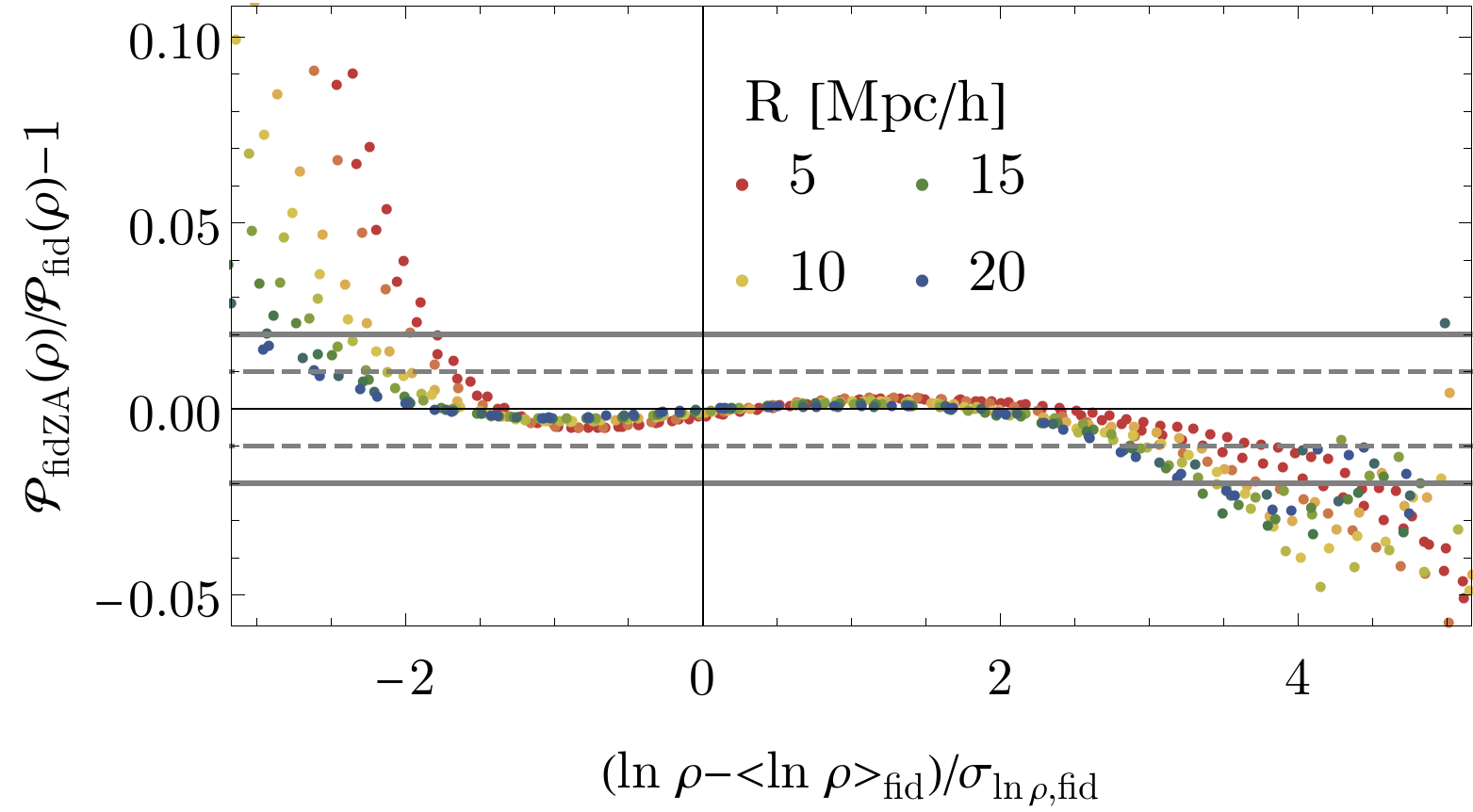}
mean\caption{Residual between the fiducial matter PDF in spheres of radius $R=5,10,15,20$ Mpc$/h$ (red, yellow, green, blue) at redshifts $z=0, 0.5, 1$ (slightly blue-shifted in color towards larger $z$), measured from the means of different simulation runs. \Cora{The dashed and solid grey lines indicate 1\% and 2\% accuracy, respectively}. (Top panel) Residual between the mean over 100 simulations with standard resolution and high-resolution. (Bottom panel) Residual between 500 realisations with initial conditions using the Zeldovich approximation and 2LPT.}
   \label{fig:PDFconvergence}
\end{figure}

In \Cora{the upper panel of} Figure~\ref{fig:PDFconvergence} we show a convergence check for the PDF, comparing the measured fiducial cosmology PDF in the standard resolution simulation with the high resolution simulation. From the plot we see that the resolutions affects the PDF of densities in spheres of radii $R=5,10,15$ Mpc$/h$ at  $6\%,3\%,1\%$ level even around the peak. This effect depends only weakly on redshift. We found that in order to obtain accurate matter PDFs, the final mesh resolution has to be about a tenth of the radius of the spheres. For a quantitative comparison of the standard deviation and the skewness, we therefore focus on radii $R=10,15,20$ Mpc$/h$ and redshifts $z=0,0.5,1$. The standard deviation $\sigma$ is underestimated by $2.5-3\%$ for radius $R=10$ Mpc/h in the lower resolution, which is mostly due to the smaller mesh resolution. We find that the lower resolution overestimates the reduced skewness $S_3$, defined in equation~\eqref{eq:S3pred}, relatively independent of scale and redshift by about $1.5-2.5\%$, mostly caused by the lower particle number (and therefore more subject to rare events).

The finite resolution effects on the matter PDF measured from simulations highlight the importance of having reliable theoretical predictions for the PDF available, allowing to cross-validate the simulation measurements and the theory. We mitigate the impact of finite resolution effects by discarding the smallest radius $R=5$ Mpc$/h$ and cutting a percentage of rare density spheres in the Fisher analysis, essentially limiting the range of logarithmic densities to the $1.5\sigma$ region around the peak. 

\Cora{In the lower panel of Figure~\ref{fig:PDFconvergence} we show residuals between the matter PDF extracted from the fiducial simulations run either from Zeldovich approximation or 2LPT initial conditions. We can see that the impact of the initial conditions is sub-percent in the region around the peak, but increases in the tails. Overall, we observe that the impact of initial conditions is smaller than the finite resolution effects and negligible in the $1.5\sigma$ region around the peak, on which we will focus in our analysis.}

\subsection{Shape of the matter PDF for fiducial cosmology}

\begin{figure}
\includegraphics[width=\columnwidth]{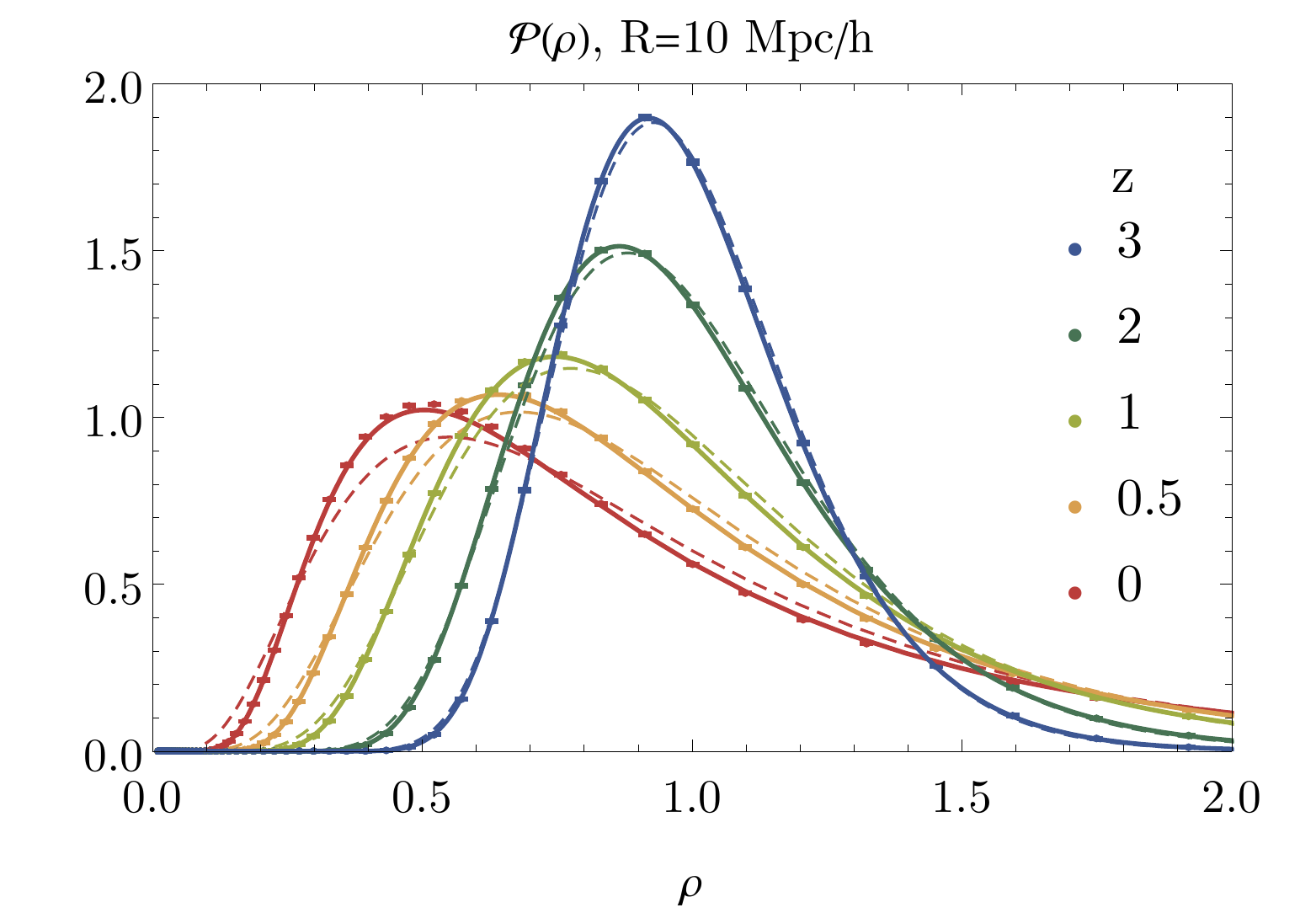}
\includegraphics[width=\columnwidth]{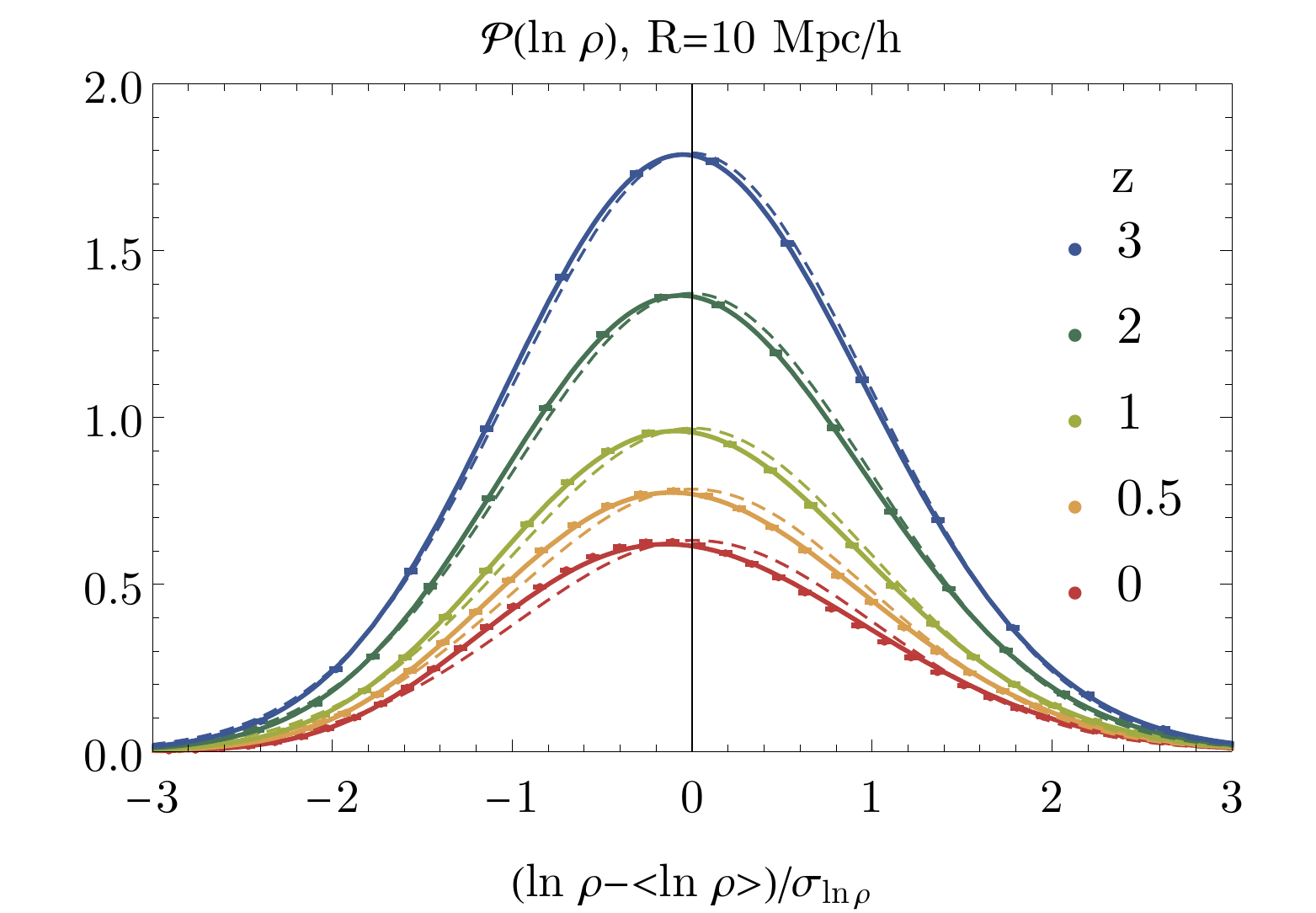}
\caption{Matter density PDF for spheres of radius $R=10$ Mpc$/h$ at redshifts $z=0, 0.5, 1, 2, 3$ (from red to blue, as indicated in the legend) for the high-resolution run of the fiducial model as measured from the mean over 100 realisations (data points) compared to the theoretical prediction (lines). We show \Cora{the density PDF} (upper panel), and the PDF of the logarithmic density, with a shifted and rescaled $x$-axis to align the peak positions and unify the widths (lower panel). \Cora{Lognormal PDFs are shown for comparison as thin dashed lines, which clearly deviate from the measurements towards lower redshifts.}}
   \label{fig:DMPDFfidtheovssim}
\end{figure}

\begin{figure}
\includegraphics[width=\columnwidth]{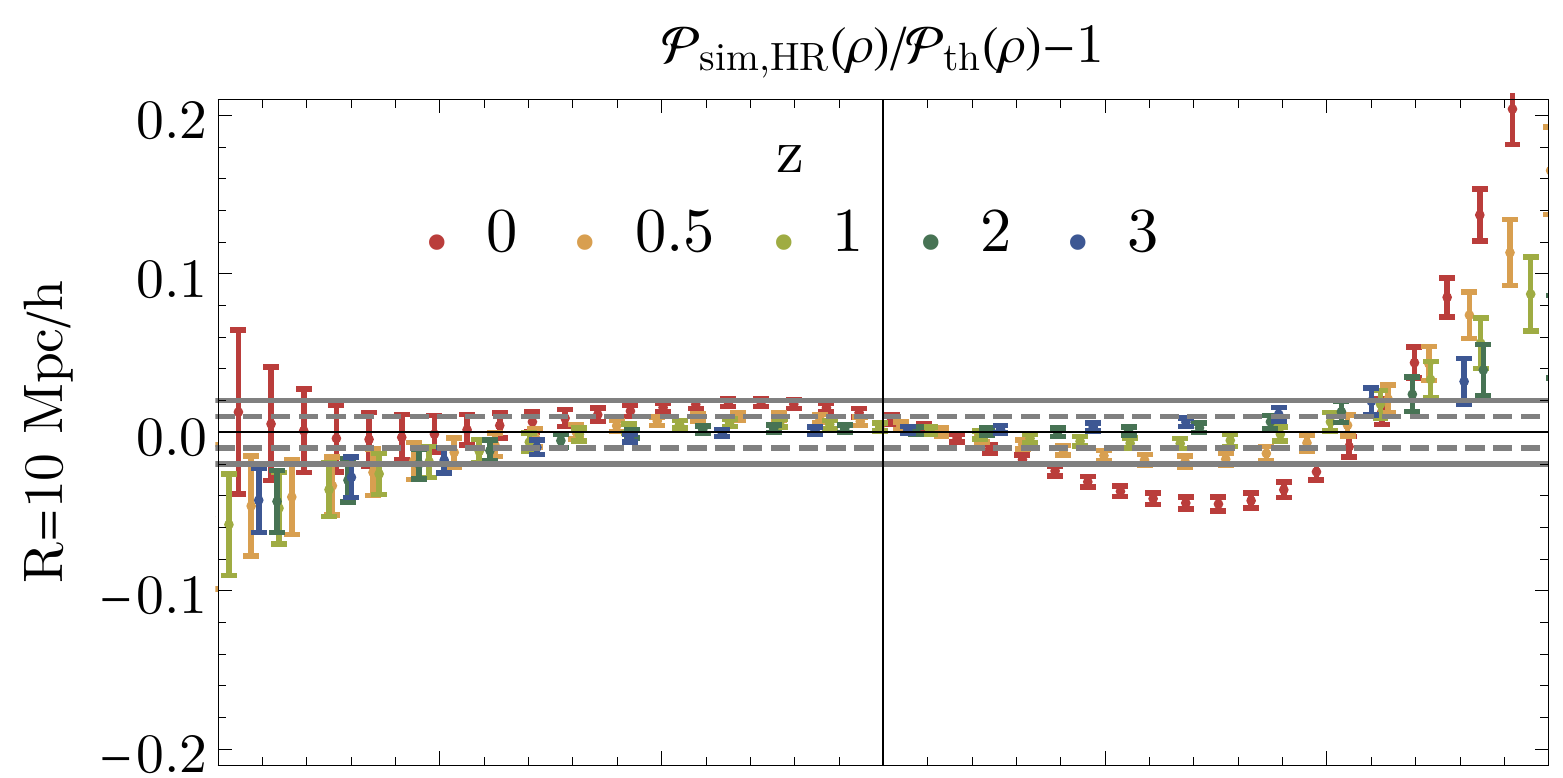}
\includegraphics[width=\columnwidth]{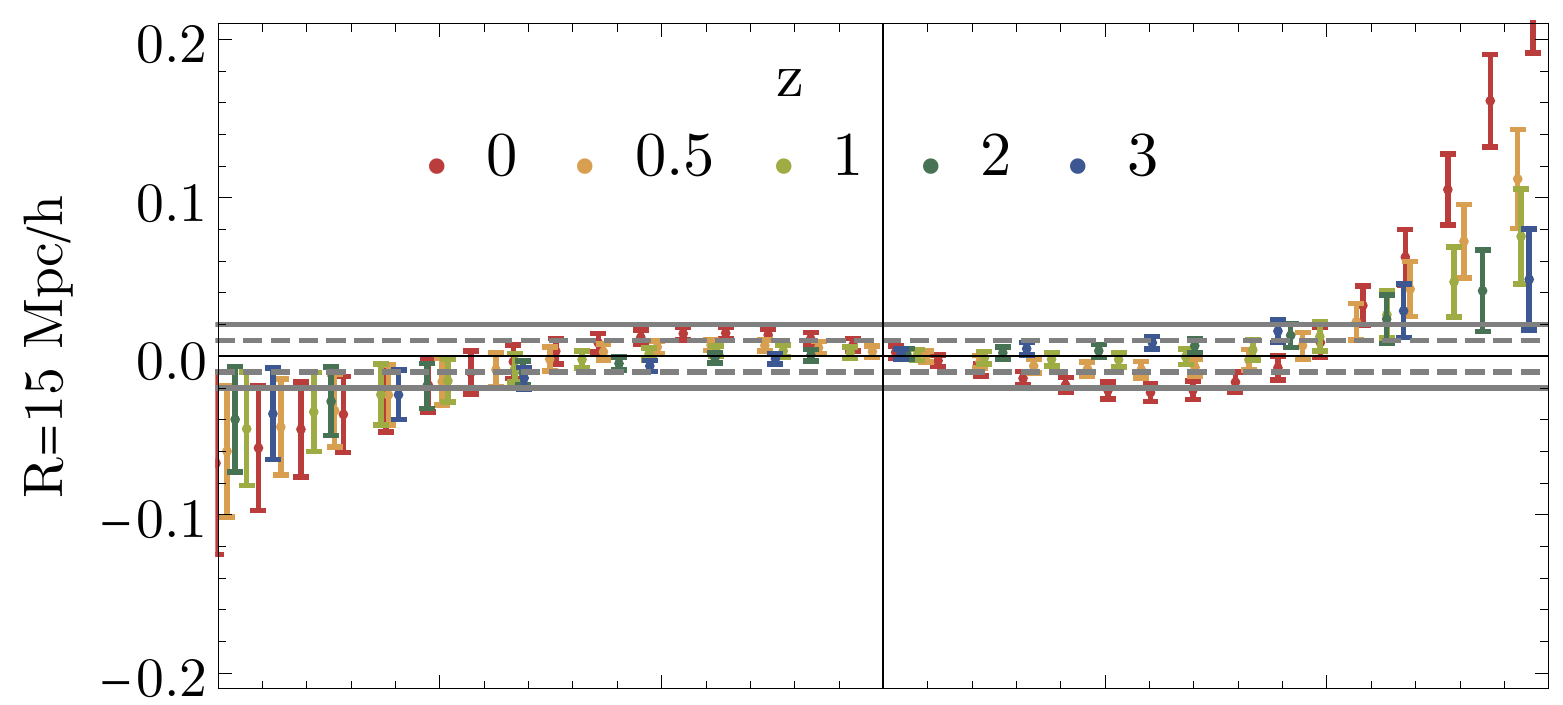}
\includegraphics[width=\columnwidth]{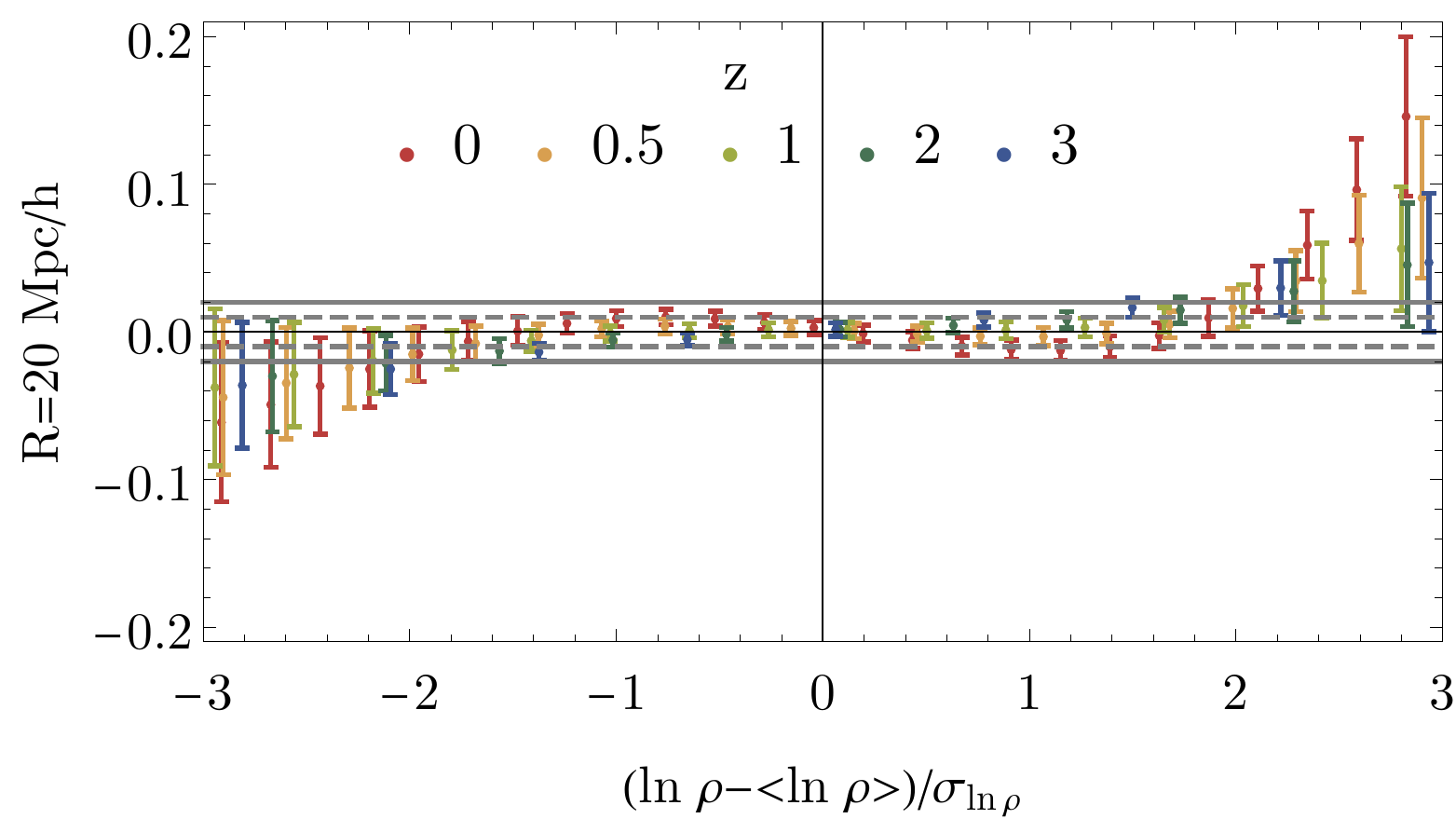}
\caption{Residuals between the measured and predicted matter PDF in spheres of radius $R=10,15,20$ Mpc$/h$ (top to bottom) for redshifts $z=0, 0.5, 1, 2, 3$ (red to blue, as indicated in the legend).}
   \label{fig:resDMPDFfidtheovssim}
\end{figure}

In Figure~\ref{fig:DMPDFfidtheovssim} we show a comparison of the theoretical prediction for the PDF and the measurement from the mean over 100 realisations of the high-resolution simulation for the fiducial cosmology. We find that the theoretical prediction for the PDF, with the measured nonlinear variance as an input, performs very well. To fairly compare the performance of the prediction at different radii and redshift, we plot residuals in Figure~\ref{fig:resDMPDFfidtheovssim} as a function of the deviation of the log-density from its mean in units of the standard deviation. \Cora{Data points with error bars indicate the mean and standard deviation of the PDF bin measurements across the 100 realisations.}
Residuals are at a few percent level in the  2-$\sigma$ region around the mean for all scales and redshifts at which the nonlinear variance, listed in Table~\ref{tab:variances}, is sufficiently below unity ($\sigma_{\rm NL}^2 \lesssim 0.5$). As expected, the agreement between the theory and the simulation improves with decreasing the nonlinear variance, which can be achieved by increasing either redshift or the smoothing radius. 

Let us note that when going from high (blue curves) to low (red) redshifts, the PDF is skewed towards underdensities as expected since voids occupy more volume while overdensities become more concentrated. \Cora{While this evolution is mainly driven by the growth of the skewness, one needs to implement a large-deviation argument \citep{Bernardeau92,Bernardeau94,LDPinLSS} to get the correct PDF shape by accounting for a complete (and therefore meaningful) hierarchy of cumulants. This has be be contrasted with an Edgeworth-like expansion which truncates this cumulant hierarchy and therefore necessarily fails to reproduce the tails of the distribution. In fact, the inclusion of higher-order cumulants is essential even to capture the full shape of the PDF around the peak, as we demonstrate in an accompanying paper where we compare the constraining power of the moments to the central region of the PDF \citefuture{(Friedrich et al. 2019)}.}

\subsection{Change of PDF shape with cosmological parameters}
\label{sec:differences}

In the following, we compare the differences of the matter PDFs for changes in $\Lambda$CDM parameters and total neutrino mass, as predicted by our theoretical model and as measured in the simulation. We show differences rather than ratios to highlight deviations in the shape of the PDF around the peak, where the signal to noise is highest and most constraining power is located. Even when excluding the tails, the full shape of the PDF still carries significant non-Gaussian information, as we will show in Figure~\ref{fig:diffPDFsimvstheo_Om_Ob_ns_h} below. 
Additionally, the differences between PDFs for different parameters \Cora{determine the response of the PDF to changing cosmology that enter} in the Fisher analysis presented in Section~\ref{sec:Fisher}. \Cora{Data points with error bars indicate the mean and standard deviations of the differences in the PDFs measured across 500 realisations.} 

\subsubsection{$\Lambda$CDM cosmological parameter dependence}
\begin{figure}
\includegraphics[width=1\columnwidth]{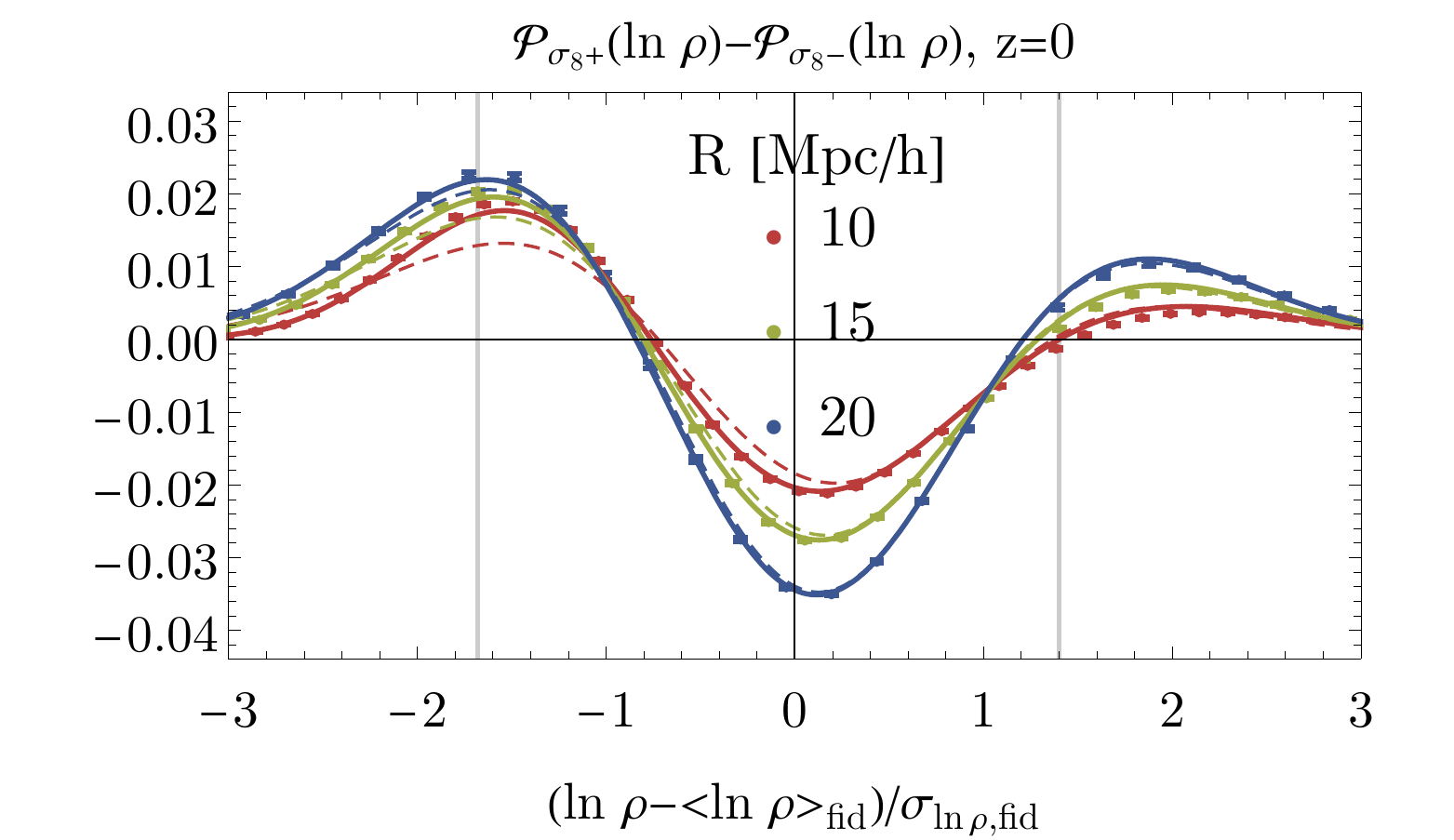}\\
   \caption{Differences between the PDFs with changed $\sigma_8$ at redshift $z=0$ with radii $R=10,15,20$ Mpc/h as predicted from our formalism (solid lines), the lognormal approximation (thin dashed lines) and measured  in the derivative simulations with standard resolution (data points). \Cora{The gray vertical lines indicate the region that is used for the Fisher analysis in Section~\ref{sec:Fisher}.}
   }
   \label{fig:diffPDFsimvstheo_sig8}
\end{figure}

\begin{figure}
\includegraphics[width=1\columnwidth]{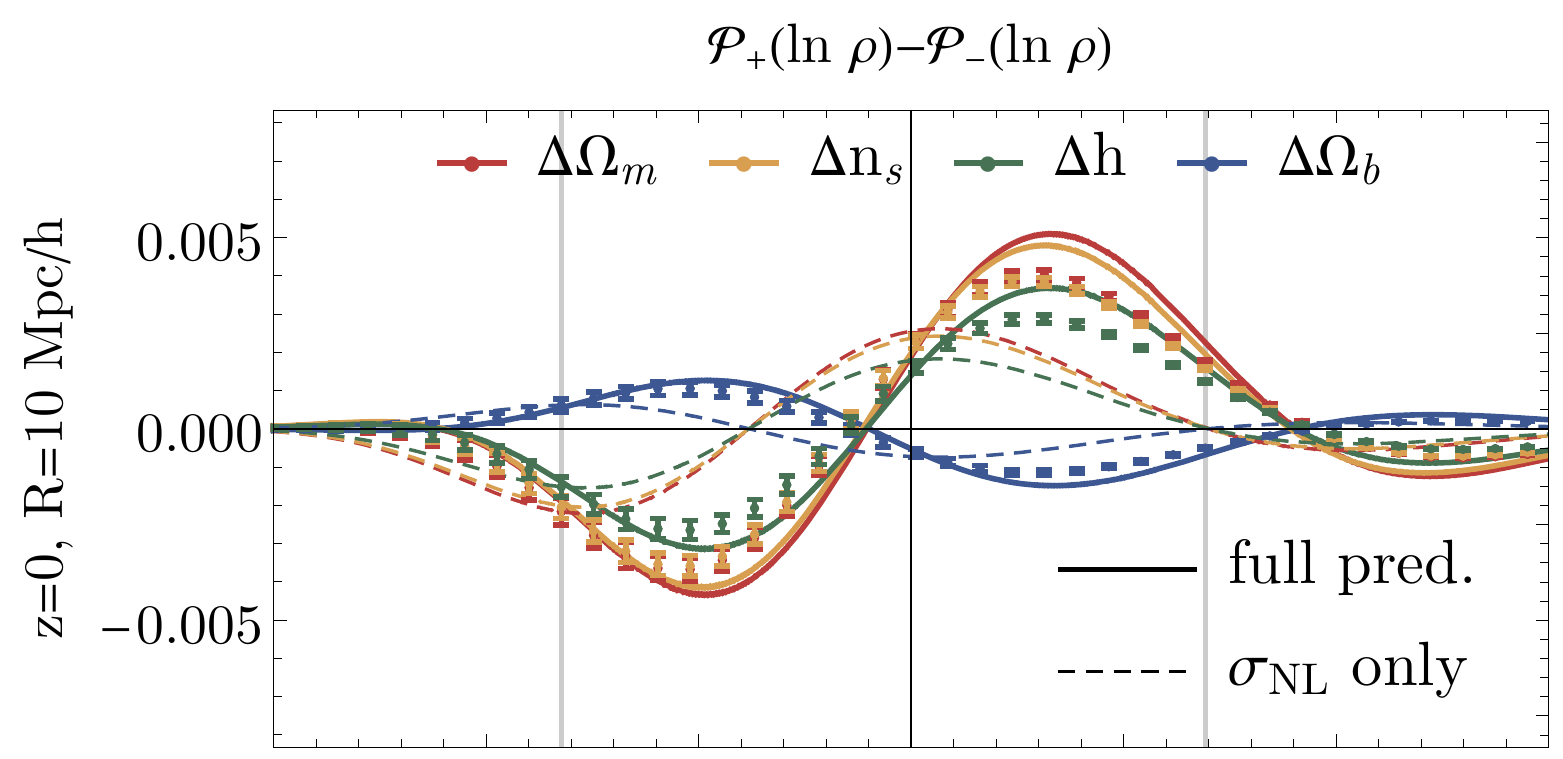} 
\includegraphics[width=1\columnwidth]{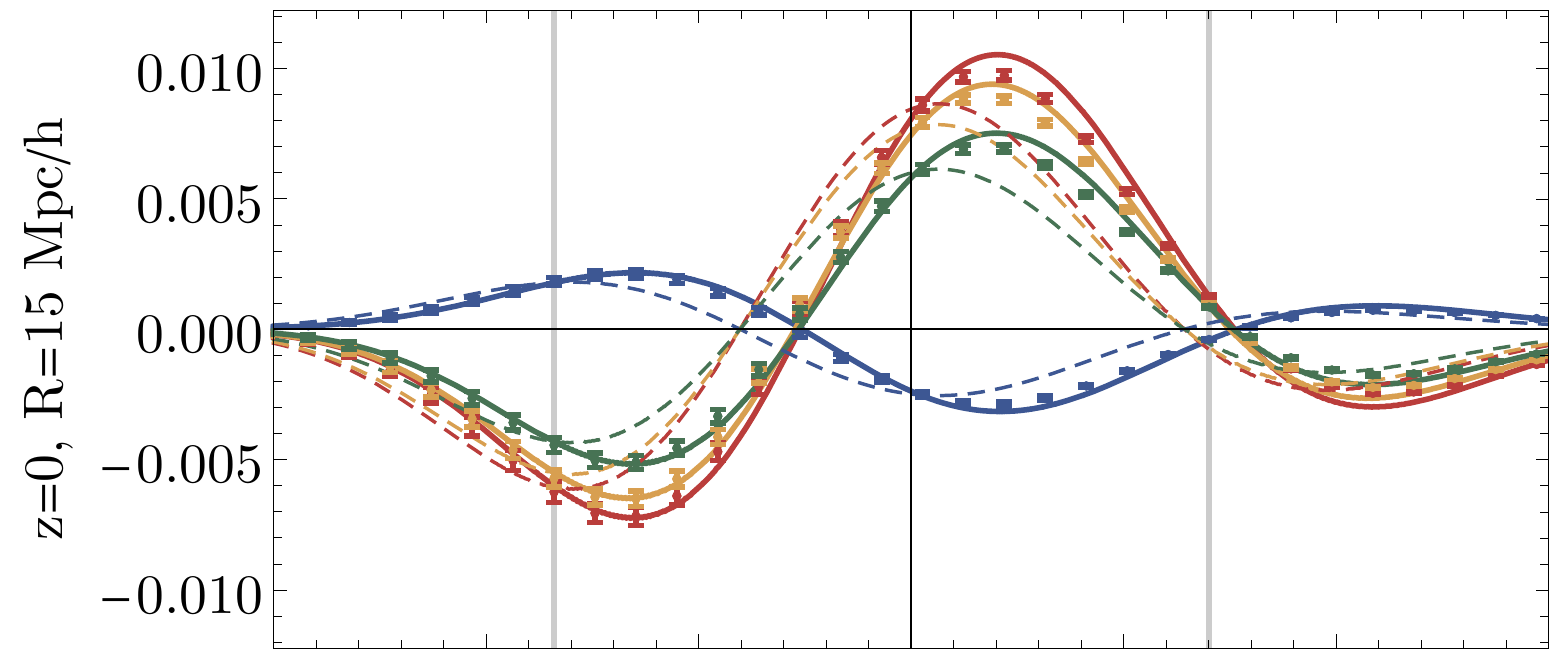}  \\
\includegraphics[width=1\columnwidth]{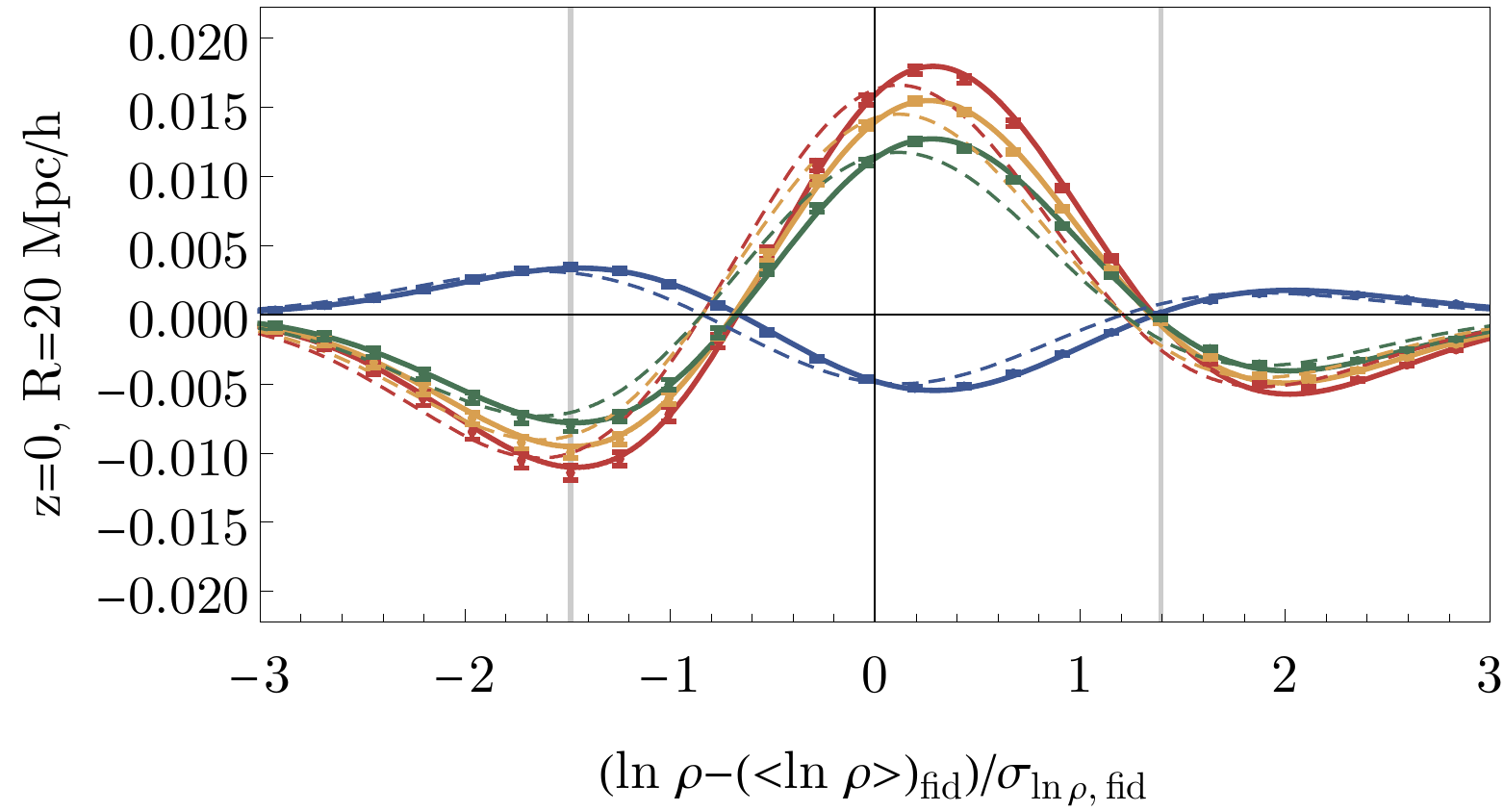}  

   \caption{Differences between the matter PDFs for changes in $\Omega_m$ (red), $\Omega_b$ (blue), $n_s$ (orange) and $h$ (green) as predicted (solid lines) and measured in the simulations (data points) at redshift $z=0$ with radii $R=10,15,20$ Mpc/h (from top to bottom). To highlight the impact of cosmology on the non-Gaussian shape on small scales, we add predictions that only account for the change in the nonlinear variance (thin dashed lines). \Cora{The gray vertical lines indicate the region that is used for the Fisher analysis in Section~\ref{sec:Fisher}.}} 
   \label{fig:diffPDFsimvstheo_Om_Ob_ns_h}
\end{figure}

In Figures~\ref{fig:diffPDFsimvstheo_sig8}~and~\ref{fig:diffPDFsimvstheo_Om_Ob_ns_h}
we show differences between the matter PDFs for changes in $\Lambda$CDM parameters. For the predictions shown as solid lines, we use the theoretical model from equations~\eqref{eq:PDFpred} along with the predicted cosmology dependence of the log-variance from equation~\eqref{eq:siglogcos}, normalised with the measured variance for the fiducial simulations. Increasing the clustering amplitude $\sigma_8$ leads to an increase in the nonlinear variance, which broadens the PDF and hence decreases the peak height, as can be seen by the dip around the origin in Figure~\ref{fig:diffPDFsimvstheo_sig8}. Since the PDF is normalised, this dip around the peak is compensated by an increase of the PDF in regions that are more significantly underdense or overdense. For comparison, we also show the expected differences assuming a lognormal matter PDF (thin dashed lines), which disagree with the simulation measurements in particular in the underdense regions.

At small scales $R\sim 10$ Mpc$/h$, changing cosmological parameters other than $\sigma_8$ mostly cause an additional skewness, which manifests in an asymmetry between densities on different sides of the peak of the PDF. This is in line with the expectation that changes in the amplitude of the variance are small around $R=8$ Mpc$/h$ due to fixed $\sigma_8$, and the main effect is the change of the scale-dependence shown in Figure~\ref{fig:sigmalincomparison} that modifies the skewness according to equation~\eqref{eq:S3pred}. At larger scales, the asymmetry disappears because the main effect is a few percent change in the variance, resembling the impact of changing $\sigma_8$ shown in Figure~\ref{fig:diffPDFsimvstheo_sig8}.  To highlight the cosmological information in the PDF shape, we contrast the full theoretical prediction including the impact of the cosmological parameters on the scale-dependence of the variance (solid lines) to a mere change in the nonlinear variance (thin dashed lines) in Figure~\ref{fig:diffPDFsimvstheo_Om_Ob_ns_h}. This demonstrates the importance of the reduced skewness, which responds to a change in cosmological parameters according to equation~\eqref{eq:S3pred}. Indeed, the full hierarchy of cumulants affects the PDF shape even in the central region excluding the tails, as we discuss in an accompanying paper focused on the impact of primordial non-Gaussianity \citefuture{(Friedrich et al. 2019)}. For completeness, we show the corresponding ratios of the PDFs when varying $\Lambda$CDM parameters other than $\sigma_8$ in Figure~\ref{fig:resPDFsimvstheo_Om_Ob_ns_h} in Appendix~\ref{app:PDFplots}.

\subsubsection{Total neutrino mass dependence}
\begin{figure}
\includegraphics[width=\columnwidth]{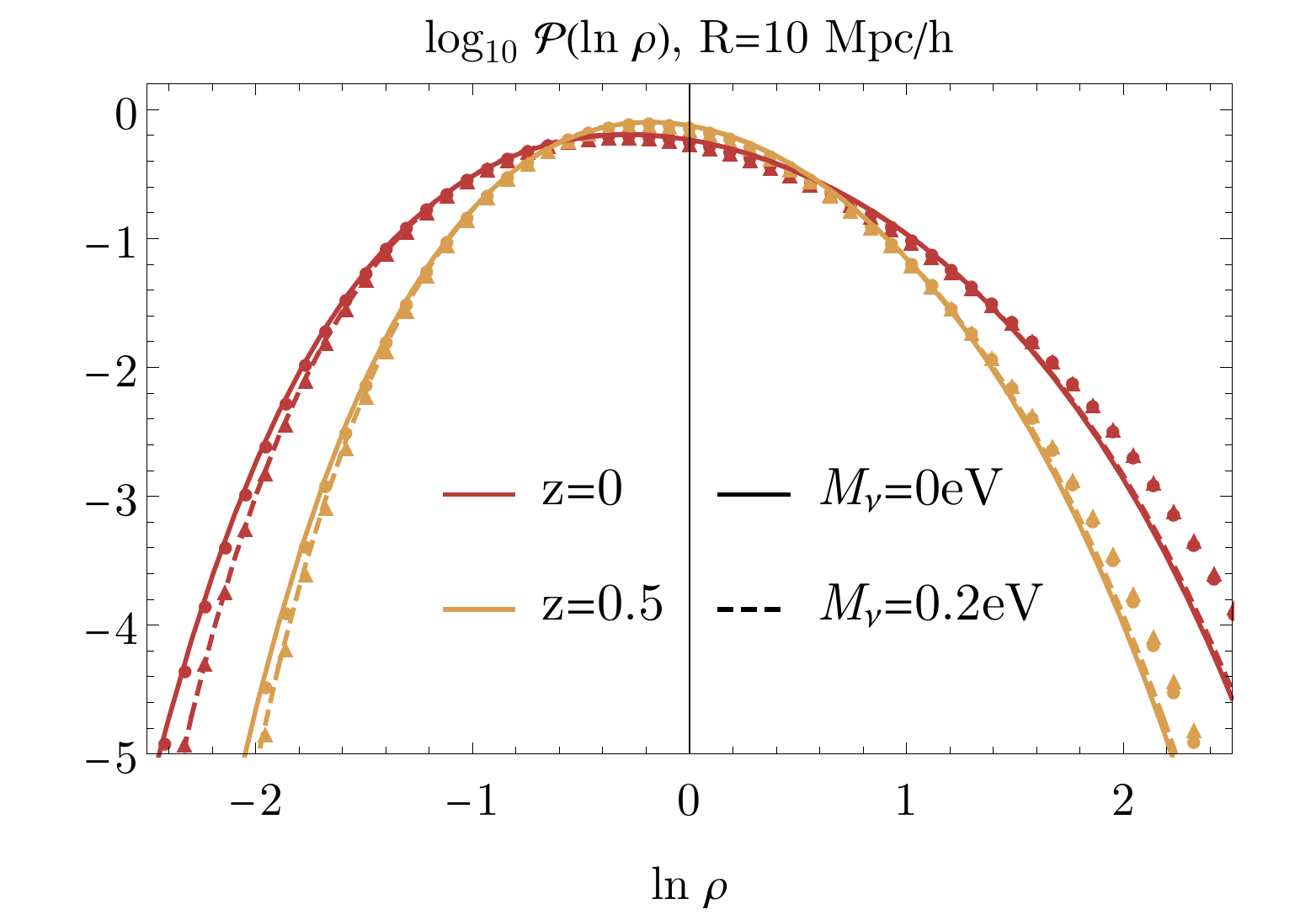}
\includegraphics[width=\columnwidth]{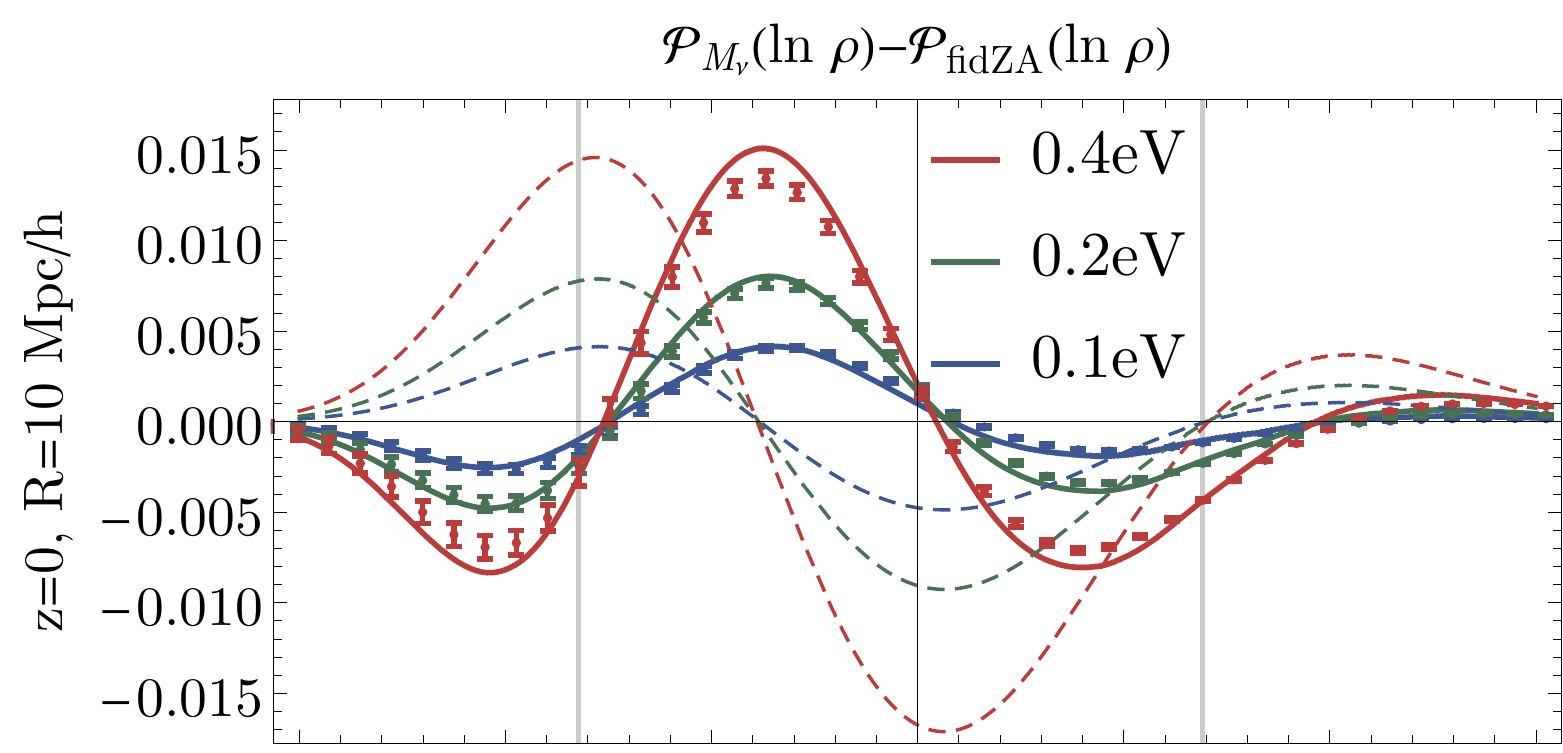}
\includegraphics[width=\columnwidth]{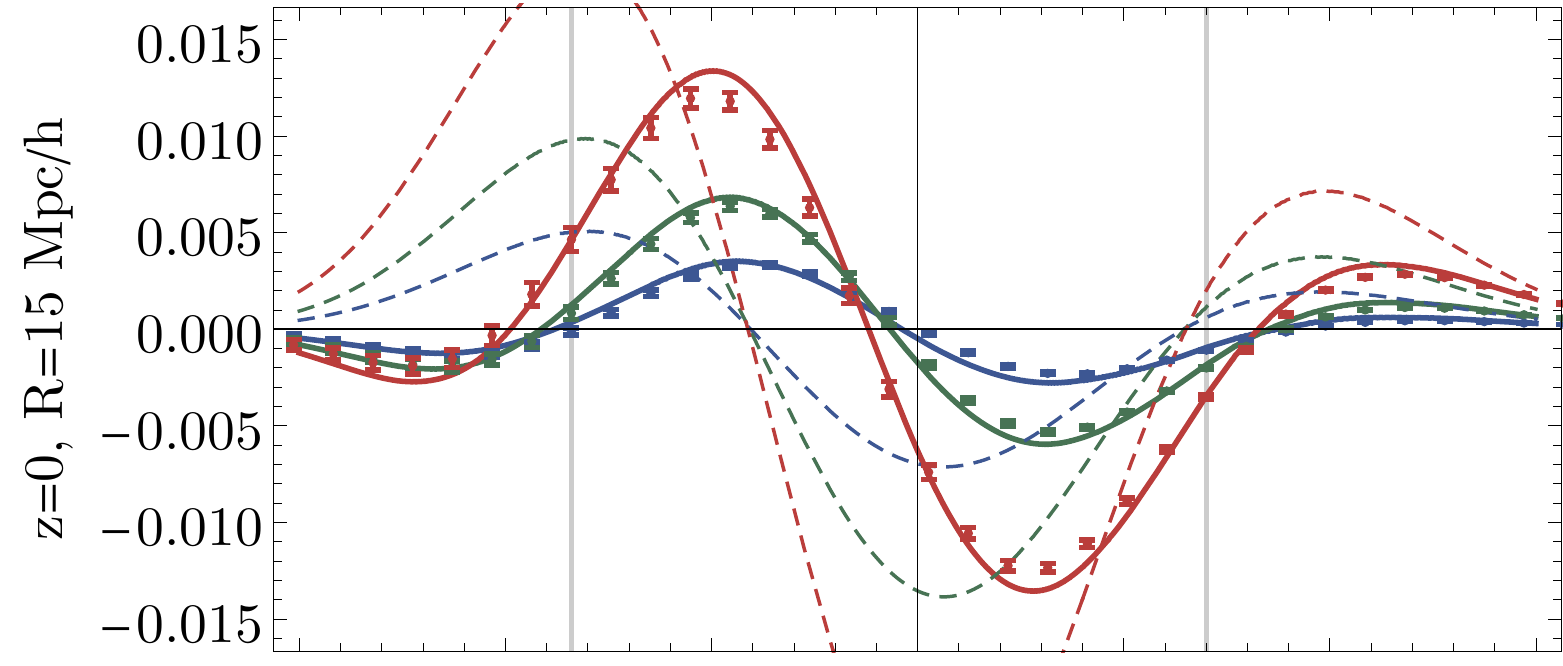}
\includegraphics[width=\columnwidth]{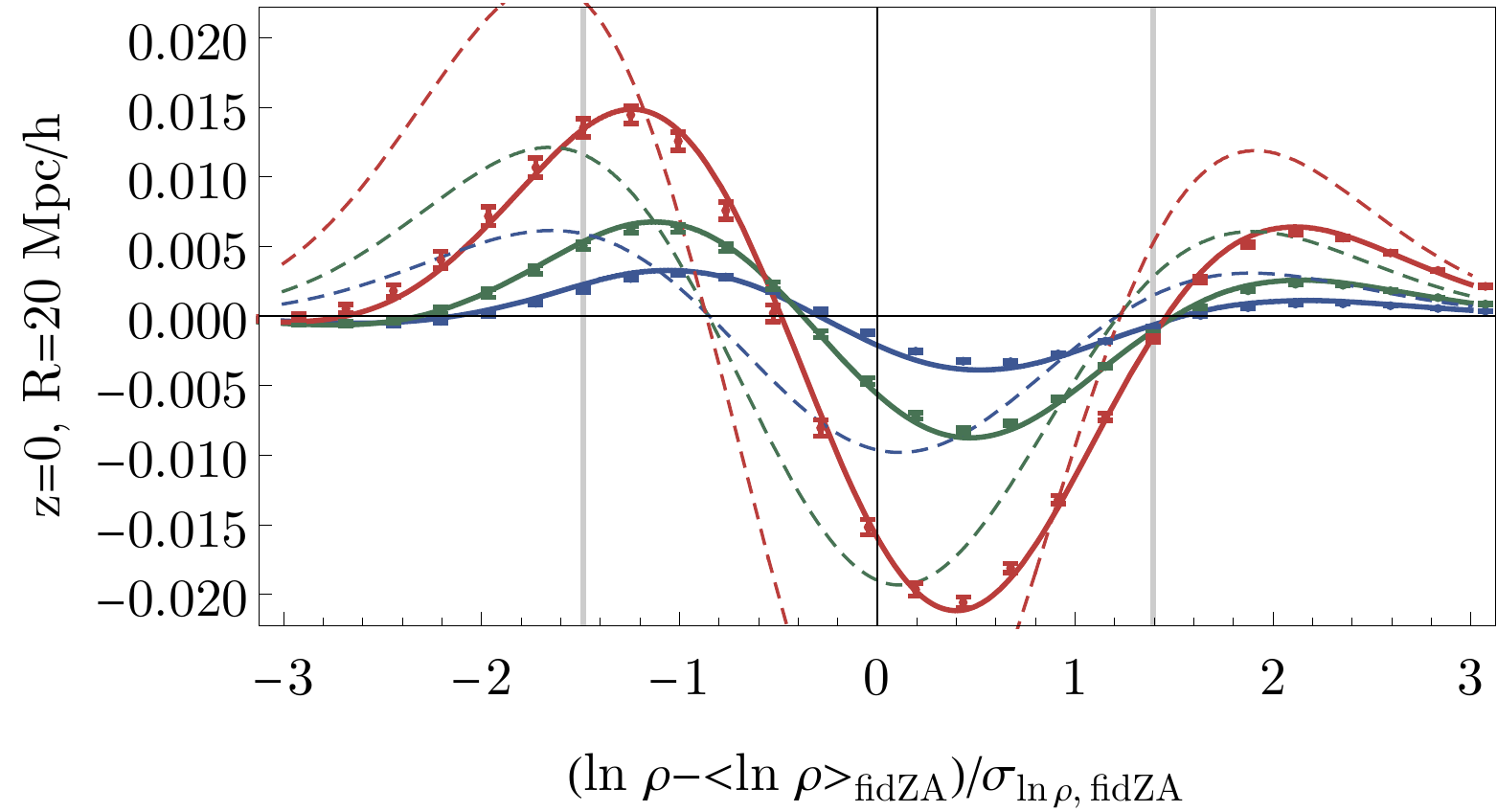}

\caption{(Upper panel) Total matter PDF in spheres of radius $R=10$ Mpc$/h$ for redshifts $z=0, 1$ for the fiducial model without massive neutrinos and with $M_\nu=0.2$ eV, as indicated in the legend. (Lower panels) The fractional difference of the PDF as measured (data points) and predicted (solid lines) for massive neutrinos with $m_\nu=0.1$eV (blue), $m_\nu=0.2$eV (green) and $m_\nu=0.4$eV (red) and the fiducial model (with equal $\sigma_8$) as a function of density at redshift $z=0$ and radii $R=10,15,20$ Mpc$/h$ (top to bottom). To highlight that the imprint of massive neutrinos in the matter PDF is distinct from a change in $\sigma_8$, we add predictions that only account for the change in the nonlinear variance (thin dashed lines). \Cora{The gray vertical lines indicate the region used for the Fisher analysis in Section~\ref{sec:Fisher}.}}
   \label{fig:diffPDFsimvstheo_Mnu}
\end{figure}

In Figure~\ref{fig:diffPDFsimvstheo_Mnu} we show differences in the total matter PDF for the massive neutrino models compared to the fiducial model. Note that both simulations have been run using initial conditions generated from the Zeldovich approximation. We find that massive neutrinos affect the shape of the PDF in a distinct way, that is well predicted by our model (solid lines) and not degenerate with a change in $\sigma_8$, for which differences are displayed in Figure~\ref{fig:diffPDFsimvstheo_sig8}. \Cora{To highlight this, we contrast our model with naive predictions only accounting for the change in the nonlinear variance as thin dashed lines. For the smallest radius $R=10$ Mpc$/h$, we can see a significant suppression in underdense regions, as expected from Figure~\ref{fig:changerho}. Additionally, the skewness is enhanced by the presence of massive neutrinos due to a combination of the enhancement by the scale-dependent variance demonstrated in Figure~\ref{fig:sigmalincomparisonMnu} and the change of variables~\eqref{eq:rhomappingnu}. This leads to a characteristic signature that can be even distinguished by eye from the shapes induced by changing other $\Lambda$CDM parameters shown in Figure~\ref{fig:diffPDFsimvstheo_Om_Ob_ns_h}. For larger radii, this signature gets concealed, because the range of probed densities becomes smaller and the differences in the variances grow.} The scale-dependence of the impact of massive neutrinos on the matter PDF has a significant advantage over the mildly nonlinear matter power spectrum, where the clustering amplitude $\sigma_8$ and the total neutrino mass $M_\nu$ are largely degenerate \citep{Paco_18a}, as we show in Figure~\ref{fig:Fisher_sig8Mnu_PDFvsPk}. \Cora{The main effect in the matter PDF is caused by the presence of a massive neutrino background in underdense regions and a partial clustering of massive neutrinos encoded in the scale-dependent bias from  equations~\eqref{eq:rhomappingnu}).}

For completeness, we provide  a residual plot between the total matter PDF in the presence of massive neutrinos and the fiducial model in  Figure~\ref{fig:resPDFsimvstheo_Mnu}. \Cora{This plot demonstrates that our theoretical model for massive neutrinos achieves a similar accuracy as the predictions for changes in the $\Lambda$CDM parameters.}

\section{Fisher forecast for $\nu\Lambda$CDM cosmology}
\label{sec:Fisher}

In this section, we quantify the information content of the matter PDF on the full set of $\Lambda$CDM cosmological parameters and the total neutrino mass using a Fisher matrix formalism together with the large suite of Quijote simulations.

After briefly reviewing the basis of the Fisher analysis in Section~\ref{sec:Fisher-bas}, \Cora{we explain our data vector in Section~\ref{sec:datavector}.} We discuss the covariance matrix in Section~\ref{sec:covariance} and the derivatives with respect to cosmological parameters in Section~\ref{sec:derivatives}. In Section~\ref{sec:complement}, we determine a suitable combination of smoothing radii and redshifts for the matter density PDF, and establish the complementarity between the matter PDF and the matter power spectrum on mildly nonlinear scales. The final constraints on the full set of $\nu\Lambda$CDM parameters are presented in Figure~\ref{fig:Fisher_cosmo_neutrino_PDF+Pk}.

\subsection{Basics for the Fisher analysis}
\label{sec:Fisher-bas}
The Fisher matrix on a set of cosmological parameters, $\vec{\theta}$, given a (combination of) statistics $\vec{S}$ is defined as
\begin{equation}
F_{ij}= \sum_{\alpha,\beta}\frac{\partial S_\alpha}{\partial \theta_i}C^{-1}_{\alpha \beta}\frac{\partial S_\beta}{\partial \theta_j}~,
\label{eq:Fisher}
\end{equation}
where $S_i$ is the element $i$ of the statistic $\vec{S}$ and $C$ is the covariance matrix, defined as
\begin{equation}
\label{eq:covariance}
C_{\alpha \beta} = \langle (S_\alpha-\bar{S}_\alpha)(S_\beta - \bar{S}_\beta) \rangle\,,\quad \bar{S}_\alpha = \langle S_\alpha \rangle~.
\end{equation}
We multiply the inverse of the covariance matrix measured in the simulation by the Kaufman-Hartlap factor \citep{Kaufman67,Hartlap06}, $h=(N_{\rm sim} - 2 - N_{S})/(N_{\rm sim} - 1)$, to correct for a potential bias for the inverse of the maximum-likelihood estimator of the covariance depending on the ratio of the length of the data vector $N_{S}$ to the number of simulations $N_{\rm sim}$. Since in our case the number of simulations for covariance estimation is very large \Cora{(15,000) compared to the maximal length of the data vector (218 for our three-redshift analysis of the PDF at three scales and the mildly nonlinear power spectrum)}, this factor will be close to unity throughout. 
\Cora{Additionally, we mimic a BOSS-like effective survey volume by multiplying the covariance with the ratio of the considered survey volume $V$ and the simulation volume  $V_{\rm sim}$ .}

The Fisher matrix allows us to determine the error contours on a set of cosmological parameters under the assumption that the likelihood is Gaussian. The inverse of the Fisher matrix gives the parameter covariance. The error on the parameter $\theta_i$, marginalised over all other parameters, is given by
\begin{equation}
\delta\theta_i\geq \sqrt{\left(F^{-1}\right)_{ii}} \,.
\end{equation}

The Fisher analysis relies on three ingredients, 
\begin{enumerate}
    \item the \Cora{chosen} summary statistics \Cora{that enter} the data vector,
    \item their covariance matrix, and
    \item their derivatives with respect to cosmological parameters.
\end{enumerate}
As discussed in \cite{Quijote}, the Quijote simulations are designed to numerically evaluate those three pieces for different summary statistics, including matter power spectra and matter density PDFs. 
There are 15000 simulations at fiducial cosmology available to estimate covariances, along with 500 simulations each for increasing/decreasing every single $\Lambda$CDM parameter. To assess the impact of massive neutrinos, there are 500 simulations run from Zeldovich approximation \Cora{(instead of 2LPT)} initial conditions for fiducial cosmology, and for total neutrino masses of $M_\nu=0.1,0.2,0.4$eV. For the Fisher analysis, we assume a total effective cosmic volume of ($6$ Gpc$/h$)$^3$ spread equally across the 3 lowest redshifts in the simulation $z=0,0.5,1$. The total volume roughly corresponds to the effective volume of the BOSS galaxy survey \citep{BOSS16} and about one tenth of Euclid.

\Cora{Note that such a Fisher analysis has a number of limitations: it only yields realistic error bars if measurements of the considered data vectors have Gaussian noise and if 
the responses of these data vectors to changing cosmological parameters are close to linear.
We checked that the distribution of individual bins of the PDFs measured in the Quijote sims are sufficiently close to a Gaussian distribution to expect a small impact on the total width of our forecasted contours. Additionally, realistic data analyses might have to account for systematic effects by marginalising over nuisance parameters. However, the main focus of this study is to explore the complementarity between the mildly nonlinear power spectrum and the matter density PDF as cosmological probes. Hence, we expect these limitations to have limited impact on our findings. }

\subsection{Data vector for matter PDFs and power spectra}
\label{sec:datavector}
In our case the data vector $\vec{S}$ is built from  the values of the matter PDF in density bins. First, we consider a single redshift $z$ and radius $R$. Then, we combine multiple redshifts and radii. In all cases, we only use those bins for the Fisher analysis, where the cumulative probability distribution function (CDF) is between 0.03 and 0.9. This amounts to removing 3\% of the lowest densities and 10\% of the highest densities. We choose this approach in order to limit the impact of finite resolution effects that are most severe for rare events, as shown in Figure~\ref{fig:PDFconvergence}, while still capturing the PDF shape around the peak, which is located in underdense regions. \Cora{We chose an asymmetric cut in the CDF, because the PDF rises more steeply towards the peak in underdense regions (see Figure~\ref{fig:DMPDFfidtheovssim}).} Additionally, it takes into account that the theoretical modelling of the matter density PDF and the impact of tracer bias or projections will be more challenging in the tails \Cora{(particularly for overdensities)} and correspondingly degrade the signal to noise from those regions. 
Finally, we will include the matter power spectrum in Fourier bins up to a given $k_{\rm max}$ in the data vector. The matter power spectrum is linearly binned in $k$-space in steps of the fundamental frequency $k_F=2\pi/L_{\rm box}$ \Cora{and the bin center is determined by averaging over all modes in the interval}, exactly as in \cite{Quijote}.

In the next section, we determine the covariance matrix of the matter PDF at different scales and its cross-covariance with the matter power spectrum. The final ingredient are the partial derivatives of the summary statistics with respect to the cosmological parameters, which are discussed in section~\ref{sec:derivatives}. 

\subsection{Covariance of matter PDF bins}
\label{sec:covariance}

We estimate the covariance using 15,000 realisations of the Quijote simulation with fiducial cosmology. While the covariance matrix defined in equation~\eqref{eq:covariance} is the quantity that enters the Fisher analysis, for visualisation purposes it is useful to normalise this matrix on the diagonal. For that purpose, we consider the correlation matrix, defined as 
\begin{equation}
\text{Corr}_{ij} = \frac{C_{ij}}{\sqrt{C_{ii}C_{jj}}}\,,
\end{equation}
where $C$ is the covariance matrix.  

\begin{centering}
\begin{figure}
  \includegraphics[width=1\columnwidth]{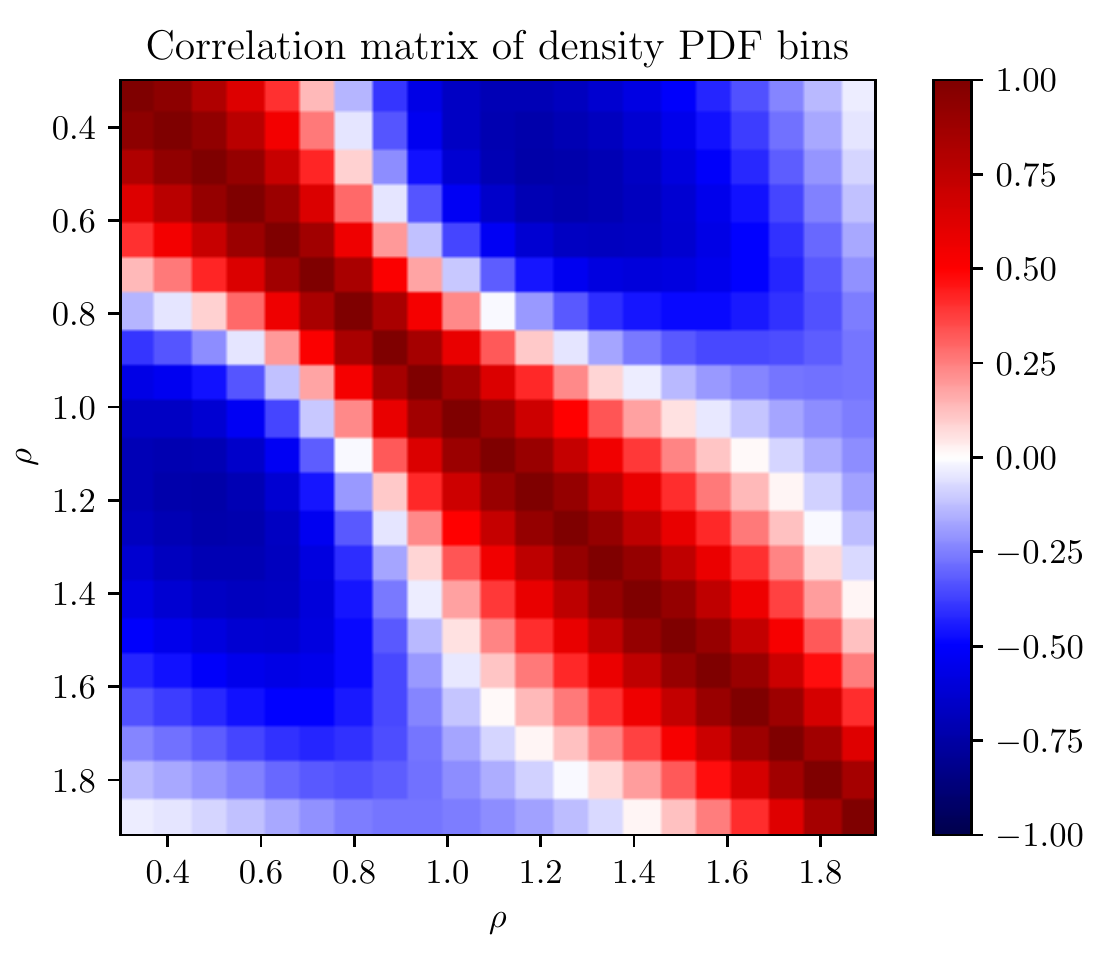}
   \caption{Correlation matrix for the bins of the matter density PDF at radius $R=10$ Mpc$/h$ and redshift $z=0$.}
  \label{fig:correlation_matrix}
\end{figure}
\end{centering}

In Fig. \ref{fig:correlation_matrix} we show the correlation matrix of the matter density PDF at $z=0$ and $R=10$ Mpc$/h$ showing the correlation between different density bins used for the Fisher analysis. We see that, as expected, neighbouring bins are positively correlated, while intermediate underdense and overdense bins are anticorrelated with each other. Note that the tails of the PDF, which are excluded in the plot, are strongly correlated with each other and anti-correlated with the peak.
\Cora{Note that this is completely in line with the correlation matrix predicted by the large-deviation formalism \citep[see Appendix C in][]{Codis16a}, which can be decomposed into a shot noise contribution, a cosmic variance term due to the finite volume of the survey and a term describing the spatial correlation of spheres. This last contribution dominates when enough spheres are considered and is proportional to the product of the sphere bias of the respective density bins, multiplied by the average dark matter correlation function at the typical separation of the spheres. Because the sphere bias is negative for underdensities and positive for overdensities, this product is positive when the cross-correlations of overdensities (or underdensities) is considered and negative for the cross-correlations of underdensities and overdensities.} 

In an accompanying paper \citefuture{(Friedrich et al. 2019)}, we show that the correlation matrix measured in zero-mean shifted lognormal realisations \citep{Hilbert2011, Xavier2016} closely resembles the simulation result. This is an encouraging result, as it provides a simple way to estimate covariances and is used in current analyses \citep{Friedrich18,Gruen18}. \Cora{In particular, this could be used to estimate the impact of super-sample covariance \citep{Takada13,Chan18,Barreira18}, which is not captured in our analysis using the full periodic simulation boxes.}

Note that, when assuming a diagonal covariance matrix for the matter PDF at one scale and radius, corresponding parameter errors are significantly underestimated. For constraints on the clustering amplitude $\sigma_8$ and matter density $\Omega_m$, a single PDF with diagonal covariance underestimates the contour area by a factor of 5. On the other hand, when combining the PDF at two different radii and assuming a block-diagonal covariance matrix, parameter errors are overestimated. Combining two PDFs as if they were independent using a block-diagonal covariance leads to a wrong orientation of the error ellipse and an overestimation of its area by a factor of about 2. 
This highlights the importance of an accurate covariance matrix and potentially valuable information in cross-correlations of summary statistics.

\begin{centering}
\begin{figure}
    \includegraphics[width=1\columnwidth]{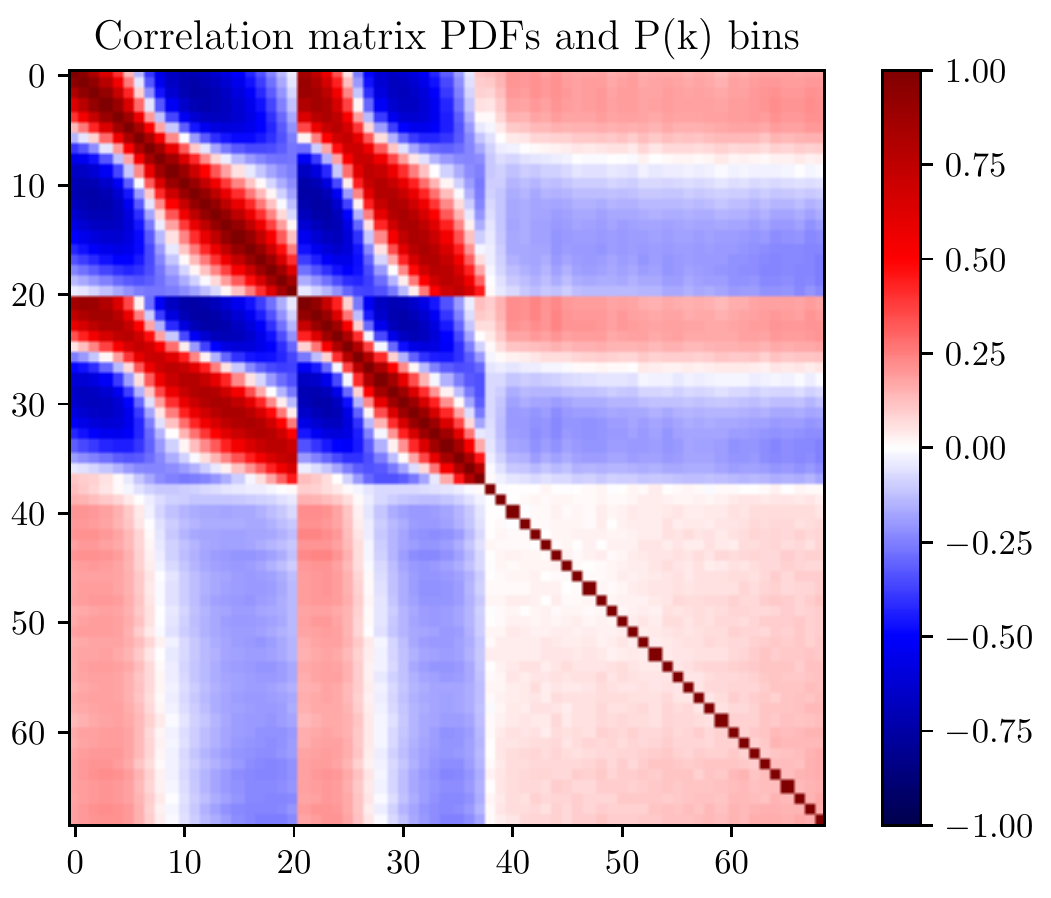}
   \caption{Correlation matrix for the matter PDF $\mathcal P_R(\rho)$ at radii $R=10, 15$ Mpc$/h$ and the mildly nonlinear power spectrum $P(k)$, both at redshift $z=0$. \Cora{The density PDF bins correspond to a range of densities as shown in Figure~\ref{fig:correlation_matrix}. The power spectrum  is shown in 31 bins of the fundamental frequency $k_f\simeq 0.0063 h/$Mpc up to $k_{\rm max}=0.2 h/$Mpc.}}
  \label{fig:correlation_matrix_PDFPk}
\end{figure}
\end{centering}

Figure~\ref{fig:correlation_matrix_PDFPk} shows the cross-correlation matrix between the matter PDF at two different scales and the mildly nonlinear power spectrum. First, we observe that PDFs at different scales are relatively strongly correlated with each other. This is expected, as the matter density fields smoothed at radii of $R=10$ and $R=15$ Mpc$/h$ are qualitatively similar, as about of a third of the mass in a sphere of radius $R=15$ Mpc$/h$ comes from a sub-sphere of $R=10$ Mpc$/h$. \Cora{The cross-correlations between PDF bins of different radii  look very similar to the bin correlations for the individual PDFs, because the clustering of spheres encoded in the sphere bias changes mildly with radius \citep[see Figure 5 in][]{Uhlemann17Kaiser}.} The variance of the PDF at the smaller scale probes a slightly wider range in the nonlinear power spectrum, as we demonstrate in Figure~\ref{fig:sigmaNLintegrand}. In contrast, there is hardly any cross-correlation between the individual $k$-bins of the matter power spectrum since we focus on mildly nonlinear scales. The correlation between the PDFs and the power spectrum is overall small, suggesting that the two probes are complementary and their combination can increase the constraining power. \Cora{The correlation between the power spectrum and the PDF bins varies very mildly with the considered $k$-bin, because it reflects the correlation of the variance with the PDF bins. As can be inferred from Figure~\ref{fig:diffPDFsimvstheo_sig8}, the variance is negatively correlated with the overall height of the PDF, but positively correlated with the tails. Due to our conservative cut in overdense regions (chosen to mitigate the impact of resolution effects and nonlinear tracer bias), the positively correlated overdense tail does not appear in the correlation matrix for the Fisher analysis.} 

For our multi-redshift analysis, we assume no cross-correlation between different redshift slices and build a block-diagonal covariance matrix. We adopt this approach as it is common practice in cosmological analysis of galaxy clustering in current surveys like DES \citep{Krause17}, and because correlations between different redshift bins are expected to be negligible for non-neighbouring bins of size $\Delta z=0.1$ as intended for Euclid \citep{Bailoni17}. Additionally, estimating those correlations from our simulation suite would not lead to realistic results, as the snapshots are extracted from identical runs and hence highly correlated.

\subsection{Derivatives with respect to cosmological parameters}
\label{sec:derivatives}

The third ingredient of the Fisher matrix~\eqref{eq:Fisher} are the derivatives of the matter PDFs with respect to cosmological parameters. They can be obtained from the differences between the matter PDFs with one cosmological parameter varied each, which have been discussed in Section~\ref{sec:differences}. For $\Lambda$CDM parameters, $\theta\in \{\Omega_{\rm m}, \Omega_{\rm b}, h, n_s, \sigma_8\}$), we compute partial derivatives from two-point finite differences
\begin{equation}
\frac{\partial \vec{S}}{\partial \theta}\simeq\frac{\vec{S}(\theta+d\theta)-\vec{S}(\theta-d\theta)}{2d\theta}\,.
\label{eq:derivatives_LCDM}
\end{equation}
For variations in the total neutrino mass $M_\nu$, we use finite difference formulas to estimate the derivative at the left endpoint $M_\nu=0$ using two, three and four points 
\begin{align}
\label{eq:derivatives_Mnu}
\frac{\partial \vec{S}}{\partial M_\nu} &\simeq \frac{\vec{S}(M_\nu)-\vec{S}(M_\nu=0)}{M_\nu}\nonumber\,,\\
\frac{\partial \vec{S}}{\partial M_\nu} &\simeq \frac{-\vec{S}(2M_\nu) + 4\vec{S}(M_\nu) - 3\vec{S}(M_\nu=0)}{2M_\nu}\,,\\
\frac{\partial \vec{S}}{\partial M_\nu} &\simeq \frac{\vec{S}(4M_\nu) - 12\vec{S}(2M_\nu) + 32\vec{S}(M_\nu) - 21\vec{S}(M_\nu=0)}{12M_\nu}\,.\nonumber
\end{align}
By default, we rely on the 4-point derivative and use the other formulas for consistency checks.
While for technical reasons, the simulations have been run with zero fiducial total neutrino mass, we know already that there is a lower limit to the total neutrino mass of about $M_\nu \gtrsim 0.056eV$ \citep{Lesgourgues06}. Hence, we will quote constraints in terms of $\Delta M_\nu$ and avoid to fold in a hard prior to enforce a positive neutrino mass.

\subsection{Constraining $\Lambda$CDM and massive neutrinos}
\label{sec:complement}

In the following we present constraints on key $\Lambda$CDM parameters and the total neutrino mass using the matter PDF at different redshifts and radii. We compare its constraining power with results from the matter power spectrum and combine the two large-scale structure probes.

We have verified the convergence of our results, as constraints do not change if the covariance and derivatives are computed from a smaller number of realisations, or the massive neutrino derivatives are evaluated using lower order approximations.

\subsubsection{Understanding constraints from the PDF at redshift zero}

\begin{figure}
\includegraphics[width=1\columnwidth]{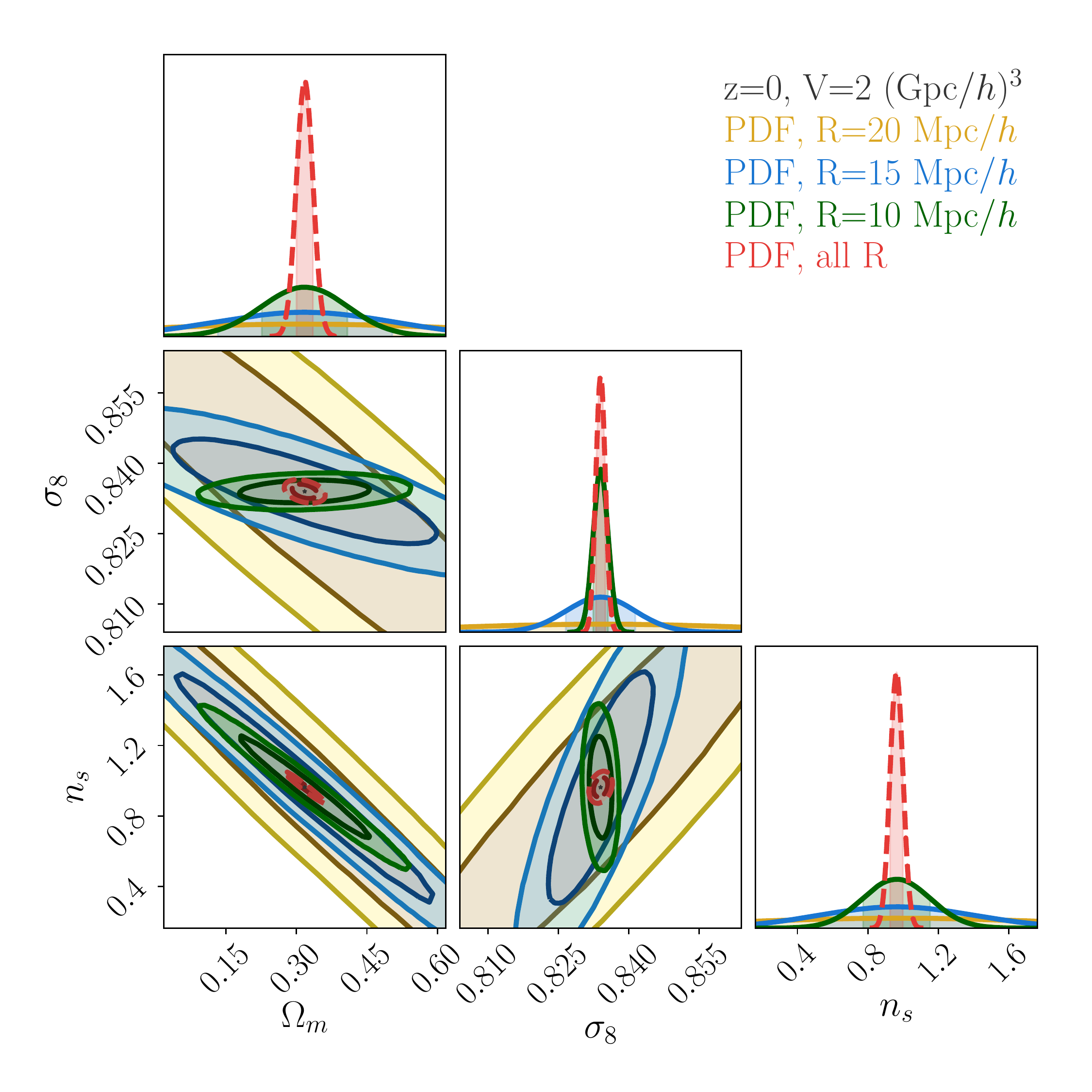}  \\
   \caption{Fisher forecast constraints on $\{\sigma_8,\Omega_m, n_s\}$ (with fixed $\Omega_b$ and $h$) from the matter PDF at redshift $z=0$ using a single radius $R=20$ Mpc$/h$ (yellow), $R=15$ Mpc$/h$ (blue), $R=10$ Mpc$/h$ (green) or three radii combined (red dashed).} 
   \label{fig:Fisherz0R_3param}
\end{figure}

\begin{figure}
\includegraphics[width=1\columnwidth]{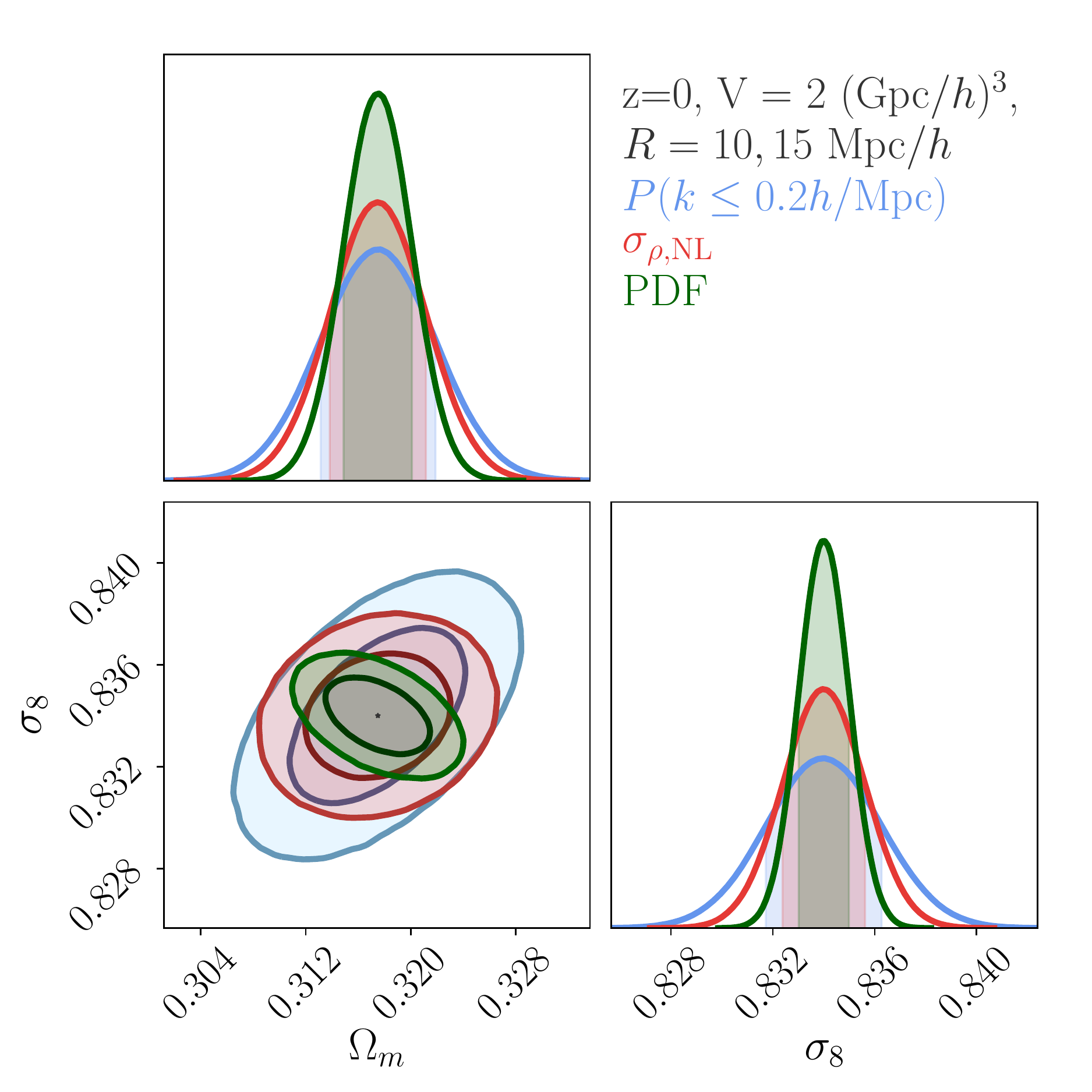}\\
   \caption{Fisher forecast for constraints on $\{\sigma_8,\Omega_m\}$ at redshift $z=0$ from the matter PDF (green) or the nonlinear variance (red) using two radii $R=10$ and $R=15$ Mpc$/h$, and the matter power spectrum up to $k_{\rm max}=0.2 h/$Mpc (blue). 
   By considering the shape of the PDF the area of the contours shrinks by a factor of about $2.5$ compared to the variance and by more than a factor of $3.5$ compared to the mildly nonlinear power spectrum.} 
   \label{fig:Fisherz0_PDFvssig}
\end{figure}

A Fisher forecast for constraints on the three $\Lambda$CDM parameters $\{\Omega_m,\sigma_8,n_s\}$ using one redshift $z=0$ and different  radii $R=10,15,20$ Mpc/h, as well as their combination is presented in Figure~\ref{fig:Fisherz0R_3param}. The degeneracy between the matter density $\Omega_m$ and the spectral index $n_s$ arises because the shape of the PDF around its peak is sensitive to the overall `tilt' of the scale-dependent linear variance, as explained in Section~\ref{sec:scaledepvar}. Since the magnitude of the induced additional tilt depends on scale, the degeneracy direction between $\Omega_m$ and $n_s$ slightly rotates when increasing the radius. When increasing the radius of the spheres, most of the change in the PDF shape is due to a change in the nonlinear variance, which induces a degeneracy between $\sigma_8$ and $\{\Omega_m,n_s\}$, as expected from Figure~\ref{fig:diffPDFsimvstheo_Om_Ob_ns_h}. When combining the matter PDF at different radii, those degeneracies are broken and the constraining power on $\Omega_m$ and $n_s$ is significantly enhanced.

In Figure~\ref{fig:Fisherz0_PDFvssig} we demonstrate that the shape of the PDF in the central region, even excluding the tails, contains more information than the nonlinear variance of the smoothed density field. We use two radii for this analysis, because the variance measured at $N$ scales can only constrain $N$ parameters. We compare the Fisher forecasts at redshift $z=0$ for the two $\Lambda$CDM parameters $\{\sigma_8,\Omega_m\}$, with all other parameters fixed, using the matter PDF (green) and the nonlinear variance\footnote{Note that the nonlinear variance was measured directly from the grid of smoothed densities instead of from the histogram that gives the PDF.} (red). By considering the shape of the PDF rather than just the variance, the area of the contours shrinks by a factor of $2.5$. For comparison, we show Fisher constraints for the matter power spectrum on mildly nonlinear scales up to $k_{\rm max}=0.2 h$/Mpc (blue), whose constraints are weaker  by a factor of 3.5. This demonstrates that the matter PDF measured at two scales contains more information on $\Omega_m$ and $\sigma_8$ than both the nonlinear variance and the matter power spectrum, when all other parameters are fixed.

\subsubsection{Increasing constraining power with multiple redshifts}

\begin{figure}
\includegraphics[width=1\columnwidth]{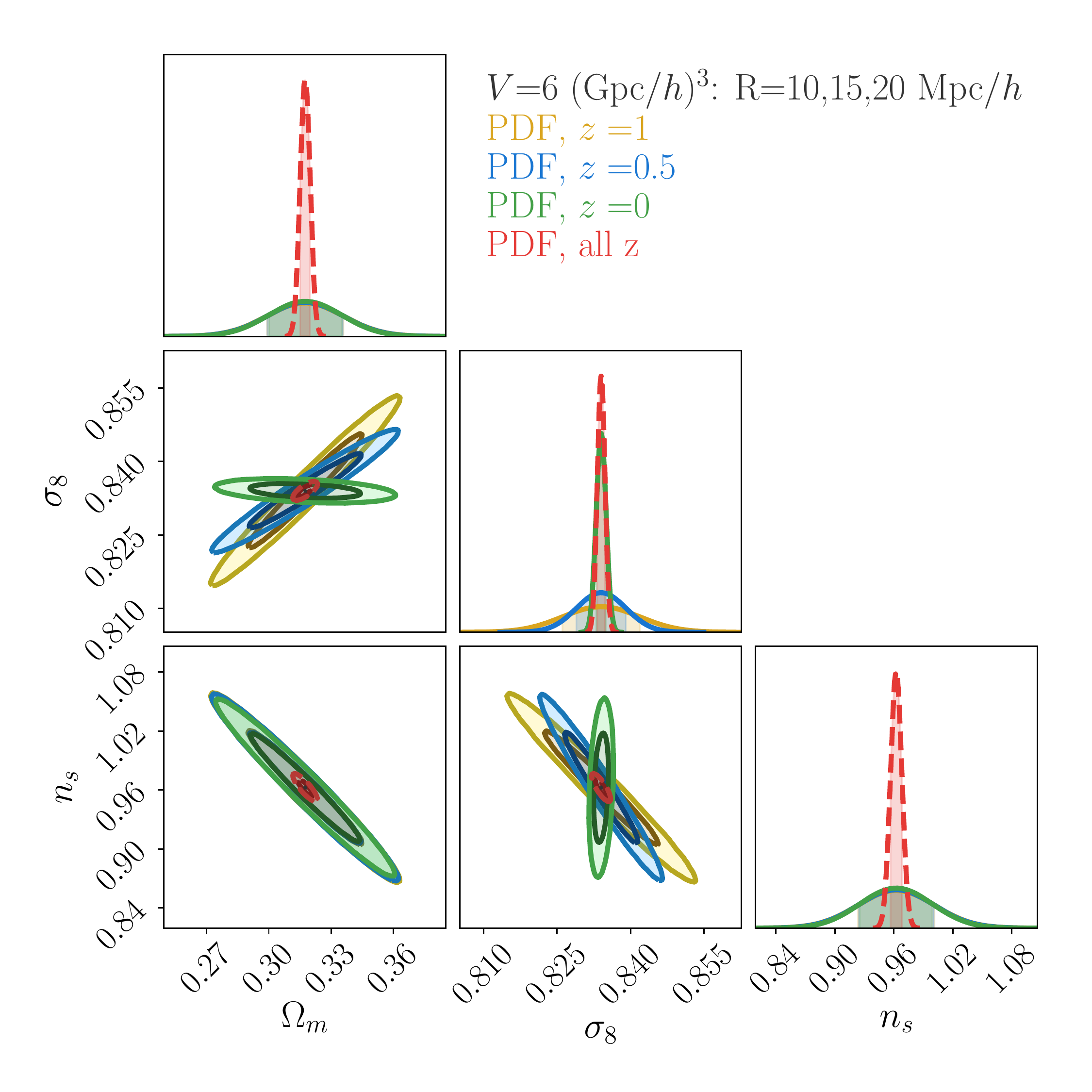}  \\
   \caption{Fisher forecast for PDF constraints on $\{\sigma_8,\Omega_m ,n_s\}$ (with fixed $\Omega_b$ and $h$) from the matter PDF with three radii $R=10,15,20$ Mpc$/h$ at redshifts $z=0$ (green) $0.5$ (blue) and $1$ (yellow)  Mpc$/h$, each for one third of the volume, and combined to the total volume (red dashed). Note that the green contours here correspond to the red contours in Figure~\ref{fig:Fisherz0R_3param}} 
   \label{fig:degeneracybreakingdiffz}
\end{figure}

From theoretical grounds, we expect that the PDF at nonzero redshift will have a degeneracy between the clustering amplitude $\sigma_8$ and the matter density $\Omega_m$, which affects the linear variance through the growth $D(z)$ according to $\sigma_L(z,R)\propto D(z)\sigma_8 $. In Figure~\ref{fig:degeneracybreakingdiffz} we show that measuring the matter PDF at different redshifts breaks this degeneracy between $\sigma_8$ and $\Omega_m$. A split in redshift slices also helps to disentangle a change in matter density $\Omega_m$, that changes both the scale-dependence and the growth of the linear variance, from a change in the spectral index $n_s$. We consider three redshifts $z=0,0.5,1$, which are of particular interest for upcoming galaxy surveys like Euclid and LSST. For simplicity, we split the total volume $V$ evenly across the supposedly independent redshift slices, such that the constraints shown for the three individual redshifts use only one third of the total volume. 

\begin{figure}
\includegraphics[width=1\columnwidth]{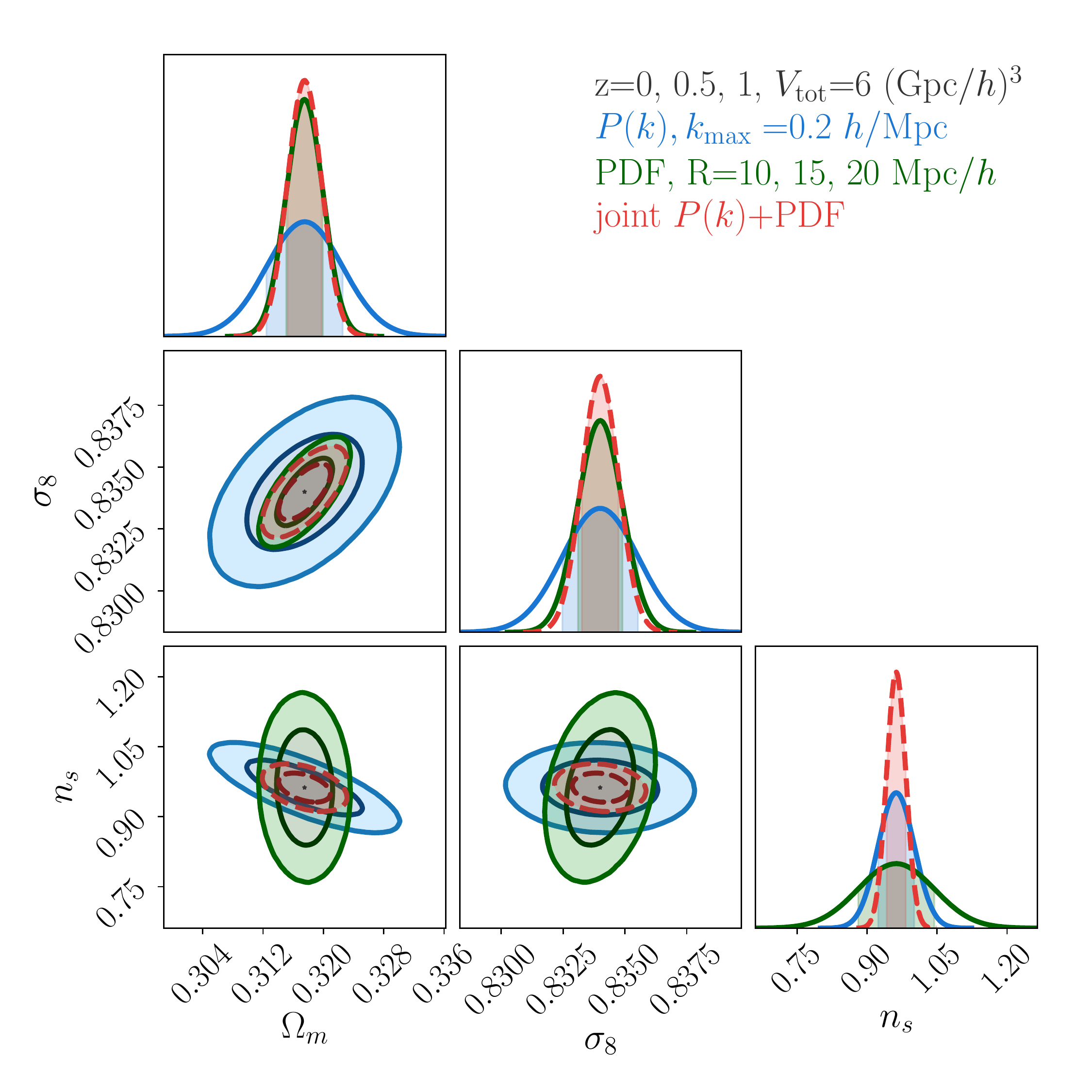}  \\
\centering
    \begin{tabular}{c|ccc}
        \hline
		 & $\Delta \Omega_m$ & $\Delta \sigma_8$& $\Delta n_s$ \\ 
		\hline
		$P(k)$ & $\pm 0.00503$ & $\pm 0.00153$ & $\pm 0.0386$ \\
		PDF & $\pm 0.00243 $ & $\pm 0.000895$ & $\pm 0.081$ \\ 
		joint & $\pm 0.00225$ & $\pm 0.000740$ & $\pm 0.0205$ \\
    \end{tabular}
\caption{Fisher forecast for marginalised constraints on $\{\sigma_8,\Omega_m ,n_s\}$ from an analysis at three redshifts $z=0,0.5,1$ for the matter PDF at three radii $R=10,15,20$ Mpc$/h$ (green), the matter power spectrum up to $k_{\rm max}=0.2 h/$Mpc (blue) and both probes combined (red dashed).}
\label{fig:FisherLCDMsel_PDF+Pk}
\end{figure}

\vspace{-0.2cm}
\subsubsection{Combining the matter PDF and power spectrum}
Using the matter power spectrum and the matter PDF at mildly nonlinear scales as complementary probes,
one can enhance the constraining power by combining both observables. To demonstrate this, we first focus on the five $\Lambda$CDM parameters. In Figure~\ref{fig:FisherLCDMsel_PDF+Pk} we show constraints on $\{\Omega_m,\sigma_8,n_s\}$ (marginalised over $\Omega_b$ and $h$) obtained from the matter PDF at three radii and the mildly nonlinear matter power spectrum, both analysed at three redshifts. \Cora{The $1\sigma$ constraints quoted in the Table are obtained by marginalising over all other parameters.} We find that the matter PDF is strong at constraining the clustering amplitude $\sigma_8$ and matter density $\Omega_m$, as expected from Figure~\ref{fig:Fisherz0_PDFvssig}. Constraints on those two parameters improve only by about 10-15\% when adding the matter power spectrum. However, the matter power spectrum is stronger at constraining $n_s$ by about a factor of two, and adding the PDF improves constraints by another factor of 2. As expected, the matter power spectrum shape can more easily distinguish between changes in $\Omega_b$, $h$ and $n_s$, which all lead to a similar signature in the matter PDF as demonstrated in Figure~\ref{fig:sigmalincomparison}.

\begin{figure}
\includegraphics[width=1\columnwidth]{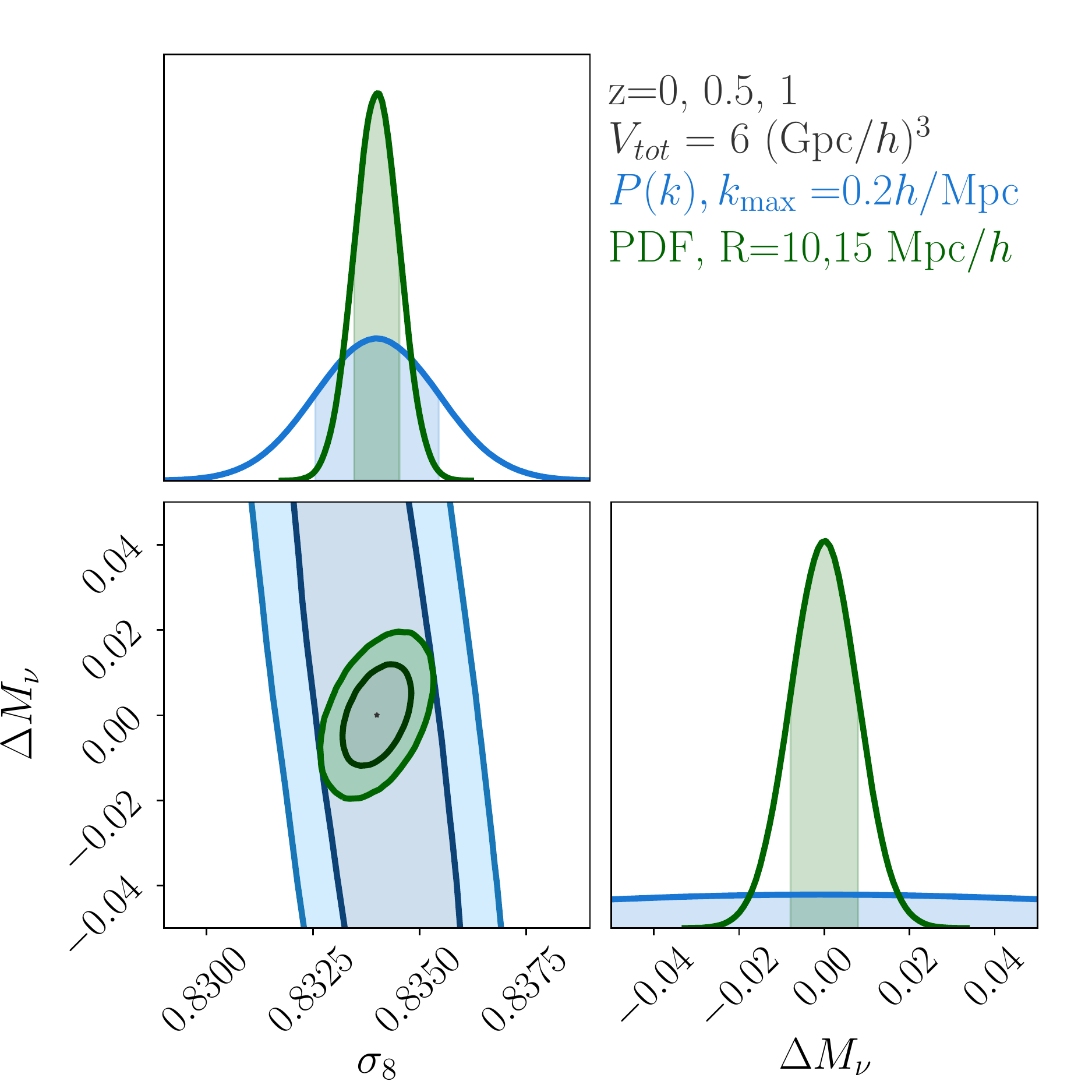}  \\
   \caption{Fisher forecast for PDF constraints on $\{\sigma_8,M_\nu\}$ (fixing all other parameters) from a joint analysis at redshifts $z=0,0.5,1$ for the matter power spectrum up to $k_{\rm max}=0.2 h/$Mpc (blue) and the matter PDF at two radii $R=10,15$ Mpc$/h$ (green).} 
   \label{fig:Fisher_sig8Mnu_PDFvsPk}
\end{figure}

\begin{figure*}
\includegraphics[width=2\columnwidth]{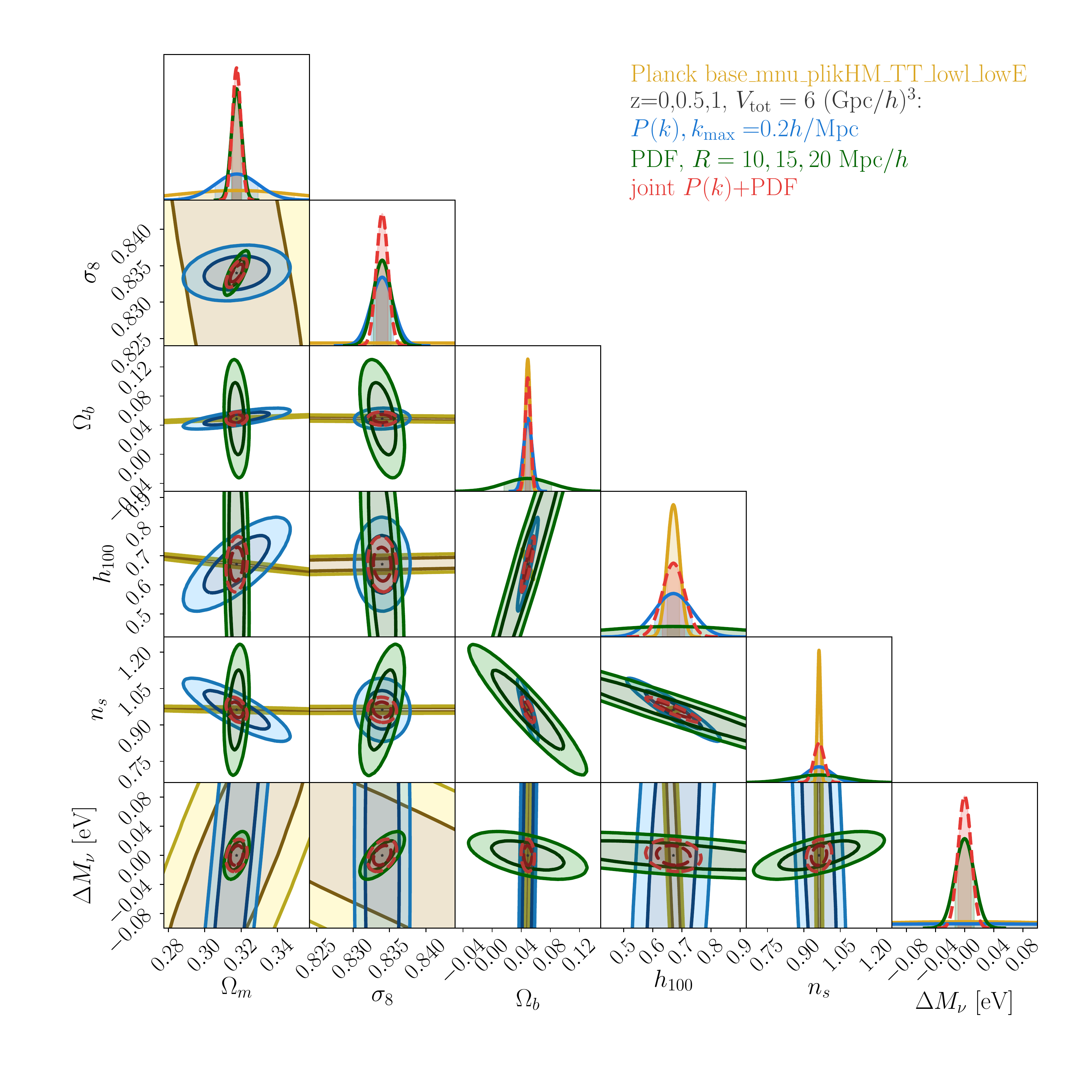}  \\
    \centering
    \label{tab:model_params}
    \large
    \begin{tabular}{l|cccccc}
        \hline
		 & $\Delta\Omega_m$ & $\Delta\sigma_8$ & $\Delta\Omega_b$ & $\Delta h_{100}$ & $\Delta n_s$ & $\Delta M_{\nu} [eV]
		$ \\ 
		\hline
		Planck base\_mnu\_plikHM\_TT\_lowl\_lowE & $\pm 0.0315$ & $\pm 0.0372$ & $\pm 0.00317 $ & $\pm 0.0212$ & $\pm 0.006214 $ & $\pm 0.181$ \\
		\hline
		$P(k), z=0,0.5,1, k_{\rm max}=0.2 h/$Mpc & $\pm 0.0117$ & $\pm 0.00153$ & $ \pm 0.00575 $ & $\pm 0.0644$ & $\pm 0.0520$ & $\pm 0.267$ \\
		\hline
		PDF, $ z=0,0.5,1, R=10,15,20$ Mpc$/h$ & $\pm 0.00276 $ & $ \pm 0.00122$ & $\pm 0.0324$ & $\pm 0.267$ & $\pm 0.108$ & $\pm 0.0131$ \\ 
		\hline
		joint $P(k)$ + PDF & $\pm 0.00232$ & $\pm 0.000792$ & $\pm 0.00370$ & $\pm 0.0380$ & $\pm 0.0208$ & $ \pm 0.00887 $
    \end{tabular}
   \caption{Fisher forecast for constraints on $\nu\Lambda$CDM parameters at redshifts $z=0,0.5,1$ using the matter PDF for three radii $R=10,15,20$ Mpc$/h$ (green), the matter power spectrum $P(k)$ up to $k_{\rm max}=0.2 h/$Mpc (blue) or both in combination (red dashed). The combination with the power spectrum improves the constraints on the $\Lambda$CDM parameters, while the information on the total neutrino mass $M_\nu$ comes almost exclusively from the PDF. For comparison, we also show contours from Planck (yellow) obtained from the chains for free neutrino mass from temperature and low multipole polarisation data only (base\_mnu\_plikHM\_TT\_lowl\_lowE).} 
   \label{fig:Fisher_cosmo_neutrino_PDF+Pk}
\end{figure*}

\subsubsection{Joint $\Lambda$CDM and neutrino mass constraints}
\label{sec:nuLCDM}

Having established the complementarity of the matter PDF and matter power spectrum at mildly nonlinear scales for $\Lambda$CDM parameters, we now include the total mass of massive neutrinos $M_\nu=\sum m_\nu$ as additional parameter. 

For the matter power spectrum, the total neutrino mass is known to be largely degenerate with the amplitude of matter fluctuations $\sigma_8$ \citep{Paco_18a}. In Figure~\ref{fig:Fisher_sig8Mnu_PDFvsPk}, we contrast constraints from the matter power spectrum up to $k_{\rm max}=0.2 h/$Mpc (blue) to the matter PDF at two radii $R=10, 15$ Mpc$/h$ (green), both using three redshifts and fixing all parameters except for $\sigma_8$ and $M_\nu$. We observe that while the matter power spectrum presents a strong degeneracy \Cora{(which is not alleviated by extending the range to $k_{\rm max}=0.5 h/$Mpc)} the matter PDF can easily disentangle the two parameters, as expected from the imprint of scale-dependent neutrino clustering shown in Figure~\ref{fig:diffPDFsimvstheo_Mnu}. Indeed, the impressive constraining power of the matter PDF for the total neutrino mass is hardly diminished by opening up all $\Lambda$CDM parameters.

 In Figure~\ref{fig:Fisher_cosmo_neutrino_PDF+Pk} we show constraints on the full set of $\nu\Lambda$CDM parameters from an analysis in three redshift slices at $z=0,0.5,1$ with a combined volume of $6$ (Gpc$/h$)$^3$. We show one and two-sigma contours obtained using the matter power spectrum up to $k_{\rm max}=0.2 h/$Mpc (blue), the matter PDF at three radii $R=10,15,20$ Mpc$/h$ (green) and their combination (dashed, red). We choose the scales to be in a regime where theoretical predictions, based on perturbation theory for the matter power spectrum, and large-deviation statistics with spherical collapse for the density PDF, can be expected to be accurate. For comparison, we show results from a Gaussian likelihood approximation of the Planck CMB data with free neutrino mass using only temperature and low multiple polarisation data (yellow). While the matter power spectrum has virtually no sensitivity to neutrino mass, as demonstrated in Figure~\ref{fig:Fisher_sig8Mnu_PDFvsPk} and Figure~15 in \cite{Quijote}, it helps to improve constraints on $\Lambda$CDM parameters and combining it with the matter PDF tightens neutrino mass constraints by 30\% to less than 0.01 eV. Combining the matter PDF and matter power spectrum improves constraints for the matter density $\Omega_m$ by a factor of 5, and for the clustering amplitude $\sigma_8$ by a factor of 2 compared to power spectrum only.

\Cora{The improvement of neutrino mass constraints by considering the matter PDF compared to the matter power spectrum is much bigger than the one expected from adding the matter bispectrum. As demonstrated in \citet{Coulton19bispecMnu}, the tomographic weak lensing convergence bispectrum has a similar degeneracy between $M_\nu$ and the clustering amplitude $\sigma_8$ (or a combination of $A_s$ and $\Omega_m$) than the power spectrum. This suggest that the additional constraining power of the PDF is rooted in its ability to detect differences in clustering between underdense and overdense regions, which are sensitive to neutrino mass as shown in Figure~\ref{fig:changerho} but get mixed up in $N$-point spectra. While focused on different scales, recent simulation results for massive neutrino constraints from the weak lensing convergence PDF in the presence of shape noise \citep{Liu19WLPDF} indeed show a turning of the degeneracy direction between $M_\nu$ and the clustering amplitude compared to the power spectrum.} \Cora{On the other hand, the nonlinear redshift-space halo bispectrum was recently shown to significantly improve constraints from the redshift space halo power spectrum due to the shape-dependent imprint of massive neutrinos measured in simulations \citep{Hahn19bispecMnu}.}

\subsection{Applicability to survey data}
\label{sec:applicability}

\Cora{In this study, we have investigated the statistical power of the3D matter density PDF as a cosmological probe. In realistic observational data we cannot access those 3D matter density in cells directly. Observables that probe the matter density field fall into two broad categories: tracer densities and weak lensing fields.}

One possibility is to \Cora{avoid tracer bias altogether by extracting} the weak lensing convergence or shear field, which probes the total projected matter field. \Cora{In the spirit of our analysis, one could} parametrise the cosmology dependence for PDFs of the weak lensing convergence \citep{Valageas00,Clerkin16,Patton17,Barthelemy19}, aperture mass \citep{BernardeauValageas00,Reimberg18} or cosmic shear \citep{Takahashi11,Friedrich18}. Recently, the convergence PDF \Cora{in tomographic redshift slices} has been predicted from large-deviation statistics and cylindrical collapse \citep{Barthelemy19}, using ingredients similar to the ones discussed here.  \Cora{While weak lensing is insensitive to tracer bias, the presence of baryons could affect the total matter field on small scales.
According to \cite{Foreman19}, baryonic effects do not change the scaling relations between the matter bispectrum and the power spectrum on mildly nonlinear scales. This suggests that baryonic effects on the PDF could potentially be modelled through their impact on the nonlinear variance, while leaving the reduced skewness $S_3$ untouched. Shape noise effects that encapsulate the uncertainties on the intrinsic shape of galaxies can be modeled by convolving the weak lensing PDF with a Gaussian filter of appropriate width \citep{Liu19WLPDF,Barthelemy19}. }

A second option is to extract biased tracer densities from discrete counts of galaxies \citep{Yang11,Bel16,Hurtado-Gil17}, \Cora{Lyman-alpha absorption in quasar spectra \citep{Munshi12,Lidz06}} or \Cora{21cm} emission of neutral hydrogen \citep{Leicht19}. The impact of tracer bias on the shape of the PDF can be modeled using scatter plots between the matter and tracer density in cells extracted from simulations \citep{Manera11,Jee12}, measured moments \citep{Salvador18} or abundance-matching inspired techniques that directly operate on the PDFs \citep{Sigad00,Szapudi04}. As shown in \cite{Uhlemann18bias}, suitable weightings by halo mass or galaxy luminosity can substantially reduce the scatter around the mean bias relation, and redshift space distortions can potentially be absorbed in the bias model.

When focusing solely on the tracer density PDF, a partial degeneracy between the linear bias coefficient $b_1$ and the amplitude of matter fluctuations $\sigma_8$ arises. Breaking this degeneracy can be achieved by considering the density-dependent clustering of spheres \citep[called sphere bias][]{Bernardeau96,Codis16b,Uhlemann17Kaiser,Uhlemann18bias}, which can also be used to quantify the cosmic error induced by extracting counts-in-cells statistics from a finite number of tracers in a finite volume \citep{Colombi95,SzapudiColombi96,Szapudi99,Codis16b}.

\Cora{A middle ground in between measuring weak lensing and modelling bias are the so-called density-split statistics \citep{Gruen16troughs, Friedrich18, Gruen18} that measure the galaxy density PDF and use lensing measurements to relate it to the matter density PDF quantile-by-quantile. \citet{Gruen18} have successfully applied this technique in an analysis of observational data taken from the Dark Energy Survey and SDSS, thus demonstrating that tracer bias can be dealt with in a PDF-based analyses. With photometric galaxy surveys, we do not have access to the undistorted 3D density field, but instead probe line-of-sight projections of the density field with a certain, irreducible uncertainty in the radial (redshift) direction. However, the formalism employed in this work can be extended to account for such projections \citep{Bernardeau95ang,Uhlemann18cyl} and has already been successfully applied to observational data in \citet{Friedrich18, Gruen18}.}

\section{Conclusion}
\label{sec:conclusion}

In this work, we determined the information content of the matter density PDF with regards to all $\Lambda$CDM parameters and the total neutrino mass. Based on a theoretical model for the matter PDF from large-deviation statistics, we analysed the impact of cosmological parameters on the ingredients that determine the shape of the matter PDF. We demonstrated that the $\Lambda$CDM parameter dependence of the matter PDF can be predicted from the scale-dependence of the linear variance, the growth of structure, spherical collapse and the nonlinear variance at the considered radius. For the first time, we modelled the impact of massive neutrinos on the total matter PDF, finding that their distinct imprint on the shape 
\Cora{is due to a massive neutrino background affecting underdensities and their partial clustering along with the cold dark matter plus baryon component. In all cases, we find an excellent agreement between the theoretically predicted and numerically measured response of the matter PDF to changing cosmological parameters.}

Finally, we performed a Fisher analysis and demonstrated that measuring the PDF in multiple redshift slices and at different radii breaks parameter degeneracies and tightens constraints. In Figure~\ref{fig:Fisher_cosmo_neutrino_PDF+Pk}, we demonstrated the significant constraining power of the matter PDF for the matter density $\Omega_m$, the clustering amplitude $\sigma_8$ and the total neutrino mass $M_\nu$, highlighting its complementarity to the matter power spectrum and cosmic microwave background data from Planck. Combining the total matter density PDFs at three radii and the matter power spectrum up to mildly nonlinear scales in three redshift slices with a total BOSS-like volume of $6$ (Gpc$/h$)$^3$ gives a marginalised constraint on the total neutrino mass of order 0.01 eV. \Cora{This would allow to place a 5$\sigma$ constraint on the minimum sum of the neutrino masses with a rather small volume. Additionally, the inclusion of the PDF} improves constraints on $\{\Omega_m, \sigma_8, n_s\}$ by a factor of 5, 2, 2.5 compared to the matter power spectrum alone, see Table~\ref{fig:Fisher_cosmo_neutrino_PDF+Pk}. 

This is an exciting prospect for density-split statistics \citep{Friedrich18,Gruen18}, whose combined analysis of counts- and lensing-in-cells allows to constrain bias and stochasticity along with cosmological parameters\Cora{, as discussed in Section~\ref{sec:applicability}.} 

\subsection*{Outlook: primordial non-Gaussianity \& dark energy}

In this work we focus on Gaussian initial conditions, while an accompanying paper \citefuture{(Friedrich et al. 2019)} generalises the theoretical model to include arbitrary non-Gaussian initital conditions. In particular, the focus is on the imprint of a general primordial bispectrum  and illustrate effects of orthogonal and equilateral primordial non-Gaussianity in the matter PDF, complementing previous results for local non-Gaussianity \citep{Uhlemann18pNG}. \Cora{We find that the amplitude of primordial non-Gaussianity $f_{\rm NL}$ can be constrained from the matter PDF at two scales even when marginalising over their variances.}

\begin{figure}
\includegraphics[width=1.\columnwidth]{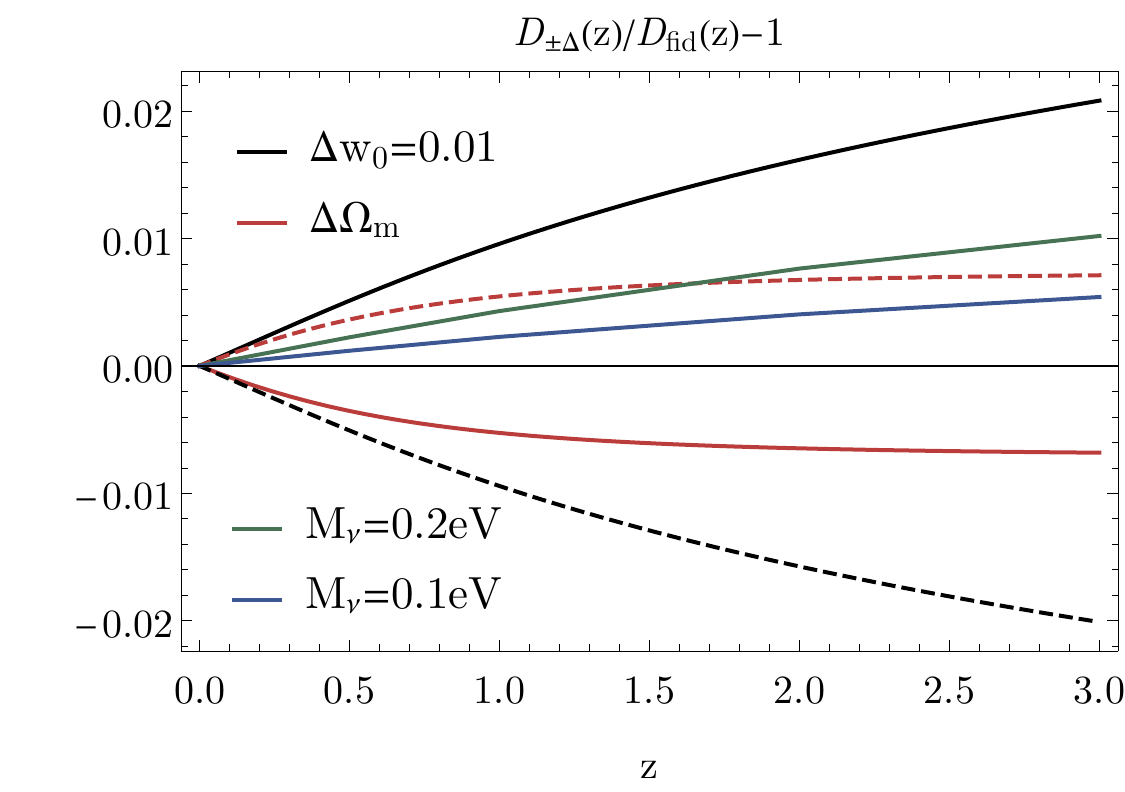}
   \caption{Fractional changes in the linear growth factor $D(z)$ for cosmologies with varying $\Omega_m=\pm 0.01$ (red solid/dashed) and $w_0=\pm 0.01$ (black solid/dashed). We also show the impact of massive neutrinos on the growth of the linear variance for cold dark matter plus baryons at radius $R=15$ Mpc$/h$ for a total neutrino mass of $M_\nu=0.1$eV (blue solid) and 0.2eV (green solid).} 
   \label{fig:Dofzcomparison}
\end{figure}

While it is beyond the scope of this paper, we plan to include changes in the dark energy equation of state in a future analysis.
From our theoretical model, we expect that dark energy affects the matter PDF trough a \Cora{redshift-dependent} change in the variance driven by the growth of structure, see equation~\eqref{eq:Dcosmo}. In principle, this allows us to constrain the dark energy equation of state in a multi-redshift analysis \citep{Codis16b}. In Figure~\ref{fig:Dofzcomparison} we compare  modifications in the linear growth induced by a constant change in the dark energy equation of state (purple) to a change in the matter density (black lines) and the imprint total neutrino mass (red, blue, green). Since the characteristic imprint of massive neutrinos in the PDF is mainly driven by a scale-dependent bias instead of a change in the nonlinear variance induced by the growth, we expect it to be distinguishable from dark energy. When focusing on the time-dependence of the variance at low redshifts, a constant change in the dark energy equation of state could be degenerate with a change in matter density. However, since the matter density also changes the scale-dependence of the linear variance, one could hope to jointly constrain both parameters without loosing too much constraining power.

\section*{Acknowledgements}  
CU kindly acknowledges funding by the STFC grant RG84196 `Revealing the Structure of the Universe'. OF gratefully acknowledges support by the Kavli Foundation and the International Newton Trust through a Newton-Kavli-Junior Fellowship and by Churchill College Cambridge through a postdoctoral By-Fellowship. We thank Ken Osato and Takahiro Nishimichi for making their codes for computing nonlinear matter power spectra publicly available. Part of the work of FVN has been supported by the Simons Foundation. SC's work is partially supported by the SPHERES grant ANR-18-CE31-0009 of the French {\sl Agence Nationale de la Recherche} and by Fondation MERAC. The authors thank Alexandre Barreira, Francis Bernardeau, Daniel Gruen, ChangHoon Hahn, Christophe Pichon and Blake Sherwin for discussions and comments on the draft.
 
\bibliographystyle{mnras}
\bibliography{LSStructure}

\appendix

\section{Approximating the variance}
\label{app:linPSvar}

\subsection{Approximating the cosmology-dependent linear variance}
\label{app:linvarapprox}
\begin{figure}
\includegraphics[width=1.\columnwidth]{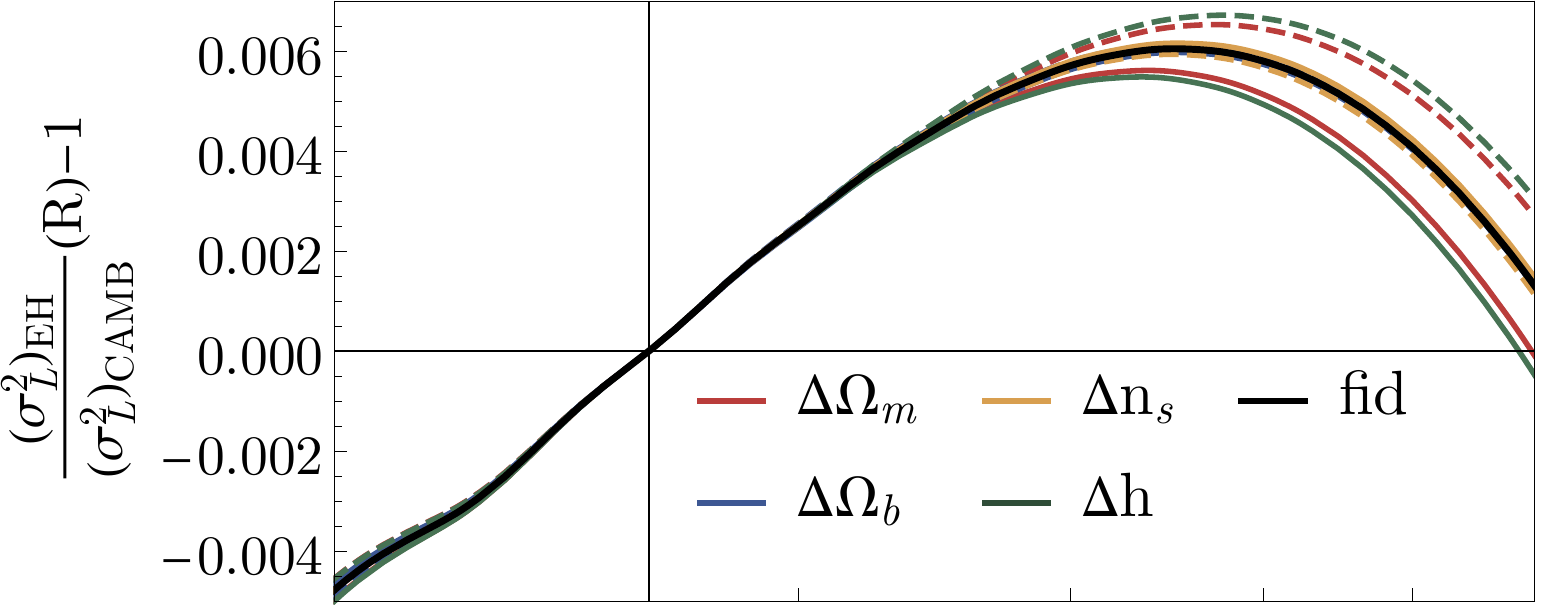}\\
\includegraphics[width=1.\columnwidth]{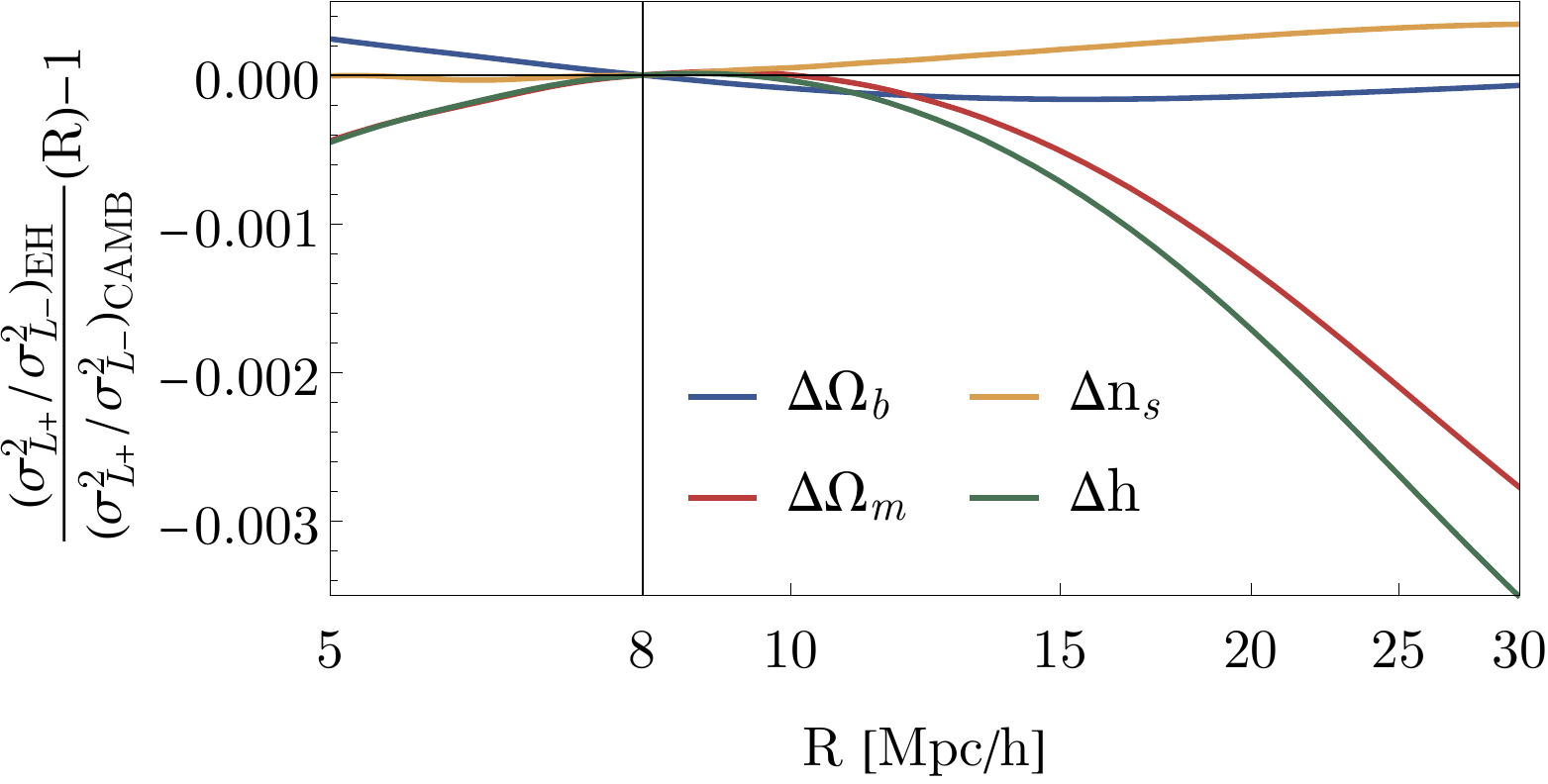}
   \caption{Comparison between the linear variance $\sigma_L^2(R)$ computed from the Eisenstein-Hu approximation vs. CAMB for the fiducial cosmology (black) and the derivative cosmologies with varying $\Omega_m$ (red), $\Omega_b$ (blue), $n_s$ (yellow) and $h$ (green) with positive sign (solid lines) and negative sign (dashed lines) as indicated in Table~\ref{tab:models} with fixed $\sigma_8$. The lower panel shows the fractional difference of the ratio of the linear variance between the positive and negative sign derivative cosmologies. \Cora{We find that derivatives computed using the Eisenstein-Hu have sub-percent accuracy with respect to CAMB.}} 
   \label{fig:sigmalinEisensteinHu}
\end{figure}

In Figure~\ref{fig:sigmalinEisensteinHu} we show how well the Eisenstein-Hu transfer function \citep{EisensteinHu98} captures the cosmology dependence of the linear variance.

For a cosmology with dark energy beyond a cosmological constant, the growth function describing the time-dependence of the variance according to equation~\eqref{eq:siggrowth} can be modelled as \citep{Glazebrook}, 
\begin{align}
\label{eq:Dcosmo}
D(z)&=\frac{5\Omega_m}{2} \frac{H(a)}{H_0}\int_0^a \frac{{\rm d} a' H_0^3}{a'^3 H^3(a')}\,, \\
\frac{H(a)}{H_0} &= \sqrt{\frac{\Omega_m}{a^3}+ \Omega_\Lambda \exp\left(3 \int_0^z \frac{1+w(z')}{1+z'} {\rm d} z'  \right)}\,, \label{eq:Hcosmo}
\end{align}
 with the dark matter density, $\Omega_m$, the dark energy density, $\Omega_\Lambda$, the Hubble constant at zero redshift, $H_0$, the expansion factor $a\equiv1/(1+z)$ and the dark energy equation of state $w(z)$. In Figure~\ref{fig:Dofzcomparison} in Section~\ref{sec:conclusion} we compare the impact of a changed equation of state parameter $w=w_0=-1\pm 0.01$ on the growth with the change of changing $\Omega_m \pm 0.01$.

\begin{figure}
\includegraphics[width=1.\columnwidth]{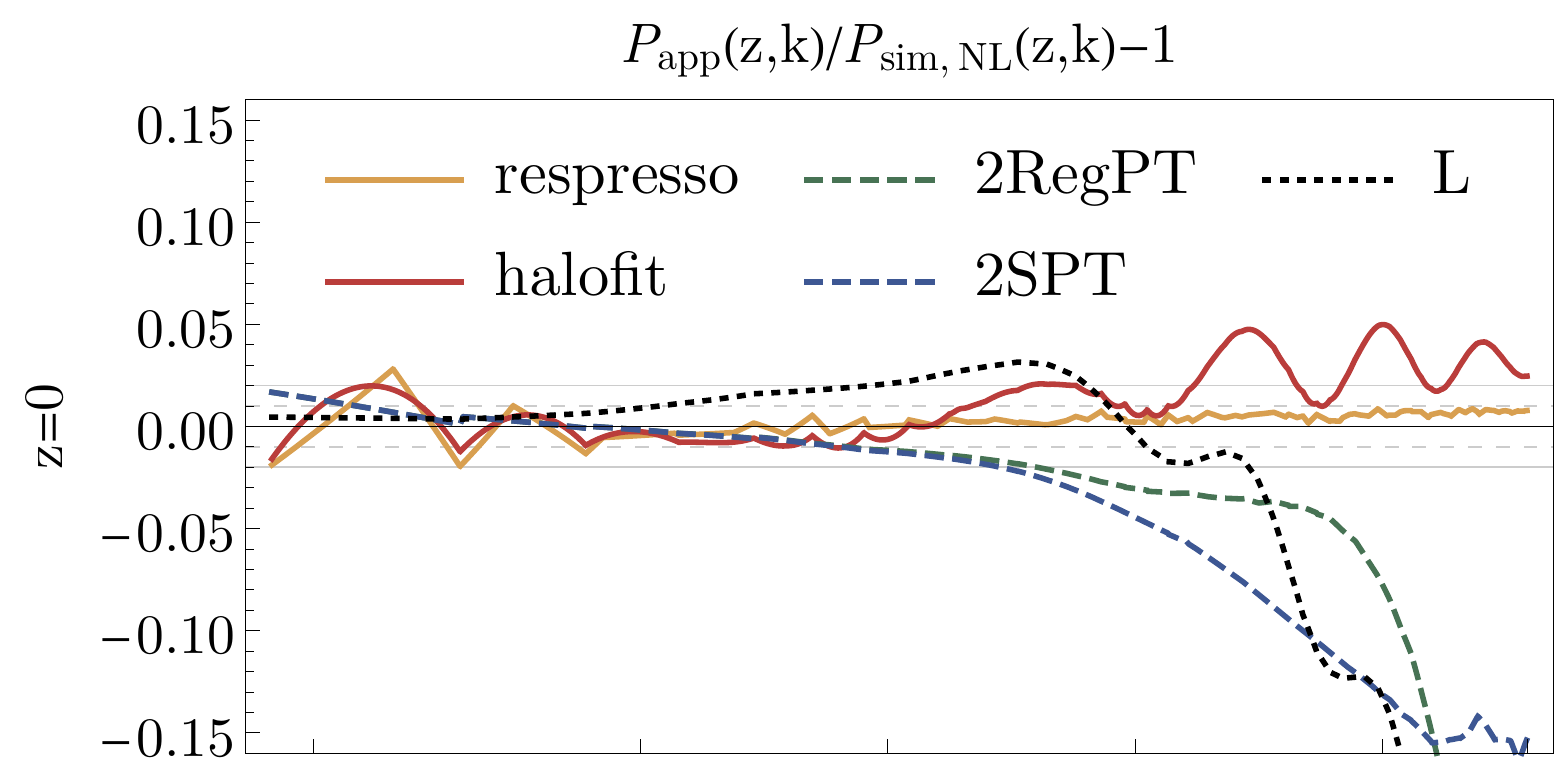}
\includegraphics[width=1.\columnwidth]{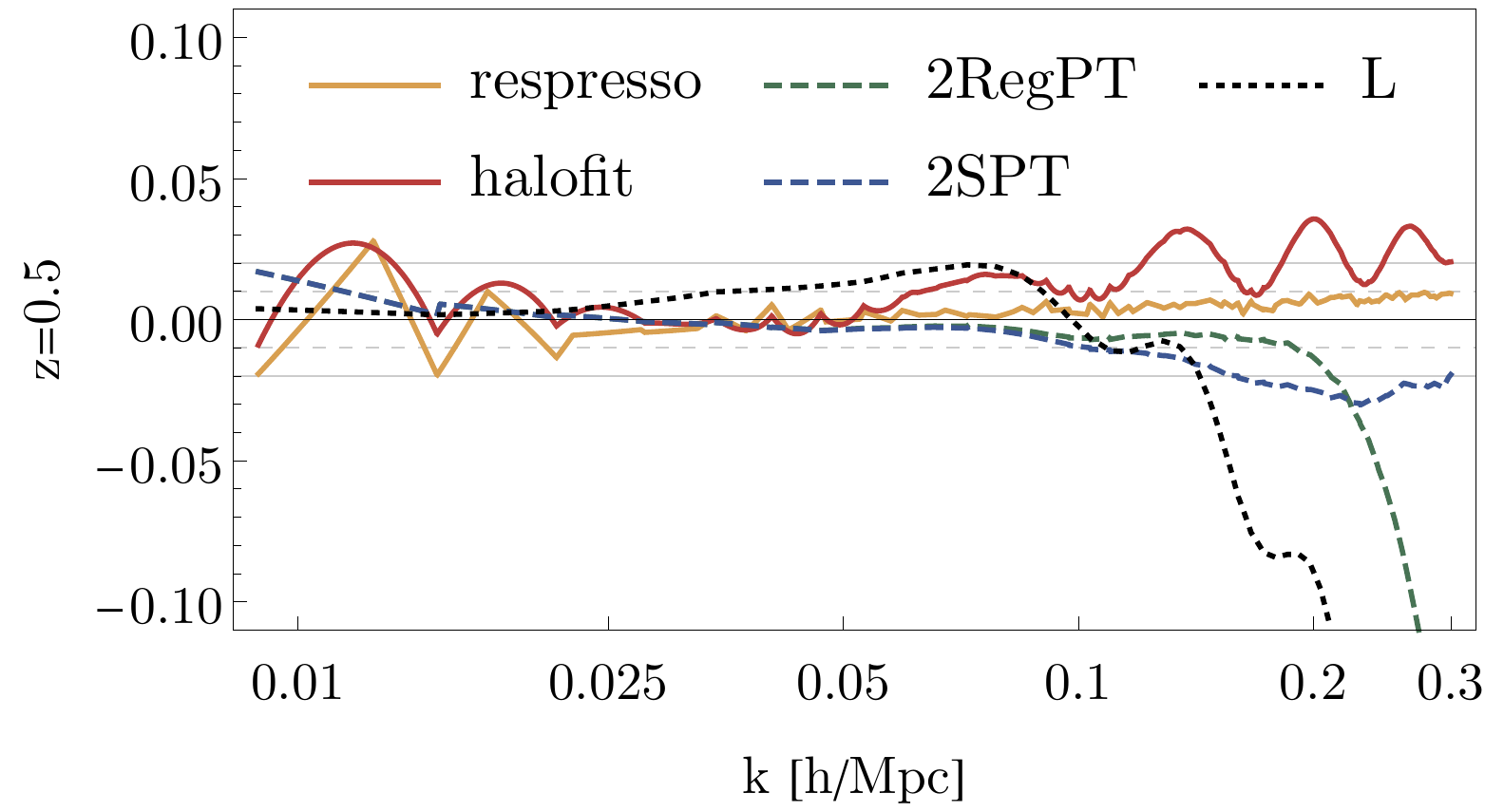}
   \caption{Comparison between the measured nonlinear power spectrum averaged over 150000 realisations of the fiducial simulation at redshift $z=0$ (upper panel) and $z=0.5$ (lower panel) with predictions using respresso (brown),  halofit (red), 2-loop perturbation theory in RegPT (green dashed) or SPT (blue dashed) and linear theory (black dotted). The thin gray lines indicate 2\% error (solid) and 1\% error (dashed).} 
   \label{fig:PNLcomparison}
\end{figure}

\begin{figure}
\includegraphics[width=1.\columnwidth]{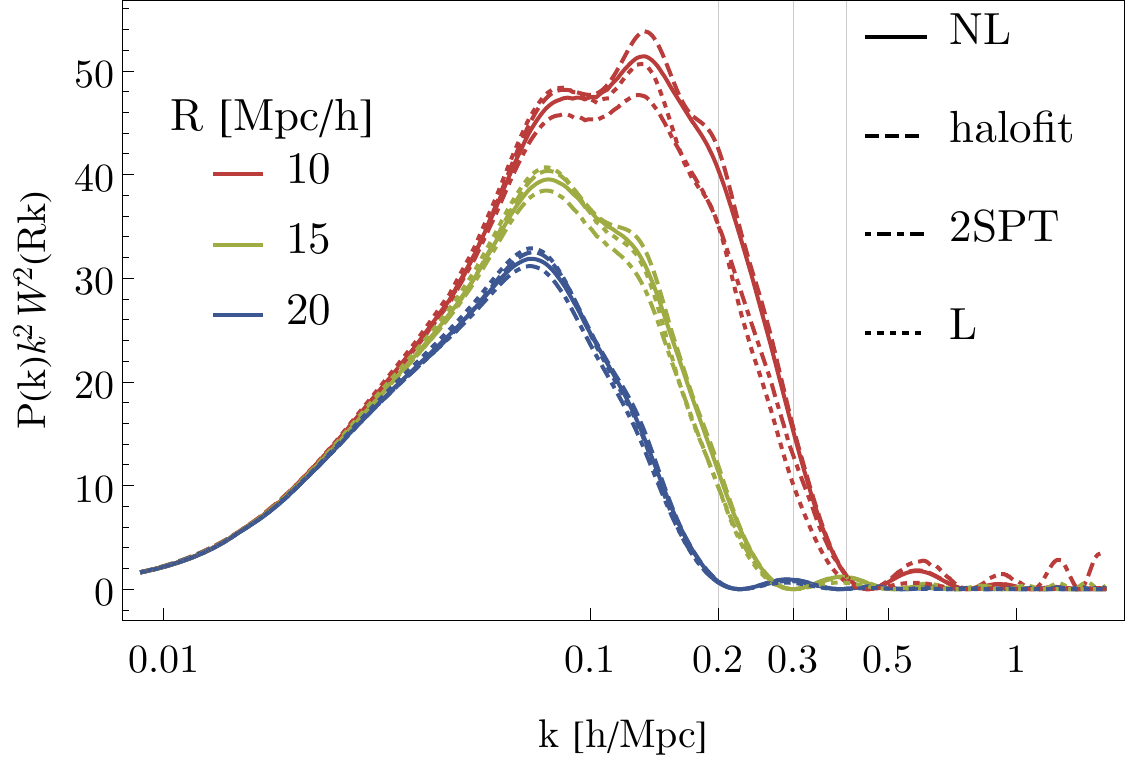}
   \caption{The integrand for the nonlinear variance from equation~\eqref{eq:defSigma2nonlin} at scales $R=10,15,20$ Mpc$/h$ at redshift $z=0$ using different expressions for the power spectrum as indicated in the legend.} 
   \label{fig:sigmaNLintegrand}
\end{figure}

\subsection{Approximating the nonlinear variance}
In Figure~\ref{fig:PNLcomparison} we compare the measured nonlinear power spectrum from the Quijote simulations to standard fitting function (respresso and halofit) as well as perturbative methods at 2-loop order (SPT and RegPT) at redshifts $z=0$ and $z=0.5$.

\Cora{In Figure~\ref{fig:sigmaNLintegrand} we show the integrand that enters the computation of the nonlinear variance for different radii.}

\newpage
\section{Residuals of matter density PDFs}
\label{app:PDFplots}

\begin{figure}
\includegraphics[width=1\columnwidth]{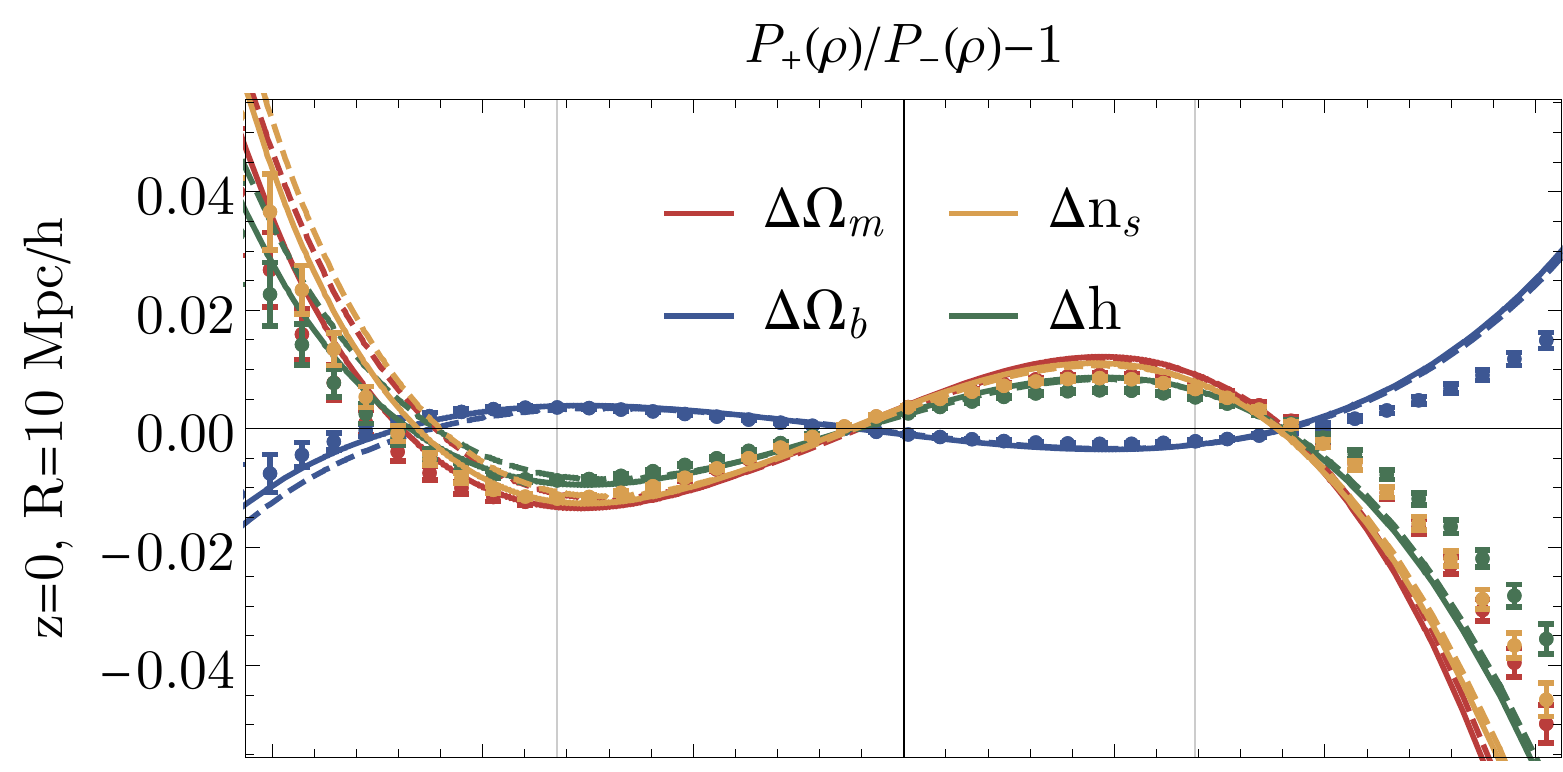} 
\includegraphics[width=1\columnwidth]{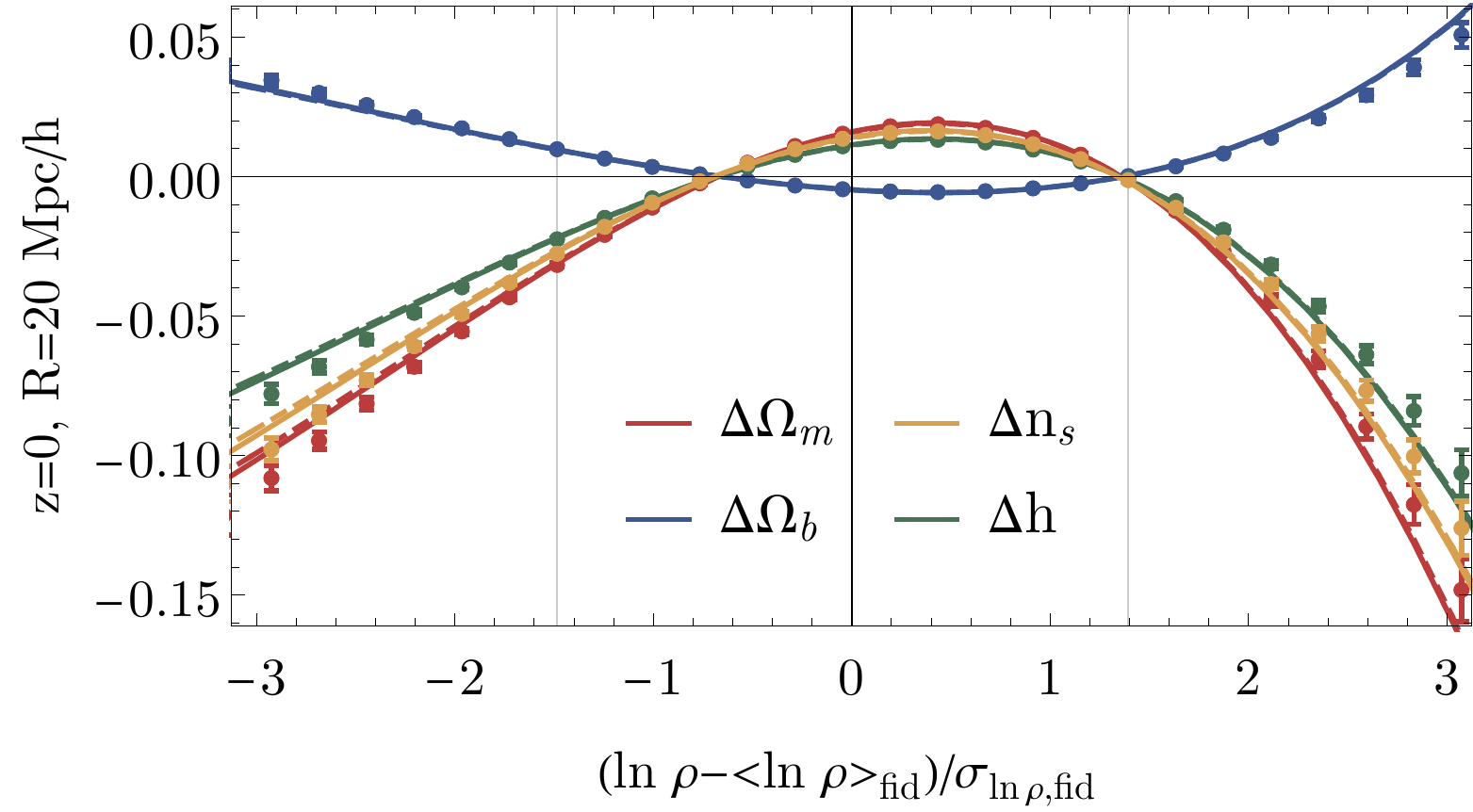}  

   \caption{The measured ratios of the PDFs for the derivative simulations at $z=0$ with radius $R=10,20$ Mpc/h (top and bottom, points with error bars) compared to the predictiona using the measured nonlinear variance at the reference scale as input parameter (solid lines) or predicting the nonlinear variance from the measured nonlinear variance of the fiducial model (dashed lines) for changes in $\Omega_m$ (red), $\Omega_b$ (blue) $n_s$ (yellow) and $h$ (green).} 
   \label{fig:resPDFsimvstheo_Om_Ob_ns_h}
\end{figure}

\begin{figure}
\includegraphics[width=1\columnwidth]{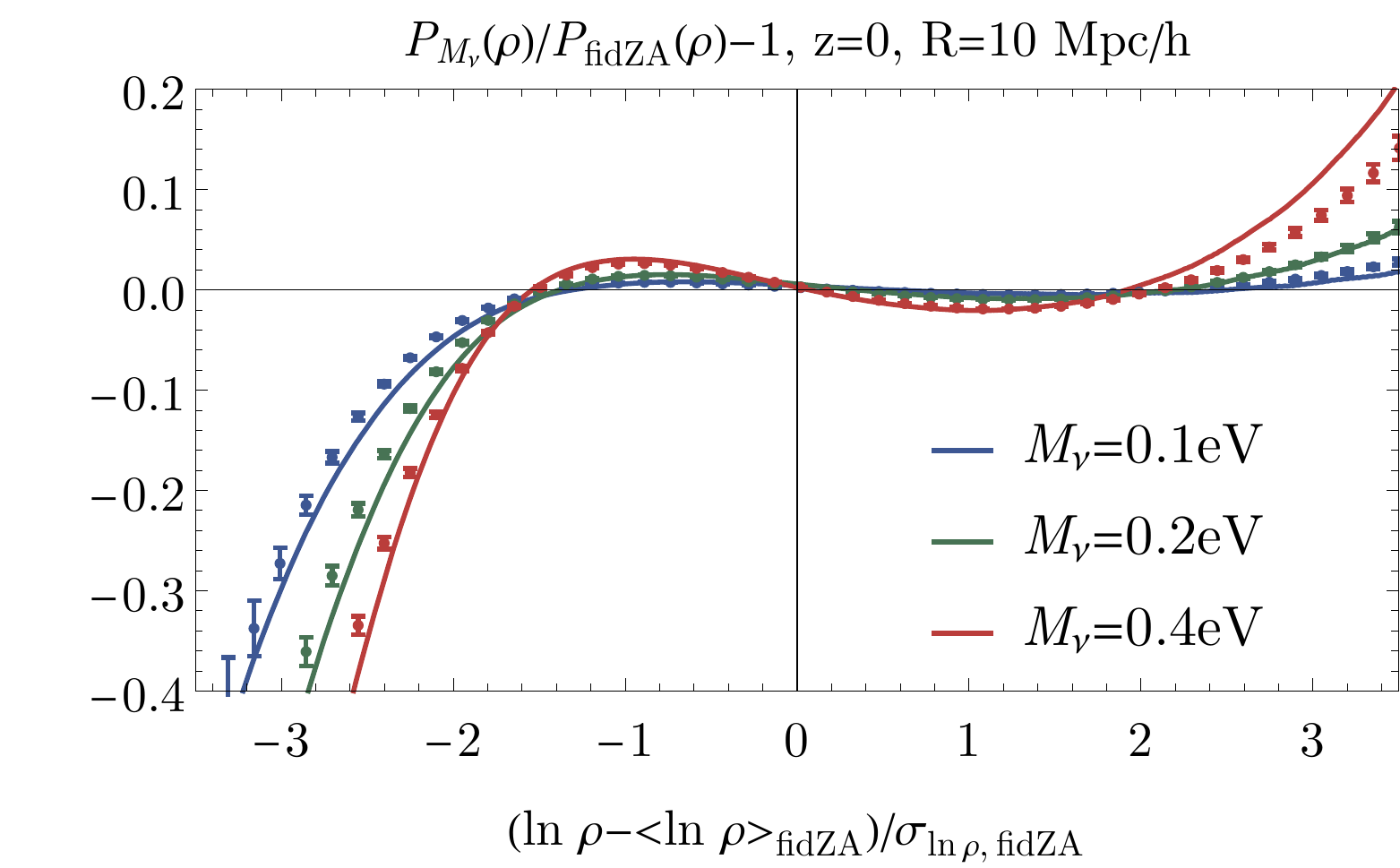}\\
   \caption{Residuals between the total matter density PDFs with massive neutrinos for a total mass $M_\nu=0.1,0.2,0.4$ eV (blue, green, red) at redshift $z=0$ and radius $R=10$ Mpc/h.} 
   \label{fig:resPDFsimvstheo_Mnu}
\end{figure}

In Figures~\ref{fig:resPDFsimvstheo_Om_Ob_ns_h}~and~\ref{fig:resPDFsimvstheo_Mnu}, we show the ratios of matter PDFs when varying $\Lambda$CDM parameters and the total neutrino mass. 

\end{document}